\documentclass[prd,superscriptaddress,twocolumn,amsfonts,amssymb,amsmath,showpacs]{revtex4-2}
\usepackage{bm}
\usepackage{amsfonts}
\usepackage{latexsym}
\usepackage[latin1]{inputenc}
\usepackage{graphicx}
\usepackage{amsmath}
\usepackage{palatino}
\usepackage{caption}
\usepackage{subcaption}
\usepackage{mathpazo}
\usepackage{textcomp}
\usepackage{epsfig}
\usepackage{float}
\usepackage{booktabs}
\usepackage{dcolumn}
\usepackage{booktabs}
\usepackage{multirow}
\usepackage{hyperref}
\hypersetup{colorlinks,citecolor=blue}
\usepackage{amsmath}
\usepackage{xcolor}
\usepackage{orcidlink}
\usepackage{commath}

\usepackage{multirow}

\def\ear{\end{eqnarray}}
\def\beq{\begin{equation}}             \def\earn{\nonumber \end{eqnarray}}
\def\eeq{\end{equation}}               
\def\bear{\begin{eqnarray}}

\begin{document}

\title{Seed models in scalar field cosmology}

\author{I. V. Fomin\orcidlink{0000-0003-1527-914X}}
\email{ingvor@inbox.ru}
\affiliation{Bauman Moscow State Technical University, Russia}


\begin{abstract}
The correspondence of single-field cosmological models based on Einstein gravity to modern observational data is considered. A method is proposed to determine possible types of dynamics based on extreme values of the scalar field. It is shown that within the framework of this approach, it is possible to obtain a limited class of known inflationary models at early times. It is also shown that on large times the proposed approach leads to the $\Lambda$CDM--model in order to describe the dynamics of the second accelerated expansion of the universe. An interpretation of the considered models is presented as a starting point for constructing verifiable cosmological models based on modified gravity theories.
\end{abstract}

\maketitle

\section{Introduction}\label{SEC1}

Modern cosmological models imply the existence of the inflationary stage of accelerated expansion of the early universe during early
times~\cite{Starobinsky:1980te,Guth:1980zm,Linde:1981mu,
Albrecht:1982wi,Linde:1983gd,Linde:1984ir,Belinsky:1985zd,Piran:1986dh,Baumann:2014nda,Chervon:2019sey}. In the simplest models of cosmological inflation based on Einstein gravity, the source of accelerated expansion of the early universe, the source of radiation and matter, and the source of cosmological perturbations that generate both large-scale structure and anisotropy and polarization of the relic radiation is a single scalar field, which is characterized by the type of its evolution and potential~\cite{Mukhanov:1990me,Riotto:2002yw,Brandenberger:2003vk,Straumann:2005mz}.

Currently, there are many inflationary scenarios that differ in the characteristics of the scalar field~\cite{Martin:2013tda,Gron:2018rtj}. Despite the fact that current observational constraints on the values of the parameters of cosmological perturbations allow one to exclude some inflationary models, there remains a fairly large number of different scenarios based on Einstein gravity that satisfy the observational data~\cite{Martin:2013tda,Gron:2018rtj}.
Also, within the framework of various single-field quintessential models, both stages of the accelerated expansion of the universe can be explained (see, for example, in~\cite{Dimopoulos:2001ix,Tsujikawa:2013fta,deHaro:2021swo}).
At the same time, a large number of cosmological models with different types of fields are considered, both based on Einstein gravity~\cite{Johri:2003rh,Singh:2003vx,Astashenok:2012tv,Gomez-Valent:2024tdb,Cai:2009zp,Chervon:2013btx,Chervon:2019nwq,Paliathanasis:2018vru,
Fomin:2021snm,Gibbons:2002md,Barbosa-Cendejas:2017pbo,Nautiyal:2018lyq,Aguirregabiria:2004rc,Babichev:2007dw,Bose:2008ew,Dinda:2023mad,FerreiraJunior:2023qxi} and modified or extended gravity theories~\cite{Sotiriou:2008rp,Nojiri:2010wj,DeFelice:2010aj,DeFelice:2011bh,DeFelice:2011zh,DeFelice:2011jm,
Capozziello:2011et,Nojiri:2017ncd,Ishak:2018his,CANTATA:2021asi,Bahamonde:2021gfp,Odintsov:2023weg}.
Thus, a current problem in studying cosmological models is the development of some general methods that allow one to determine additional conditions on the verifiable models.

Dynamical system analysis makes it possible to limit the number of cosmological models based on the analysis of the stability of solutions of dynamical equations for various potentials of the scalar field (see, for example, in~\cite{Copeland:2006wr,Bahamonde:2017ize,Boehmer:2022wln,Alho:2023pkl,Chakraborty:2024zga,Duchaniya:2024vvc}).
Another way to constrain the possible solutions of the cosmological dynamic equations is to use the Noether symmetry and different conservation quantities ~\cite{deRitis:1990ba,Capozziello:1996bi,Chimento:2002gb,Capozziello:2013bma,Tsamparlis:2018nyo,Giacomini:2021xsx,Bajardi:2022ypn,Bhaumik:2022adi,Piedipalumbo:2023dzg}.
In addition, the approach to solving the problem of correspondence of the inflationary model's parameters to the observational data by optimization methods was proposed in~\cite{Sousa:2023unz}. As a result, various types of optimal potentials of the scalar field were obtained.

However, when constructing and analyzing cosmological models, questions arise about how different inflationary models are related to each other or what types of cosmological dynamics can be realized. The question of the need to modify simple single-field cosmological models based on Einstein gravity is also relevant.

In this paper a new method for analyzing single-field cosmological models based on Einstein gravity is proposed. As a result of applying this approach, the relationship between different well-known inflationary models and their verification by modern observational constraints were considered. Also, based on this approach, a description of the post-inflationary stages of the evolution of the universe was considered as well.

The paper is organized as follows. In Sec. \ref{SEC3} a description of the single-field models of cosmological inflation with a canonical scalar field based on Einstein gravity is presented. A proposed method for analyzing these cosmological models was presented as well. In Sec. \ref{SEC4}, possible types of cosmological dynamics were defined from the extreme values of a scalar field. The corresponding relationships between the slow-roll parameters were also considered.
In Sec. \ref{SECPOT}, a reconstruction of the scalar field potentials was performed for comparison with known inflationary models.
In Sec. \ref{OBSSEC} observational constraints on the parameters of cosmological inflationary models in terms of the slow-roll parameters were presented.
In Sec. \ref{VERSEC} the verification of inflationary models by Planck~\cite{Planck:2018vyg,Galloni:2022mok} and Atacama Cosmology Telescope (ACT)~\cite{ACT:2025fju,ACT:2025tim} observational constraints on the basis of relationships between slow-roll parameters was considered.
It was shown that all inflationary scenarios under consideration do not comply with combined Planck and ACT observational constraints. Nevertheless, it was noted, that exponential SUSY inflation and superconformal $\alpha$-attractor inflation are consistent with the Planck observational constraints.
However, in Sec. \ref{EXACTSEC} it was shown that these inflationary models imply a future singularity and a negative cosmological constant significantly larger than current observational estimates within the $\Lambda$CDM-model.
In Sec. \ref{LATESEC} the possible post-inflationary dynamics of the universe based on the proposed approach was investigated.
By considering the same extreme values of the scalar field for any time of the universe's evolution, it was shown that the current stage of accelerated expansion is described in accordance with the $\Lambda$CDM-model.
Sec. \ref{SEC7} contains a discussion of the results obtained. As a main conclusion, it is proposed to consider the cosmological models under consideration as seed ones for the construction and analysis of more complex verifiable models based on modified gravity theories.

\section{Inflationary models based on Einstein gravity}\label{SEC3}

Within the framework of Einstein gravity theory, inflationary models with single scalar field in the system of units $8\pi G=M^{-2}_{P}=c=\hbar=1$ are described by  action~\cite{Baumann:2014nda,Chervon:2019sey}
\begin{equation}
\label{E}
S=\frac{1}{2}\int d^{4}x\sqrt{-g}[R-g^{\mu\nu}\partial_{\mu}\phi\partial_{\nu}\phi -2V(\phi)],
\end{equation}
where $R$ is the scalar curvature, $\phi$ is a scalar field, $V(\phi)$ is the potential of a scalar field, and $g^{\mu\nu}$ is a metric tensor of a space-time.

We will consider the spatially flat Friedmann-Robertson-Walker (FRW) space-time with metric
\begin{equation}
\label{FRW}
ds^2=-dt^2+a^2(t)\,\delta_{ij} dx^i dx^j,
\end{equation}
where $a=a(t)$ is the scale factor, and $t$ is cosmic time.

The cosmological dynamic equations for action (\ref{E}) and metric (\ref{FRW}) can be written as~\cite{Baumann:2014nda,Chervon:2019sey}
\begin{eqnarray}
 \label{DE1}
&&3H^{2}=\frac{1}{2}\dot{\phi}^{2}+V(\phi)\equiv\rho_{\phi},\\
\label{DE2}
&&-3H^{2}-2\dot{H}=\frac{1}{2}\dot{\phi}^{2}-V(\phi)\equiv p_{\phi},\\
\label{DE3}
&& \ddot{\phi} + 3H\dot{\phi} +V'_{\phi}= 0,
\end{eqnarray}
where $\rho_{\phi}$ and $p_{\phi}$ are the energy density and the pressure of a scalar field, $V'_{\phi}=dV/d\phi$, and the Hubble parameter $H=\dot{a}/a$.

It is well known that in this system there are only two independent equations~\cite{Chervon:2019sey}.

Independent dynamic equations can be written as
\begin{eqnarray}
\label{E1m}
&& V(\phi)=3H^{2}+\dot{H}, \\
\label{E2m}
&&X=\frac{1}{2}\dot{\phi}^{2}=-\dot{H}.
\end{eqnarray}
where $H=\dot{a}/a$ is the Hubble parameter, and $X$ is the kinetic energy of a scalar field.

In addition, equations (\ref{E1m})--(\ref{E2m}) can be presented as
\begin{eqnarray}
\label{E1mF}
&& V(\phi)=3H^{2}-2\left[H'_{\phi}\right]^{2}, \\
\label{E2mF}
&&\dot{\phi}=-2H'_{\phi},
\end{eqnarray}
based on expression
\begin{eqnarray}
\label{ISB}
&&\dot{H}=\frac{dH}{d\phi}\dot{\phi}=H'_{\phi}\dot{\phi},
\end{eqnarray}
in accordance with the Ivanov--Salopec--Bond approach~\cite{Chervon:2017kgn}.

The simplest solutions of equations (\ref{E1m})--(\ref{E2m}) can be written as
\begin{eqnarray}
\label{DESITTER}
&&H=const,~~~\phi=const,~~~V=const.
\end{eqnarray}

These solutions correspond to the pure de Sitter stage with $H=const\neq0$ and the exponential scale factor $a(t)\sim\exp(Ht)$, where non-zero flat potential $V=const\neq0$ can be interpreted as a cosmological constant~\cite{Baumann:2014nda,Chervon:2019sey}.
On the other hand, solutions $H=0$, $\phi=0$ and $V=0$ correspond to a stationary universe $a=const$.
Thus, the study of cosmological models can be considered as the analysis of the deviations from a stationary universe, and deviations from the pure de Sitter stage as well.

Due to dynamic equations (\ref{E1m})--(\ref{E2m}) and solutions (\ref{DESITTER}) the scalar field potential can be determined up to the cosmological constant $V\rightarrow V+\Lambda$. The cosmological constant $\Lambda$ can be associated with  the vacuum energy density $\rho_{\Lambda}=\rho_{vac}$~\cite{Rugh:2000ji,Padmanabhan:2002ji,Sahni:2002kh,Nobbenhuis:2004wn,Martin:2012bt,Sola:2013gha}, where $\rho_{vac}=\Lambda$ in the chosen system of units.

We will also consider the following conditions for the Hubble parameter~\cite{Baumann:2014nda,Chervon:2019sey}:
\begin{itemize}
\item $H>0$ corresponds to the expansion of the universe;
\item $\ddot{a}>0$ or $\frac{\ddot{a}}{a}=H^{2}+\dot{H}>0$ corresponds to accelerated expansion of the universe;
\item $\dot{H}<0$ corresponds to the possibility of exit from the accelerated expansion of the early universe.
\end{itemize}

Equations (\ref{DE1}) and (\ref{DE3}) can be represented as
\begin{eqnarray}
\label{SLOWROLLP1}
&&V(\phi)=3H^{2}\left(1-\frac{\epsilon}{3}\right),\\
\label{SLOWROLLP2}
&&V'_{\phi}=-3H\dot{\phi}\left(1-\frac{\delta}{3}\right),
\end{eqnarray}
where the slow-roll parameters $\epsilon$ and $\delta$ are
\begin{eqnarray}
\label{epsilonex}
&&\epsilon=-\frac{\dot{H}}{H^{2}}=\frac{\dot{\phi}^{2}}{2H^{2}}=2\left(\frac{H'_{\phi}}{H}\right)^{2},\\
\label{deltanex}
&&\delta=-\frac{\ddot{H}}{2H\dot{H}}=-\frac{\ddot{\phi}}{H\dot{\phi}}=2\frac{H''_{\phi}}{H}.
\end{eqnarray}

Slow-roll conditions during inflation can be formulated as follows~\cite{Baumann:2014nda,Chervon:2019sey}
\begin{eqnarray}
\label{CSLOWROLL}
&&\epsilon\ll1,~~~~|\delta|\ll1,
\end{eqnarray}
for any inflationary model.

The end of the inflationary stage of the accelerated expansion of the early universe is defined by condition $\epsilon=\epsilon_{e}=1$. Also, for the positive potentials of a scalar field $V\geq0$ from (\ref{SLOWROLLP1}) it follows that the maximal value of the first slow-roll parameter $\epsilon_{max}=3$ corresponds to $V=0$.

The slow-roll parameters can be defined in different ways~\cite{Liddle:1994dx,Schwarz:2001vv,Leach:2002ar,Kinney:2005vj}, for example, as
\begin{eqnarray}
\label{CSLOWROLL2A}
&&\epsilon_{1}=-\frac{\dot{H}}{H^{2}},~~~~\epsilon_{n+1}=\frac{\dot{\epsilon}_{n}}{H\epsilon_{n}},
\end{eqnarray}
where $\epsilon=\epsilon_{1}$ and $\delta=\epsilon_{1}-\frac{1}{2}\epsilon_{2}$.

Under the slow-roll conditions (\ref{CSLOWROLL}) from (\ref{SLOWROLLP1})--(\ref{SLOWROLLP2}) one has
\begin{eqnarray}
\label{E1mSR}
&& V(\phi)\simeq3H^{2},\\
\label{E2mSR}
&&3H\dot{\phi}\simeq-V'_{\phi}.
\end{eqnarray}

In addition, equations (\ref{E1mSR})--(\ref{E2mSR}) can be represented as
\begin{eqnarray}
\label{E1SR2}
&& V(\phi)\simeq3H^{2},\\
\label{E2SR3}
&&\frac{1}{2}\dot{\phi}^{2}\simeq-\dot{H}.
\end{eqnarray}

The other way to obtain equations (\ref{E1mSR})--(\ref{E2mSR}) from the cosmological dynamic equation (\ref{DE1}) and (\ref{DE3}) is to use the following slow-roll conditions~\cite{Baumann:2014nda,Chervon:2019sey}
\begin{eqnarray}
\label{E1mSRT2}
&&\frac{1}{2}\dot{\phi}^{2}\ll V,~~~~~ \ddot{\phi}\ll 3H\dot{\phi}.
\end{eqnarray}

In this case, the slow-roll parameters can be approximately defined by the potential of a scalar field as~\cite{Chervon:2019sey}
\begin{eqnarray}
\label{FIRST4}
&&\epsilon\simeq\frac{1}{2}\left(\frac{V'_{\phi}}{V}\right)^{2},~~~
\delta\simeq\frac{V''_{\phi}}{V}-\frac{1}{2}\left(\frac{V'_{\phi}}{V}\right)^{2}.
\end{eqnarray}

The other approximate expressions for the slow-roll parameters are~\cite{Baumann:2014nda}
\begin{eqnarray}
\label{SRSB1}
&&\epsilon\simeq\frac{1}{2}\left(\frac{V'_{\phi}}{V}\right)^{2},~~~\eta\simeq\frac{V''_{\phi}}{V}.
\end{eqnarray}

Equations (\ref{E1mSR})--(\ref{E2mSR}) and expressions (\ref{FIRST4})--(\ref{SRSB1}) are often used to construct and analyze models of cosmological inflation~\cite{Baumann:2014nda,Martin:2013tda,Gron:2018rtj}.

It should also be noted that in modern cosmology the natural system of units $c=\hbar=1$ is widely used~\cite{Baumann:2014nda}. The dimensionless parameters of the inflationary models in the system of units under consideration $8\pi G=M^{-2}_{P}=c=\hbar=1$ can be defined in terms of the reduced Planck mass in the natural system of units as
\begin{eqnarray}
\label{NU}
&&\tilde{V}=V\times\left[M^{4}_{P}\right], \tilde{H}=H\times\left[M_{P}\right], \tilde{\phi}=\phi\times\left[M_{P}\right],\\
\label{NU1}
&&\tilde{t}=t\times\left[M^{-1}_{P}\right], \tilde{\Lambda}=\Lambda\times\left[M^{2}_{P}\right], \tilde{\rho}_{\Lambda}=\Lambda\times\left[M^{4}_{P}\right],
\end{eqnarray}
where $M_{P}=\left(8\pi G\right)^{-1/2}=2.4\times10^{18}$ GeV is the reduced Planck mass~\cite{Baumann:2014nda}.

For example, the value of the cosmological constant in the natural system of units is estimated as follows~\cite{Rugh:2000ji,Padmanabhan:2002ji,Sahni:2002kh,Nobbenhuis:2004wn,Martin:2012bt,Sola:2013gha}
\begin{eqnarray}
\label{COSMCONST}
&&\tilde{\Lambda}\simeq5\times10^{-84}\left(GeV\right)^{2},
\end{eqnarray}
while this value of the cosmological constant in chosen system of units is
\begin{eqnarray}
\label{COSMCONST2}
&&\Lambda=\rho_{\Lambda}=\frac{\tilde{\Lambda}}{M^{2}_{P}}\simeq9\times10^{-121}.
\end{eqnarray}

When returning to the natural system of units, one has the following value of the energy density of the cosmological constant
\begin{eqnarray}
\label{COSMCONST4}
&&\tilde{\rho}_{\Lambda}=\Lambda\times\left[M^{4}_{P}\right]\simeq3\times10^{-47}\left(GeV\right)^{4}.
\end{eqnarray}

The other parameters can also be translated from the system of units under consideration into the natural system of units by means of relations (\ref{NU})--(\ref{NU1}) in the same way.

In this paper, the analysis of the single-field cosmological models will be presented at different successive levels, namely:
\begin{enumerate}
\item Determination of possible types of cosmological dynamics;
\item Determination of the corresponding potentials of the scalar field in the slow-roll approximation and comparison potential obtained with those for known inflationary models;
\item Comparison of the parameters of cosmological perturbations in the obtained inflationary models with observational constraints;
\item Construction and analysis of exact cosmological solutions for inflationary models satisfying observational constraints;
\item Construction and analysis of the models of post-inflationary stages including the second accelerated expansion of the universe based on the obtained types of cosmological dynamics.
\end{enumerate}

This approach leads to the possibility of exclusion of single-field cosmological models that do not satisfy modern observational constraints.

\section{The types of the cosmological dynamics}\label{SEC4}

From exact expressions for slow-roll parameters (\ref{epsilonex})--(\ref{deltanex}) it follows that
\begin{eqnarray}
\label{RELATIONED}
&&\frac{\delta^{2}}{\epsilon}=-\frac{1}{\dot{H}}\left(\frac{\ddot{H}}{2\dot{H}}\right)^{2}=
2\left(\frac{\ddot{\phi}}{\dot{\phi}^{2}}\right)^{2}.
\end{eqnarray}

Slow-roll conditions (\ref{E1mSRT2}) lead to uncertainty in the estimate of ratio (\ref{RELATIONED}).
For this reason, an additional condition of minimal variation of a scalar field $\delta\phi=0$ can be considered.
This condition was proposed in~\cite{Zhuravlev:1998ff} in the context of the variational interpretation of the slow-roll inflationary regime.
We consider the other interpretation of this condition in order to obtain cosmological dynamics restricted by the extreme values of a scalar field.

\subsection{Cosmological dynamics corresponding to extreme values of the scalar field}

From \eqref{E2m}, a scalar field can be defined as
\begin{eqnarray}
\label{VAR1}
\nonumber
&&\phi[f(t)]=\int^{t_{e}}_{t_{i}}\mathcal{L}\left(t,f,\dot{f},\ddot{f},...,f^{(k)}\right)dt=\\
&&=\pm\int^{t_{e}}_{t_{i}}\sqrt{-2\dot{H}}dt,
\end{eqnarray}
where $t_{i}$ and $t_{e}$ are the times of beginning and end of the inflationary stage, $f=f(t)$ is some function of cosmic time, and $\dot{f}=df(t)/dt$.

Thus, the condition of minimal variation of a scalar field can be formulated as
\begin{eqnarray}
\label{VAR1A}
\delta\phi[f(t)]=\int^{t_{e}}_{t_{i}}\delta\mathcal{L}\left(t,f,\dot{f},\ddot{f},...,f^{(n)}\right)dt=0.
\end{eqnarray}

Due to expression (\ref{VAR1A}) corresponding Euler-Lagrange equations can be written as follows~\cite{Pons:1988tj,Baptista:2020adz}
\begin{eqnarray}
\label{VAR2N}
&&\sum^{n}_{k=0}\left(-1\right)^{k}\frac{d^{k}}{dt^{k}}\left(\frac{\partial\mathcal{L}}{\partial f^{(k)}}\right)=0.
\end{eqnarray}

In second order, for the Lagrangian
\begin{eqnarray}
\label{VARNL}
&&\mathcal{L}=\mathcal{L}\left(t,f,\dot{f},\ddot{f}\right)\equiv\sqrt{-2\dot{H}},
\end{eqnarray}
equation (\ref{VAR2N}) is reduced to the following form
\begin{eqnarray}
\label{VAR2}
&&\frac{\partial\mathcal{L}}{\partial f}-\frac{d}{dt}\left(\frac{\partial\mathcal{L}}{\partial \dot{f}}\right)+
\frac{d^{2}}{dt^{2}}\left(\frac{\partial\mathcal{L}}{\partial \ddot{f}}\right)=0.
\end{eqnarray}

In order to determine the possible types of the Lagrangian in explicit form, different parameters of cosmological dynamics can be used.
Also, the second-order Lagrangian (\ref{VARNL}) must be considered, since equations (\ref{DE1})-(\ref{DE3}) contain dynamical parameters with derivatives, which are no higher than second order.

\subsubsection{The Lagrangian in terms of the Hubble parameter}

For Lagrangian \eqref{VARNL} in terms of the Hubble parameter
\begin{eqnarray}
\label{VAR3}
&&\mathcal{L}=\mathcal{L}\left(t,H,\dot{H}\right)=\sqrt{-2\dot{H}},
\end{eqnarray}
where $f=H$, one has
\begin{eqnarray}
\label{VAR4}
&&\frac{\partial\mathcal{L}}{\partial \dot{f}}=\frac{\partial}{\partial \dot{H}}\left(\sqrt{-2\dot{H}}\right)=
-\frac{1}{\sqrt{-2\dot{H}}}.
\end{eqnarray}

From equation (\ref{VAR2}) for $\dot{H}\neq0$ we obtain
\begin{eqnarray}
\label{VAR5}
&&\ddot{H}=0,\\
\label{VAR5A}
&&H(t)=-c_{1}t+c_{2},
\end{eqnarray}
where condition $\dot{H}=-c_{1}<0$ is satisfied for any $c_{1}>0$.

The scale factor corresponding to (\ref{VAR5A}) is
\begin{eqnarray}
\label{VAR6}
&&a(t)\sim\exp\left(c_{2}t-\frac{c_{1}}{2}t^{2}\right),
\end{eqnarray}
where $c_{1}$ and $c_{2}$ are the integration constants.

The consequence of (\ref{epsilonex})--(\ref{deltanex}) and (\ref{VAR5A}) is
\begin{eqnarray}
\label{VAR5ASR}
&&\delta=0,~~~~~~~~~~\left(\delta^{2}/\epsilon\right)=0.
\end{eqnarray}

These expressions correspond to the first type of extreme values of a scalar field.

\subsubsection{The Lagrangian in terms of the e-folds number and scale factor}

Lagrangian (\ref{VARNL}) can be defined in terms of the scale factor $f=a$ as
\begin{eqnarray}
\label{VAR7}
&&\mathcal{L}=\mathcal{L}\left(t,a,\dot{a},\ddot{a}\right)=\sqrt{2\frac{\dot{a}^{2}}{a^{2}}-2\frac{\ddot{a}}{a}},
\end{eqnarray}
with corresponding non-zero derivatives
\begin{eqnarray}
\label{VAR8}
&&\frac{\partial\mathcal{L}}{\partial f}=\frac{\partial\mathcal{L}}{\partial a}=
\frac{\left(\frac{\ddot{a}}{a}-2\frac{\dot{a}^{2}}{a}\right)}{\sqrt{2\frac{\dot{a}^{2}}{a^{2}}
-2\frac{\ddot{a}}{a}}},\\
\label{VAR9}
&&\frac{\partial\mathcal{L}}{\partial \dot{f}}=\frac{\partial\mathcal{L}}{\partial \dot{a}}=
\frac{2\dot{a}}{a^{2}\sqrt{2\frac{\dot{a}^{2}}{a^{2}}-2\frac{\ddot{a}}{a}}},\\
\label{VAR10}
&&\frac{\partial\mathcal{L}}{\partial \ddot{f}}=\frac{\partial\mathcal{L}}{\partial \ddot{a}}=
-\frac{1}{a\sqrt{2\frac{\dot{a}^{2}}{a^{2}}-2\frac{\ddot{a}}{a}}}.
\end{eqnarray}

Thus, equation (\ref{VAR2}) for Lagrangian (\ref{VAR7}) can be written as
\begin{eqnarray}
\label{VAR11}
\nonumber
&&2a(\ddot{a}a-\dot{a}^{2})a^{(IV)}-3a^{2}\dddot{a}^{2}\\
&&+2\dot{a}(5a\ddot{a}-2\dot{a}^{2})\dddot{a}
+3(\dot{a}^{2}-2a\ddot{a})\ddot{a}^{2}=0,
\end{eqnarray}
where $a^{(IV)}\equiv\frac{d^{4}a(t)}{dt^{4}}$.

On the other hand, Lagrangian (\ref{VARNL}) can be considered in terms of the e-folds number~\cite{Baumann:2014nda,Chervon:2019sey}
\begin{eqnarray}
\label{EFOLDS}
&&N(t)=\ln\left(\frac{a(t)}{a_{k}}\right),~~~~\dot{N}=H,
\end{eqnarray}
where $a_{k}$ is the scale factor for a certain cosmic time.

Lagrangian in terms of the e-folds number $f=N$ is
\begin{eqnarray}
\label{EFOLDS1}
&&\mathcal{L}=\mathcal{L}\left(t,N,\dot{N},\ddot{N}\right)=\sqrt{-2\ddot{N}}.
\end{eqnarray}

In this case, the one non-zero derivative is
\begin{eqnarray}
\label{EFOLDS2}
&&\frac{\partial\mathcal{L}}{\partial \ddot{f}}=\frac{\partial}{\partial \ddot{N}}\left(\sqrt{-2\ddot{N}}\right)=
-\frac{1}{\sqrt{-2\ddot{N}}}.
\end{eqnarray}

Thus, from (\ref{VAR2}) follows the equation
\begin{eqnarray}
\label{EFOLDS3}
&&2\ddot{\xi}\xi-3\dot{\xi}^{2}=0,
\end{eqnarray}
where $\xi=\ddot{N}=\dot{H}$.

The solution of this equation in terms of the Hubble parameter can be written as
\begin{eqnarray}
\label{EFOLDS4}
&&H(t)=c_{3}+\frac{c_{6}}{c_{4}t+c_{5}}.
\end{eqnarray}

The corresponding scale factor is
\begin{eqnarray}
\label{EFOLDS5}
&&a(t)\sim\exp(c_{3} t)(c_{4}t+c_{5})^{c_{6}/c_{4}},
\end{eqnarray}
where $c_{3}$, $c_{4}$, $c_{5}$ and $c_{6}$ are the integration  constants.

In addition, the fourth-order equation (\ref{VAR11}) can be reduced to the second-order equation (\ref{EFOLDS3}) on the basis of transformation
\begin{eqnarray}
\label{VARTRANSFORM}
&&a(t)=b_{1}e^{b_{2}t}\exp\left[\int\left(\int\xi(t)dt\right)dt\right],
\end{eqnarray}
where $b_{1}$ and $b_{2}$ are some constants.

Thus, scale factor (\ref{EFOLDS5}) is the solution of equation (\ref{VAR11}) as well.

It should be noted that transformation (\ref{VARTRANSFORM}) can be obtained by the relation
\begin{eqnarray}
\label{VARTRANSFORM2}
&&\xi=\dot{H}=\frac{\ddot{a}}{a}-\frac{\dot{a}^{2}}{a^{2}}.
\end{eqnarray}

Also, without loss of generality, solutions (\ref{EFOLDS4})--(\ref{EFOLDS5}) can be redefined as follows
\begin{eqnarray}
\label{EXPL}
&&H(t)=\lambda+\frac{s}{t+t_{0}},\\
\label{EXPL2}
&&a(t)\sim\exp(\lambda t)(t+t_{0})^{s}.
\end{eqnarray}

From (\ref{EXPL}) it follows that condition
\begin{eqnarray}
\label{FOC3}
&&\dot{H}=-\frac{s}{(t+t_{0})^{2}}=-\frac{1}{s}\left(H-\lambda\right)^{2}<0,
\end{eqnarray}
is satisfied for any positive non-zero constant $s>0$.

From (\ref{epsilonex})--(\ref{deltanex}) and (\ref{EXPL}) we get the relation
\begin{eqnarray}
\label{VAR13}
&&\frac{\delta^{2}}{\epsilon}=\frac{1}{s}=const,
\end{eqnarray}
for the second type of extreme values of a scalar field.

Therefore, the relation $\left(\delta^{2}/\epsilon\right)=const$ is conserved for the extreme values of a scalar field.

\subsection{Cosmological dynamics from dependence $\dot{H}=\dot{H}(H)$}

In general case, unknown cosmological dynamics can be considered on the basis of the following dependence $\dot{H}=\dot{H}(H)$.
Taking into account non-stationarity of the universe, dependence $\dot{H}=\dot{H}(H)$ can be expanded in a series around $H=0$ as
\begin{eqnarray}
\label{QDSRzero}
&&\dot{H}_{(0)}=-\sum^{\infty}_{k=1}\zeta_{k}H^{k}=-\zeta_{1}H-\zeta_{2}H^{2}+...,
\end{eqnarray}
where $\zeta_{k}$ are the constant coefficients of this expansion, and $\dot{H}(H=0)=0$.

Taking into account the accelerated expansion of the universe, for the quasi-de Sitter regime, the expansion of dependence $\dot{H}=\dot{H}(H)$ can be expanded in a series around the de Sitter stage $H=const\neq0$ for some specific value of the constant $\lambda_{(k)}$ as
\begin{eqnarray}
\label{QDSR}
&&\dot{H}_{(k)}=-\sum^{\infty}_{k=1}\tilde{\xi}_{k(k)}(H-\lambda_{(k)})^{k},
\end{eqnarray}
where $\dot{H}(H=\lambda_{(k)})=0$, and $\tilde{\xi}_{k(k)}$ are the constant coefficients of expansion.

In addition, it is necessary to sum expansions (\ref{QDSR}) over different values of the constant $\lambda_{(k)}$.
Taking into account (\ref{QDSRzero})--(\ref{QDSR}) and binomial expansion
\begin{eqnarray}
&& (H-\lambda)^{l}= \sum_{j=0}^l {l \choose j} \left(-1\right)^{l-j} H^j\lambda^{l-j},
\end{eqnarray}
the final form of the derivative of the Hubble parameter $\dot{H}$ can be written as
\begin{eqnarray}
\label{QDSR2}
\nonumber
&&\dot{H}=\sum^{\infty}_{(k)=0}\dot{H}_{(k)}=
-\sum^{\infty}_{k=0}\xi_{k}(H-\lambda_{k})^{k}\\
&&=-\xi_{0}-\xi_{1}(H-\lambda_{1})-\xi_{2}(H-\lambda_{2})^{2}+...,
\end{eqnarray}
where $\xi_{k}$ are the new constant coefficients in expansion.

For some $k$-order term expansion (\ref{QDSR2}) implies the following equation
\begin{eqnarray}
\label{QDSRA}
&&\dot{H}=-\xi_{k}(H-\lambda_{k})^{k}.
\end{eqnarray}

For $k\neq1$ from (\ref{QDSRA}) corresponding Hubble parameter can be defined as
\begin{eqnarray}
\label{QDSRB}
&&H(t)=\lambda_{k}+\left[\xi_{k}(k-1)(t+c)\right]^{\frac{1}{1-k}},
\end{eqnarray}
where $c$ is the constant of integration.

From expressions (\ref{epsilonex})-(\ref{deltanex}) and (\ref{QDSRB}) it follows that
\begin{eqnarray}
\label{QDSRC}
&&\frac{\delta^{2}}{\epsilon}=\frac{k^{2}}{4}\left(\xi_{k}\right)^{\frac{1}{k-1}}
\left[\xi_{k}(k-1)(t+c)\right]^{\frac{k-2}{k-1}}.
\end{eqnarray}

For $k=0$ expressions (\ref{QDSRB})--(\ref{QDSRC}) correspond to cases (\ref{VAR5A}) and (\ref{VAR5ASR}) where $\xi_{0}=c_{1}>0$.
For $k=2$ expressions (\ref{QDSRB})--(\ref{QDSRC}) correspond to (\ref{EXPL}) and (\ref{VAR13}) where $\xi_{2}=\frac{1}{s}>0$.
The case $k=1$ corresponds to the intermediate regime of an accelerated expansion of the universe between ones which are defined by the extreme values of a scalar field.

Since the scalar field takes extreme values up to the second order of expansion (\ref{QDSR2}), we can neglect higher order terms. It is equivalent to condition $\xi_{k}=0$ for $k>2$.
Therefore, we can break expansion (\ref{QDSR2}) at the second order for single-filed inflationary models based on Einstein gravity theory.

Thus, taking into account the extreme values of a scalar field, equation
\begin{eqnarray}
\label{QDSRCE}
&&\dot{H}=-\xi_{0}-\xi_{1}(H-\lambda_{1})-\xi_{2}(H-\lambda_{2})^{2},
\end{eqnarray}
can be used instead of expansion (\ref{QDSR2}) in order to describe possible types of cosmological dynamics.

For de Sitter stage $H=\lambda=const$ condition $\dot{H}(H=\lambda)=0$ is satisfied under condition
\begin{eqnarray}
\label{QDSRCEREL}
&&\xi_{0}+\xi_{1}(\lambda-\lambda_{1})+\xi_{2}(\lambda-\lambda_{2})^{2}=0.
\end{eqnarray}

Thus, the possible types of cosmological dynamics can be obtained from equation (\ref{QDSRCE}).

\subsection{The possible types of the cosmological dynamics}

Equations (\ref{epsilonex})--(\ref{deltanex}) and  (\ref{QDSRCE}) correspond to the following expressions for the slow-roll parameters
\begin{eqnarray}
\label{epsilonex2}
&&\epsilon=-\frac{\dot{H}}{H^{2}}=\mu_{0}+\frac{\mu_{1}}{H}+\frac{\mu_{2}}{H^{2}},\\
\label{deltanex2}
&&\delta=-\frac{\ddot{H}}{2H\dot{H}}=\mu_{0}+\frac{\mu_{1}}{2H},
\end{eqnarray}
where
\begin{eqnarray}
\label{mu0}
&&\mu_{0}=\xi_{2},\\
\label{mu1}
&&\mu_{1}=\xi_{1}-2\lambda_{2}\xi_{2},\\
\label{mu2}
&&\mu_{2}=\lambda^{2}_{2}\xi_{2}-\lambda_{1}\xi_{1}+\xi_{0}.
\end{eqnarray}
or
\begin{eqnarray}
\label{ximu0}
&&\xi_{0}=(2\lambda_{1}\lambda_{2}-\lambda^{2}_{2})\mu_{0}+\lambda_{1}\mu_{1}+\lambda_{2},\\
\label{ximu1}
&&\xi_{1}=2\lambda_{2}\mu_{0}+\mu_{1},\\
\label{ximu2}
&&\xi_{2}=\mu_{0}.
\end{eqnarray}

Also, expressions for slow-roll parameters (\ref{epsilonex})--(\ref{deltanex}) and (\ref{epsilonex2})--(\ref{deltanex2}) lead to relations
\begin{eqnarray}
\label{epsilonex3}
&&\dot{\epsilon}=2H\epsilon(\epsilon-\delta),\\
\label{epsilonex4}
&&\ddot{\epsilon}=\frac{\dot{\epsilon}^{2}}{\epsilon}+\frac{\dot{\epsilon}(\dot{\epsilon}-2\dot{\delta})}{2(\epsilon-\delta)},\\
\label{HED}
&&H^{2}=\frac{\mu_{2}}{\epsilon-2\delta+\mu_{0}},\\
\label{deltanex3}
&&\dot{\delta}=\frac{\mu_{1}}{2}\epsilon,\\
\label{deltanex4}
&&\ddot{\delta}=\frac{\mu_{1}}{2}\dot{\epsilon}.
\end{eqnarray}

It should be noted that for decaying Hubble parameter $\dot{H}<0$ (excluding case $\mu_{1}=\mu_{2}=0$) first slow-roll parameter (\ref{epsilonex2}) is the growing function of cosmic time. This property provides the possibility of exit from inflation. The condition of the end of inflation can be defined as $\epsilon_{e}=1$ according to expression
\begin{eqnarray}
\label{ACCELERATION}
&&\ddot{a}=a\left(H^{2}+\dot{H}\right)=aH^{2}\left(1-\epsilon\right).
\end{eqnarray}

The following possible relations between slow-roll parameters can be obtained on the basis of expressions (\ref{epsilonex2})--(\ref{deltanex2}):

$\text{I}.$ Slow-roll parameters are
\begin{eqnarray}
\label{I1}
&&\epsilon=\frac{\mu_{2}}{H^{2}},~~~~\delta=0,
\end{eqnarray}
under conditions
\begin{eqnarray}
\label{I2}
&&\mu_{1}=\mu_{0}=0.
\end{eqnarray}

Hubble parameter is defined by expression (\ref{VAR5A}) due to (\ref{I1}).
Also, based on expression (\ref{QDSRCE}), Hubble parameter (\ref{VAR5A}) and scale factor can be written as
\begin{eqnarray}
\label{F1}
&&H(t)=-\xi_{0}t+\eta_{0},\\
\label{F2}
&&a(t)\sim\exp\left(\eta_{0}t-\frac{\xi_{0}}{2}t^{2}\right),
\end{eqnarray}
where $\eta_{0}$ is a positive constant.

This case corresponds to the first type of extreme values of a scalar field.

$\text{II}.$ Slow-roll parameters are
\begin{eqnarray}
\label{II1}
&&\epsilon=\delta=\mu_{0}=const,
\end{eqnarray}
under conditions
\begin{eqnarray}
\label{II2}
\mu_{1}=\mu_{2}=0.
\end{eqnarray}

For the slow-roll parameters (\ref{II1}) and equation (\ref{QDSRCE}) taking into account (\ref{mu0})--(\ref{ximu2}) following Hubble parameter and scale factor
\begin{eqnarray}
\label{EX21}
&&H(t)=\frac{1}{\mu_{0}t},\\
\label{EX22}
&&a(t)\sim t^{1/\mu_{0}},
\end{eqnarray}
can be obtained.

$\text{III}.$ For $\mu_{0}\neq0$, $\mu_{1}\neq0$ and $\mu_{2}\neq0$, from (\ref{epsilonex2})--(\ref{deltanex2}), following relation between slow-roll parameters can be defined
\begin{eqnarray}
\label{RELSRPAR}
\nonumber
&&\epsilon=\frac{4\mu_{2}}{\mu^{2}_{1}}\delta^{2}+2\left(1-\frac{4\mu_{2}\mu_{0}}{\mu^{2}_{1}}\right)\delta\\
&&-\mu_{0}\left(1-\frac{4\mu_{2}\mu_{0}}{\mu^{2}_{1}}\right),
\end{eqnarray}
where $\epsilon\neq const$ and $\delta\neq const$.

For relation (\ref{RELSRPAR}) between slow-roll parameters, equation (\ref{QDSRCE}) and (\ref{mu0})--(\ref{ximu2})
two classes of cosmological dynamics can be considered.

The first class corresponds to the trigonometric Hubble parameters and scale factors, namely
\begin{eqnarray}
\label{GEN21}
&&H(t)=B_{1}-B_{2}\tan\left(B_{3}t\right),\\
\label{GEN22}
&&a(t)\sim e^{B_{1}t}\left[\cos(B_{3}t)\right]^{B_{2}/B_{3}},\\
\label{GEN21E}
&&\epsilon=\frac{B_{2}B_{3}}{\left[B_{1}\cos(B_{3}t)-B_{2}\sin(B_{3}t)\right]^{2}},\\
\label{GEN21D}
&&\delta=-\frac{B_{3}\sin(B_{3}t)}{B_{1}\cos(B_{3}t)-B_{2}\sin(B_{3}t)},
\end{eqnarray}
and
\begin{eqnarray}
\label{GEN23}
&&H(t)=B_{1}+B_{2}\cot\left(B_{3}t\right),\\
\label{GEN24}
&&a(t)\sim e^{B_{1}t}\left[\sin(B_{3}t)\right]^{B_{2}/B_{3}},\\
\label{GEN21E2}
&&\epsilon=\frac{B_{2}B_{3}}{\left[B_{1}\sin(B_{3}t)+B_{2}\cos(B_{3}t)\right]^{2}},\\
\label{GEN21D2}
&&\delta=\frac{B_{3}\cos(B_{3}t)}{B_{1}\sin(B_{3}t)+B_{2}\cos(B_{3}t)}.
\end{eqnarray}

The presented types of the Hubble parameters correspond to relation (\ref{RELSRPAR}) under conditions
\begin{eqnarray}
\label{GEN25A}
&&\mu_{0}=\frac{B_{3}}{B_{2}},~~~\mu_{2}=\frac{B_{2}\mu^{2}_{1}
\left(B^{2}_{1}+B^{2}_{2}\right)}{4B^{2}_{1}B_{3}},
\end{eqnarray}
where $B_{1}$, $B_{2}$, $B_{3}$ are positive constants.

Expression (\ref{GEN22}) implies that $a=0$ for $B_{3}t=\frac{\pi}{2}n$ ($n=1,2,3,...$), and expression (\ref{GEN24}) leads to $a=0$ for $B_{3}t=\pi n$ ($n=0,1,2,3,...$).
Since these types of cosmological dynamics lead to periodical singularities, we will not consider them to describe relevant cosmological models.

The second class of cosmological dynamics can be defined by following Hubble parameter and scale factor
\begin{eqnarray}
\label{GEN31}
&&H(t)=\lambda-\lambda_{\ast}\frac{\left(A_{1}e^{\frac{\lambda_{\ast}m_{\ast}}{2}t}
-A_{2}e^{-\frac{\lambda_{\ast}m_{\ast}}{2}t}\right)}
{\left(A_{1}e^{\frac{\lambda_{\ast}m_{\ast}}{2}t}
+A_{2}e^{-\frac{\lambda_{\ast}m_{\ast}}{2}t}\right)},\\
\label{GEN32}
&&a(t)\sim e^{(\lambda+\lambda_{\ast})t}
\left(A_{1}e^{\lambda_{\ast}m_{\ast}t}+A_{2}\right)^{-2/m_{\ast}},\\
\label{GEN33}
&&\epsilon=\frac{2A_{1}A_{2}m_{\ast}\lambda^{2}_{\ast}e^{\lambda_{\ast}m_{\ast}t}}
{\left[A_{1}(\lambda-\lambda_{\ast})e^{\lambda_{\ast}m_{\ast}t}
+A_{2}(\lambda+\lambda_{\ast})\right]^{2}},\\
\label{GEN34}
&&\delta=\frac{m_{\ast}\lambda^{2}_{\ast}(A_{1}e^{\lambda_{\ast}m_{\ast}t}-A_{2})}
{2\left[A_{1}(\lambda-\lambda_{\ast})e^{\lambda_{\ast}m_{\ast}t}
+A_{2}(\lambda+\lambda_{\ast})\right]},
\end{eqnarray}
which correspond to relation (\ref{RELSRPAR}) for
\begin{eqnarray}
\label{GEN25}
&&\mu_{0}=-\frac{m_{\ast}}{2},~~~\mu_{2}=-\frac{\mu^{2}_{1}
\left(\lambda^{2}-\lambda^{2}_{\ast}\right)}{2m_{\ast}\lambda^{2}},~~m_{\ast}\neq0,
\end{eqnarray}
where $\lambda$, $\lambda_{\ast}$, $m_{\ast}$, $A_{1}$ and $A_{2}$ are a some constants.

In this case, it is necessary to determine additional relationship between the constant parameters of these inflationary models.

Condition of the maximal value of first slow-roll parameter 
\begin{eqnarray}
\label{EXAMN1}
&&\dot{\epsilon}(t=t_{m})=0,
\end{eqnarray}
leads to the following expression for cosmic time
\begin{eqnarray}
\label{EXAMN2}
&&t_{m}=\frac{1}{\lambda_{\ast}m_{\ast}}
\ln\left[\frac{A_{2}(\lambda+\lambda_{\ast})}{A_{1}(\lambda-\lambda_{\ast})}\right].
\end{eqnarray}

From expressions (\ref{GEN33}), (\ref{EXAMN2}) and condition $\epsilon(t=t_{m})=3$ for $V=0$ in accordance with (\ref{SLOWROLLP1}), additional relationship between the constant parameters can be written as
\begin{eqnarray}
\label{EXAMN3}
&&m_{\ast}=\frac{6}{\lambda^{2}_{\ast}}\left(\lambda^{2}-\lambda^{2}_{\ast}\right)
=6\left[\left(\frac{\lambda}{\lambda_{\ast}}\right)^{2}-1\right].
\end{eqnarray}

Also, for cosmic time (\ref{EXAMN2}) and constant parameter (\ref{EXAMN3}) from expression (\ref{GEN34}) one has $\delta(t=t_{m})=\epsilon(t=t_{m})=3$.

$\text{IV}.$ Slow-roll parameters are
\begin{eqnarray}
\label{III1}
&&\epsilon=\mu_{0}+\frac{\mu_{2}}{H^{2}}=\delta+\frac{\mu_{2}}{H^{2}},\\
\label{III2}
&&\delta=\mu_{0}=const,
\end{eqnarray}
under condition
\begin{eqnarray}
\label{III3}
\mu_{1}=0.
\end{eqnarray}

This case corresponds to Hubble parameter (\ref{GEN31}) with $\lambda=0$.

$\text{V}.$  For the partial case $\mu_{0}=0$ from (\ref{RELSRPAR}) one has
\begin{eqnarray}
\label{VI}
&&\epsilon=2\delta+\frac{4\mu_{2}}{\mu^{2}_{1}}\delta^{2}.
\end{eqnarray}

In this case, Hubble parameter and scale factor are
\begin{eqnarray}
\label{VIH5}
\nonumber
&&H(t)=\lambda_{1}-\frac{\xi_{0}}{\xi_{1}}+\eta_{2}\exp\left(-\mu_{1}t\right)\\
&&=-\frac{\mu_{2}}{\mu_{1}}+\eta_{2}\exp\left(-\mu_{1}t\right),\\
\label{VIHA5}
&&a(t)\sim\exp\left(-\frac{\mu_{2}}{\mu_{1}}t-\eta_{2}e^{-\mu_{1}t}\right).
\end{eqnarray}

$\text{VI}.$  For the partial case $\mu_{2}=0$ from (\ref{RELSRPAR}) one has
\begin{eqnarray}
\label{V}
&&\epsilon=2\delta-\mu_{0}.
\end{eqnarray}

This case corresponds to Hubble parameter (\ref{GEN31}) with $\lambda=\lambda_{\ast}$.

$\text{VII}.$
For the partial case $\mu_{0}=0$ and $\mu_{2}=0$ from (\ref{RELSRPAR}) one has
\begin{eqnarray}
\label{VII}
&&\epsilon=2\delta.
\end{eqnarray}

For this case, Hubble parameter and scale factor are
\begin{eqnarray}
\label{VII31}
&&H(t)=\frac{\mu_{1}}{\eta_{2}\mu_{1}}e^{-\mu_{1}t},\\
\label{VII32}
&&a(t)\sim\exp\left(-\frac{e^{-\mu_{1}t}}{\eta_{2}\mu_{1}}\right),
\end{eqnarray}
where $\eta_{2}$ is a some constant.

$\text{VIII}.$    For the partial case
\begin{eqnarray}
\label{VIIICOND}
&&\mu^{2}_{1}=4\mu_{2}\mu_{0},
\end{eqnarray}
as consequence of (\ref{RELSRPAR}) one has the following relation
\begin{eqnarray}
\label{VIII}
&&\frac{\delta^{2}}{\epsilon}=\mu_{0}=const.
\end{eqnarray}

This case corresponds to the second type of extreme values of a scalar field with Hubble parameter and scale factor (\ref{EXPL})--(\ref{EXPL2}) for $\left(1/s\right)=\mu_{0}$.

\section{Potentials of a scalar field in the slow-roll approximation}\label{SECPOT}

Now, we consider the potentials of a scalar field in the the slow-roll approximation following from the relations between the slow-roll parameters.

$\text{I}.$ Slow-roll parameters (\ref{I1}) and expressions (\ref{FIRST4}) correspond to the following potential
\begin{eqnarray}
\label{SRPOT1}
&&V(\phi)\simeq\frac{m^{2}\phi^{2}}{2},~~~\mu_{2}=\frac{m^{2}}{3}.
\end{eqnarray}

This is potential for chaotic inflation, where $m$ is the mass of the scalar field. Chaotic inflation with quadratic potential (\ref{SRPOT1}) were considered earlier in a large number of works (see, for example, in~\cite{Chervon:2019sey}, for an review).

$\text{II}.$ Relation (\ref{II1}) and expressions (\ref{FIRST4}) lead to the following potential
\begin{eqnarray}
\label{SRPOT2}
&&V(\phi)\simeq V_{0}e^{-\sqrt{2\mu_{0}}\phi},
\end{eqnarray}
corresponding to power-law inflation~\cite{Abbott:1984fp,Lucchin:1984yf,Sahni:1988zb},
where $V_{0}$ is a constant.

$\text{III}.$ Taking into account conditions $\epsilon\neq const$ and $\delta\neq const$, for relation (\ref{RELSRPAR}) and expressions (\ref{FIRST4}) corresponding potential can be written as
\begin{eqnarray}
\label{SRPOT4}
&&V(\phi)\simeq V_{0}\left[2\alpha_{1}-\left(\alpha_{2}e^{\sqrt{\frac{\mu_{0}}{8}}\phi}+\alpha_{3}e^{-\sqrt{\frac{\mu_{0}}{8}}\phi}\right)^{2}\right]^{2},
\end{eqnarray}
with following relation
\begin{eqnarray}
\label{SRPOT5}
\nonumber
&&\alpha_{1}\left(\alpha_{1}-2\alpha_{2}\alpha_{3}\right)\left(\mu_{2}\mu_{0}
-\frac{\mu^{2}_{1}}{4}\right)\\
&&+\mu_{0}\mu_{2}\alpha^{2}_{2}\alpha^{2}_{3}=0,
\end{eqnarray}
where $\alpha_{1}$, $\alpha_{2}$ and $\alpha_{3}$ are an arbitrary constants.

$\text{VI}.$ Relations (\ref{III1})--(\ref{III2}) and expressions (\ref{FIRST4}) lead to the following potential
\begin{eqnarray}
\label{SRPOT3}
&&V(\phi)\simeq -\frac{3\mu_{2}}{2\mu_{0}}+\frac{9\mu^{2}_{2}}{16\mu^{2}_{0}c}e^{\sqrt{2\mu_{0}}\phi}+c e^{-\sqrt{2\mu_{0}}\phi},
\end{eqnarray}
where $c$ is a constant.

This potential corresponds to the constant-roll
inflation~\cite{Motohashi:2014ppa,Motohashi:2017aob} when slow-roll conditions are satisfied.
For the partial case $\mu_{2}=0$ potential (\ref{SRPOT3}) is reduced to (\ref{SRPOT2}).
Constant-roll inflation was considered in~\cite{Motohashi:2014ppa,Motohashi:2017aob} based
on exact expressions for potential instead of (\ref{SRPOT3}).
Also in~\cite{Motohashi:2017aob}, constraints on the parameters of these models in order to satisfy slow-roll conditions were considered.

$\text{V}.$ Relation (\ref{VI}) and expressions (\ref{FIRST4}) lead to the following potential
\begin{eqnarray}
\label{SRPOT7}
&&V(\phi)\simeq V_{0}\left(\phi^{2}-\frac{8\mu_{2}}{\mu^{2}_{1}}\right)^{2},
\end{eqnarray}
corresponding to inflation with double-well potential~\cite{Mishra:2018dtg,Mazumdar:2010sa,Yamaguchi:2011kg}.

When considering the Higgs boson as inflaton~\cite{Mishra:2018dtg,Mazumdar:2010sa,Yamaguchi:2011kg}
one can rewrite potential (\ref{SRPOT7}) as follows
\begin{eqnarray}
\label{SRPOT7A}
&&V(\phi)\simeq V_{0}\left(\phi^{2}-\sigma^{2}\right)^{2},
\end{eqnarray}
where $\sigma^{2}=\frac{8\mu_{2}}{\mu^{2}_{1}}$, and $\sigma\ll\phi$ is the vacuum expectation value.

$\text{VI}.$ For relation (\ref{V}) from expressions (\ref{FIRST4}) scalar field potential can be written as
\begin{eqnarray}
\label{SRPOT6}
&&V(\phi)\simeq V_{0}\left(\alpha_{2}e^{\sqrt{\frac{\mu_{0}}{8}}\phi}+\alpha_{3}e^{-\sqrt{\frac{\mu_{0}}{8}}\phi}\right)^{4}.
\end{eqnarray}

For the case $\alpha_{2}=-\alpha_{3}$ and $\mu_{0}>0$ potential (\ref{SRPOT6}) is reduced to the following form
\begin{eqnarray}
\label{SRPOT6A}
&&V(\phi)\simeq \tilde{V}_{0}\left[\sinh\left(\sqrt{\frac{\mu_{0}}{8}}\phi\right)\right]^{n},~~~n=4,
\end{eqnarray}
corresponding to the partial case of hyperbolic inflation~\cite{Basilakos:2015sza}.

For the case $\alpha_{2}=\alpha_{3}$ and $\mu_{0}<0$ potential (\ref{SRPOT6}) is reduced to the following form
\begin{eqnarray}
\label{SRPOT6B}
&&V(\phi)\simeq \tilde{V}_{0}\left[1+\cos\left(\sqrt{\frac{|\mu_{0}|}{2}}\phi\right)\right]^{m},~~~m=2,
\end{eqnarray}
corresponding to the partial case of generalized natural inflation~\cite{Munoz:2014eqa,Kitabayashi:2023vfe}.

$\text{VII}.$ Relation (\ref{VII}) and expressions (\ref{FIRST4}) lead to the quartic potential
\begin{eqnarray}
\label{SRPOT8}
&&V(\phi)\simeq V_{0}\phi^{4}.
\end{eqnarray}

Inflation with quartic potential was considered, for example, in~\cite{Martin:2013tda,Gron:2018rtj,Mishra:2018dtg}.

$\text{VIII}.$ Relation (\ref{VIII}) and expressions (\ref{FIRST4}) corresponds to the following potential
\begin{eqnarray}
\label{SRPOT9}
&&V(\phi)\simeq V_{0}\left(2\alpha_{1}-\alpha^{2}_{3}e^{-\sqrt{\frac{\mu_{0}}{2}}\phi}\right)^{2}.
\end{eqnarray}

The case $2\alpha_{1}=\alpha^{2}_{3}=1$ and $\mu_{0}=4/3$ corresponds to potential for the Starobinsky inflation
\begin{eqnarray}
\label{SRPOT10}
&&V(\phi)\simeq V_{0}\left(1-e^{-\sqrt{\frac{2}{3}}\phi}\right)^{2}.
\end{eqnarray}

Starobinsky inflation with  potential (\ref{SRPOT10}) in the Einstein frame and corresponding extension $f(R)=R+R^2$ of the Einstein gravity theory~\cite{Starobinsky:1980te} in the Jordan frame was considered earlier, for example, in~\cite{Fomin:2020caa,Ketov:2021fww,Ivanov:2021chn,Ketov:2024klm,Pozo:2024fvo}.

Thus, the proposed approach leads to the well-known models of cosmological inflation. Moreover, the extreme values of a scalar field restrict possible inflationary scenarios. As one can see, only three types of the scalar field potentials (\ref{SRPOT1}), (\ref{SRPOT4}) and (\ref{SRPOT7A}) can be considered in the framework of the proposed approach in the slow-roll approximation. The other types of the scalar field potentials are the partial cases of (\ref{SRPOT4}) and (\ref{SRPOT7A}).

Also, exact solutions of equations (\ref{DE1})--(\ref{DE3}) can be constructed for the obtained types of cosmological dynamics. However, at the following level of analysis of the models under consideration, the possibility of their verification by observational constraints on the parameters of cosmological perturbations will be previously considered.

\section{Observational constraints on the inflationary models}\label{OBSSEC}

In accordance with the theory of cosmological perturbations, quantum fluctuations of the scalar field induce the corresponding perturbations of the space-time metric during the inflationary stage. In the linear order of cosmological perturbation theory, the observed anisotropy and polarization of cosmic microwave background radiation (CMB) can be explained by the influence of two types of perturbations, namely, scalar and tensor ones~\cite{Baumann:2014nda,Chervon:2019sey,Mukhanov:1990me,Riotto:2002yw,Brandenberger:2003vk,Straumann:2005mz}.

Observational constraints on the parameters of cosmological perturbations due to the modern observations of the anisotropy and polarisation of CMB by the Planck/BICEP2/Keck Array are~\cite{Planck:2018vyg,Galloni:2022mok}
\begin{eqnarray}
\label{PS}
&&A_S=2.1\times10^{-9},\\
\label{NS}
&&n_S=0.9649\pm 0.0042,\\
\label{R}
&&r<0.028,
\end{eqnarray}
where $A_S$ and $n_S$ are the amplitude and spectral index of scalar perturbations, $r=A_{T}/A_{S}$ is the tensor-to-scalar ratio.

The improved modern constraints on the spectral index of scalar perturbations due to the Atacama Cosmology Telescope (ACT) observations are~\cite{ACT:2025fju,ACT:2025tim}
\begin{eqnarray}
\label{NSACT}
&&n_S=0.974\pm0.0030.
\end{eqnarray}

Power spectrum of the scalar perturbations can be considered as~\cite{Planck:2018vyg}
\begin{eqnarray}
\label{PSK}
&&{\mathcal P}_{S}=A_{S}\left(\frac{k}{k_{\ast}}\right)^{n(k)},
\end{eqnarray}
where
\begin{eqnarray}
\nonumber
\label{NK}
&&n(k)=n_{S}-1+\frac{1}{2}\alpha_{S}\ln\left(\frac{k}{k_{\ast}}\right)\\
&&+\frac{1}{6}\beta_{S}\left[\ln\left(\frac{k}{k_{\ast}}\right)\right]^{2}+...
\end{eqnarray}

In this expression, first derivative of the spectral index or "running'' is defined as follows
\begin{eqnarray}
\label{RUNNING1}
&&\alpha_{S}\equiv\left(\frac{dn_{S}}{d\ln k}\right)_{k=k_{\ast}}.
\end{eqnarray}

The second derivative of the spectral index or "running of the running'' is
\begin{eqnarray}
\label{RUNNING2}
&&\beta_{S}\equiv\left(\frac{d^{2}n_{S}}{d\ln k^{2}}\right)_{k=k_{\ast}},
\end{eqnarray}
where $k_{\ast}$ is the wave number corresponding to the crossing of the Hubble radius.

Due to the Planck observations the "running'' is estimated as
\begin{eqnarray}
\label{RUNNING1P}
&&\alpha_{S}=0.0041\pm0.0067,
\end{eqnarray}
while according to the results of the ACT observations, the "running'' is estimated as
\begin{eqnarray}
\label{RUNNING1A}
&&\alpha_{S}=0.0062\pm0.0052>0.
\end{eqnarray}

When the slow-roll conditions (\ref{CSLOWROLL}) are satisfied, in the linear order of the cosmological perturbations theory the parameters of cosmological perturbations can be defined as follows~\cite{Baumann:2014nda,Chervon:2019sey}
\begin{eqnarray}
\label{A}
&&A_{S}=\frac{1}{2\epsilon_{\ast}}\left(\frac{H_{\ast}}{2\pi}\right)^{2},\\
\label{PERT}
&&n_S-1=-4\epsilon_{\ast}+2\delta_{\ast},\\
\label{PERT2}
&&r=16\epsilon_{\ast},
\end{eqnarray}
where all parameters are considered at the crossing of the Hubble radius.

Thus, the scale of inflation can be defined as
\begin{eqnarray}
\label{CROSSINGH}
&&H_{\ast}=2\pi\sqrt{2A_{S}\epsilon_{\ast}}=\frac{\pi}{2}\sqrt{2A_{S}r}\simeq\sqrt{r}\times10^{-4}.
\end{eqnarray}

Slow-roll parameters at the crossing of the Hubble radius can be written as follows
\begin{eqnarray}
\label{EPSHK}
&&\epsilon_{\ast}=\frac{r}{16},~~~
\delta_{\ast}=\frac{1}{2}\left(-1+n_{S}+\frac{r}{4}\right).
\end{eqnarray}

It should be noted, that expressions (\ref{PERT})--(\ref{PERT2}) and (\ref{EPSHK}) imply that $\epsilon_{\ast}>0$ and $\delta_{\ast}<0$ for any verifiable inflationary model.

The energy scale of inflation is defined as follows~\cite{Baumann:2014nda}
\begin{eqnarray}
\label{ENSCALEINF}
\nonumber
&&V^{1/4}_{\ast}\simeq3H^{2}_{\ast}=\left(\frac{3}{2}\pi^{2}rA_{S}\right)^{1/4}\simeq0.013\times r^{^{1/4}}\\
&&\simeq3\times10^{16}\times r^{1/4}\, GeV\sim\Lambda_{GUT}\times r^{1/4},
\end{eqnarray}
where  $\Lambda_{GUT}\sim 10^{16}$ GeV is the GUT scale.

From condition of the crossing of the Hubble radius $k=aH$ one has
\begin{eqnarray}
\label{CROSSING}
&&d\ln k=\left(H+\frac{\dot{H}}{H}\right)dt=H(1-\epsilon)dt\simeq Hdt.
\end{eqnarray}

Thus, from definitions slow-roll parameters (\ref{epsilonex})--(\ref{deltanex}) and expression (\ref{CROSSING}), "running'' (\ref{RUNNING1}) and "running of the running'' (\ref{RUNNING2}) can be written as
\begin{eqnarray}
\label{PERT3}
&&\alpha_{S}=\frac{1}{H}\left(\frac{dn_{S}}{dt}\right)=-\frac{2}{H}\left(2\dot{\epsilon}-\dot{\delta}\right),\\
\label{PERT4}
&&\beta_{S}=\frac{1}{H}\left(\frac{d\alpha_{S}}{dt}\right)=\epsilon\alpha_{S}-\frac{2}{H^{2}}\left(2\ddot{\epsilon}-\ddot{\delta}\right),
\end{eqnarray}
at the time of the crossing of the Hubble radius $t=t_{\ast}$.

Equations (\ref{epsilonex3})--(\ref{deltanex4}) and (\ref{PERT3})--(\ref{PERT4}) lead to the following expressions for "running'' and "running of the running''
\begin{eqnarray}
\label{PERT3A}
&&\alpha_{S}=\frac{\epsilon_{\ast}}{H_{\ast}}\left[8H_{\ast}(\delta_{\ast}-\epsilon_{\ast})+\mu_{1}\right],\\
\label{PERT4A}
&&\beta_{S}=-24\delta_{\ast}\epsilon^{2}_{\ast}+24\epsilon^{3}_{\ast}-2\delta_{\ast}\alpha_{S}+7\epsilon_{\ast}\alpha_{S}.
\end{eqnarray}

Therefore, under the slow-roll conditions $\epsilon_{\ast}\ll1$ and $\delta_{\ast}\ll1$, the "running of the running'' is neglected compared to "running'' $\beta_{S}\ll\alpha_{S}$.

Also, equations (\ref{HED}) and (\ref{PERT3A}) imply following expression
\begin{eqnarray}
\label{RELMU1}
&&\mu_{1}=-\frac{H_{\ast}}{\epsilon_{\ast}}\left[8\epsilon_{\ast}(\delta_{\ast}-\epsilon_{\ast})-\alpha_{S}\right],\\
\label{RELMU2}
&&\mu_{2}=H^{2}_{\ast}(\epsilon_{\ast}-2\delta_{\ast}+\mu_{0}).
\end{eqnarray}

Definition of e-folds number (\ref{EFOLDS}) leads to equation
\begin{eqnarray}
\label{C1}
&&\frac{dN}{d\epsilon}=\frac{dN}{dt}\frac{dt}{d\epsilon}=\frac{H}{\dot{\epsilon}}.
\end{eqnarray}

From (\ref{epsilonex3}) and (\ref{C1}) the difference in the e-folds number between the end of inflation and the crossing of the Hubble radius can be defined as
\begin{eqnarray}
\label{C2}
&&\Delta N=\frac{1}{2}\int^{\epsilon_{e}}_{\epsilon_{\ast}}\frac{d\epsilon}
{\epsilon(\epsilon-\delta(\epsilon))},
\end{eqnarray}
where $\epsilon_{e}=1$ and $\epsilon_{\ast}$ the first slow-roll parameter at the end of inflation and at the crossing of a Hubble radius.

Also, the difference in the e-folds number between the end of inflation and the crossing of the Hubble radius is~\cite{Liddle:2003as,German:2022sjd,DiMarco:2024yzn}
\begin{eqnarray}
\label{EFOLDSOBS}
&&\Delta N=N_{e}-N_{\ast}\simeq64+\ln\left(\frac{M_{inf}}{M_{P}}\right)\simeq50-60,
\end{eqnarray}
where $M_{inf}=10^{13}-10^{16}$ GeV is the mass scale of a scalar field.

Further, equations (\ref{E2m}) and (\ref{epsilonex3}) lead to the following expression
\begin{eqnarray}
\label{C4A}
&&\frac{d\phi}{d\epsilon}=\frac{\dot{\phi}}{\dot{\epsilon}}=\pm\frac{\sqrt{-2\dot{H}}}{2H\epsilon(\epsilon-\delta)}=\pm\frac{1}{\sqrt{2\epsilon}(\epsilon-\delta)}.
\end{eqnarray}

Thus, changing a scalar field between the end of inflation and the crossing of the Hubble radius can be written as
\begin{eqnarray}
\label{C4}
&&|\Delta\phi|=\frac{1}{\sqrt{2}}\int^{\epsilon_{e}}_{\epsilon_{\ast}}\frac{d\epsilon}{\sqrt{\epsilon}(\epsilon-\delta(\epsilon))}.
\end{eqnarray}

Due to the Lyth bound, the change of a scalar field can be restricted as~\cite{Lyth:1996im,Efstathiou:2005tq,DiMarco:2017ihz}
\begin{eqnarray}
\label{C5}
&&|\Delta\phi|\geq {\mathcal O}\left(1\right)\left(\frac{r}{0.01}\right)^{1/2},
\end{eqnarray}
where the estimate ${\mathcal O}\left(1\right)=2$ presented in~\cite{Baumann:2014nda} was used.

On the other hand, the Swampland conjectures lead to the upper bound~\cite{Scalisi:2018eaz}
\begin{eqnarray}
\label{C6}
&&|\Delta\phi|\leq -\ln H_{\ast}=\frac{1}{2}\ln\left(\frac{2}{\pi^{2}A_{S}r}\right).
\end{eqnarray}

This constraint follows from the conditions of consistency of inflationary models with quantum gravity and string theory (see, for example, in~\cite{Palti:2019pca,vanBeest:2021lhn,Obied:2018sgi,Kehagias:2018uem,Denef:2018etk,Scalisi:2018eaz} for details).

The obtained expressions allow one to verify inflationary models by the constraints on cosmological parameters only on the basis of the known relationships between the slow-roll parameters.

\section{Verification of inflationary models by observational constraints}\label{VERSEC}

Now, we consider the correspondence of the obtained inflationary models to the observational constraints.
For this aim, it is sufficient to use the relations between the slow-roll parameters (\ref{I1})--(\ref{VIII}) only.
Also, one can define the dependence $r=r(1-n_{S})$ in explicit form for the inflationary models with different relations between the slow-roll parameters.
This fact leads to the possibility of various types of analysis of the correspondence of inflationary models with observational data.

\subsection{Models without additional constant parameters in dependence $r=r(1-n_{S})$}\label{wac}

Firstly, inflationary models without additional constant parameters in dependence $r=r(1-n_{S})$ will be considered.

$\text{1}.$ For the model with slow-roll parameters (\ref{I1}) from (\ref{PERT})--(\ref{PERT2}) it follows that
\begin{eqnarray}
\label{VER1}
&&r=4\left(1-n_{S}\right),
\end{eqnarray}
and from (\ref{NS}), (\ref{NSACT}) for $n_{S(max)}=0.977$ one has the following minimal value of the tensor-to-scalar ratio $r_{min}\simeq0.09$, which does not correspond to the observational constraint (\ref{R}).

$\text{2}.$ For the model with slow-roll parameters (\ref{II1}) from (\ref{PERT})--(\ref{PERT2}) it follows that
\begin{eqnarray}
\label{VER2}
&&r=8\left(1-n_{S}\right),
\end{eqnarray}
and from (\ref{NS}), (\ref{NSACT}) one has the following minimal value of the tensor-to-scalar ratio $r_{min}\simeq0.19$, which does not correspond to observational constraint (\ref{R}).

Also, condition $\epsilon=\delta=\mu_{0}=const$ does not imply an exit from inflation by a violation of the slow-roll conditions.

$\text{3}.$ For the model with relation between slow-roll parameters (\ref{VII}) from (\ref{PERT})--(\ref{PERT2}) it follows that
\begin{eqnarray}
\label{VER3}
&&r=\frac{16}{3}\left(1-n_{S}\right).
\end{eqnarray}

Taking into account (\ref{NS}), (\ref{NSACT}) one has following minimal value of the tensor-to-scalar ratio $r_{min}\simeq0.13$, which does not correspond to observational constraint (\ref{R}).

$\text{4}.$ For the model with relation between slow-roll parameters (\ref{VI}) from (\ref{EPSHK}) and (\ref{RELMU1})--(\ref{RELMU2}) it follows that
\begin{eqnarray}
\label{VER4}
\nonumber
&&1-n_{S}=\frac{r}{16}\\
&&-\frac{r^{2}\left(3r-16+16n_{S}\right)\left(-4+4n_{S}+r\right)^{2}}{64 \left(r^{2}+8\left(n_{S}-1\right)r-32\alpha_{S}\right)^{2}}.
\end{eqnarray}

Taking into account (\ref{NS}) and (\ref{RUNNING1}) one has the following minimal value of the tensor-to-scalar ratio $r_{min}\simeq0.13$, which does not correspond to observational constraint (\ref{R}).
In addition, this result can be obtained under the condition $2\delta\gg\frac{4\mu_{2}}{\mu^{2}_{1}}\delta^{2}$ in expression (\ref{VI}).

Thus, for this model, the dependence of tensor-to-scalar ratio from the spectral index of the scalar perturbations can be written as
\begin{eqnarray}
\label{VER41}
&&r\simeq\frac{16}{3}\left(1-n_{S}\right).
\end{eqnarray}

Therefore, these inflationary models do not correspond to observational constraints on the values of the parameters of cosmological perturbations.

\subsection{Models with additional constant parameters in dependence $r=r(1-n_{S})$}\label{wac2}

Secondly, inflationary models with additional constant parameters in dependence $r=r(1-n_{S})$ will be considered.

$\text{1}.$ For the model with relation between slow-roll parameters (\ref{V}) from (\ref{PERT})--(\ref{PERT2}) it follows that
\begin{eqnarray}
\label{VER5RS}
&&r=\frac{16}{3}\left(1-n_{S}+\mu_{0}\right),
\end{eqnarray}
where $|\mu_{0}|\ll1$.

Relation (\ref{V}) leads to the following dependence
\begin{eqnarray}
\label{VER5DEL}
&&\delta=\frac{1}{2}\left(\epsilon+\mu_{0}\right).
\end{eqnarray}

Expressions (\ref{C2}) and (\ref{VER5DEL}) imply that
\begin{eqnarray}
\label{VER5}
&&\Delta N=\frac{1}{\mu_{0}}\ln\left[\frac{\epsilon_{\ast}(1-\mu_{0})}{\epsilon_{\ast}-\mu_{0}}\right].
\end{eqnarray}

Also, from (\ref{PERT2}), (\ref{VER5RS}) and (\ref{VER5}) tensor-to-scalar ratio and spectral index of scalar perturbations can be written as
\begin{eqnarray}
\label{VER5R}
&&r=\frac{16\mu_{0}\exp(\mu_{0}\Delta N)}{\exp(\mu_{0}\Delta N)+\mu_{0}-1},\\
\label{VER5NS}
&&1-n_{S}=\frac{\mu_{0}\left(2\exp(\mu_{0}\Delta N)-\mu_{0}+1\right)}{\exp(\mu_{0}\Delta N)+\mu_{0}-1}.
\end{eqnarray}

From expressions (\ref{V}) and (\ref{EPSHK}) at the crossing of the Hubble radius the constant $\mu_{0}$ can be defined as
\begin{eqnarray}
\label{VER5A}
&&\mu_{0}=2\delta_{\ast}-\epsilon_{\ast}=-1+n_{S}+\frac{3r}{16}.
\end{eqnarray}

Further, the dependence $\Delta N=\Delta N(r)$ with the values of the spectral index of the scalar perturbations restricted by observational constraints (\ref{NS}) can be considered.

After substituting (\ref{EPSHK}) and (\ref{VER5A}) into (\ref{VER5}) one has
\begin{eqnarray}
\label{VER5C}
&&\Delta N=\frac{-80\ln\left(2\right)+16\ln\left(\frac{r \left(-32+16n_{S}+3 r\right)}{r-8+8n_{S}}\right)}{-16+16n_{S}+3 r}.
\end{eqnarray}

\begin{figure}[ht]
	\centering
		\includegraphics[width=0.45\textwidth]{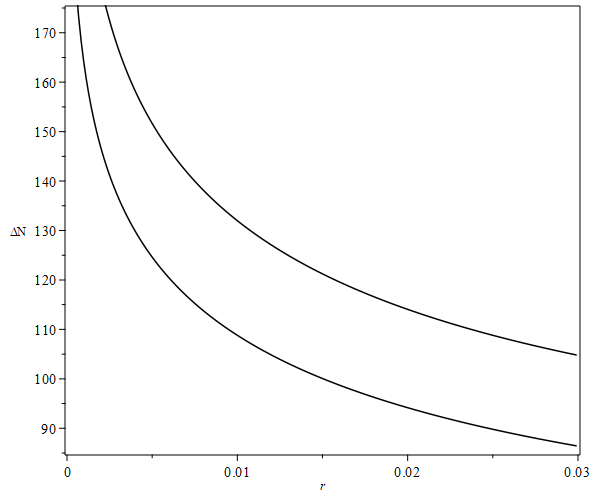}
	\caption{The dependence $\Delta N=\Delta N(r)$ for the spectral index of the scalar perturbations $n_S=0.9649\pm 0.0042$ (Planck). The values of $\Delta N$ lie within the region bounded by the curves.}
	\label{FIG1}
\end{figure}

\begin{figure}[ht]
	\centering
		\includegraphics[width=0.45\textwidth]{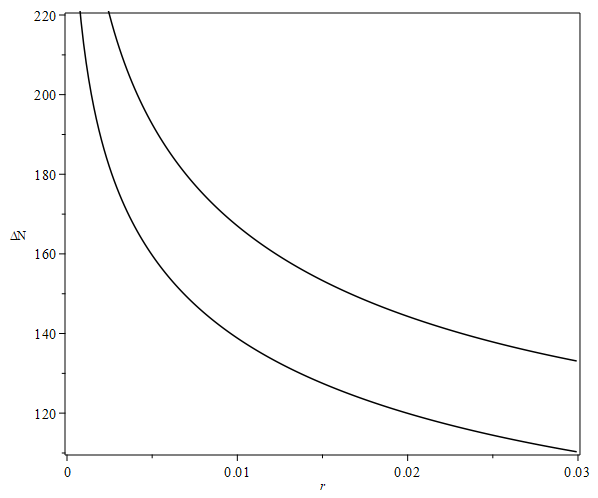}
	\caption{The dependence $\Delta N=\Delta N(r)$ for the spectral index of the scalar perturbations $n_S=0.9649\pm 0.0042$. The values of $\Delta N$ lie within the region bounded by the curves.}
	\label{FIG1A}
\end{figure}

\begin{figure}[ht]
	\centering
		\includegraphics[width=0.45\textwidth]{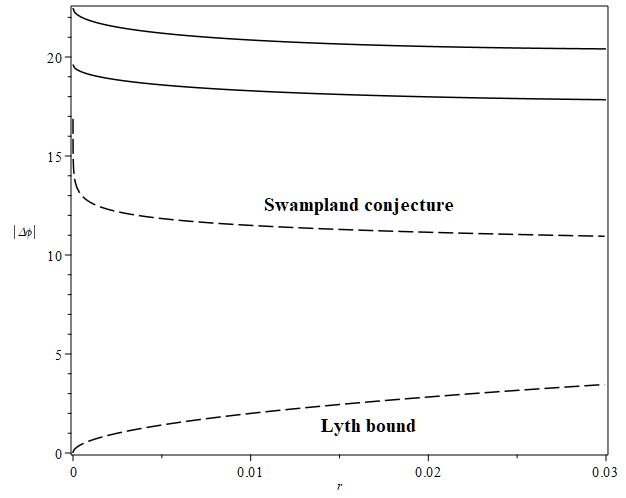}
	\caption{The dependence $|\Delta\phi|=|\Delta\phi|(r)$ for the spectral index of the scalar perturbations $n_S=0.9649\pm 0.0042$. The values of $|\Delta\phi|$ lie within the region bounded by the solid curves.}
	\label{FIG1F}
\end{figure}

\begin{figure}[ht]
	\centering
		\includegraphics[width=0.45\textwidth]{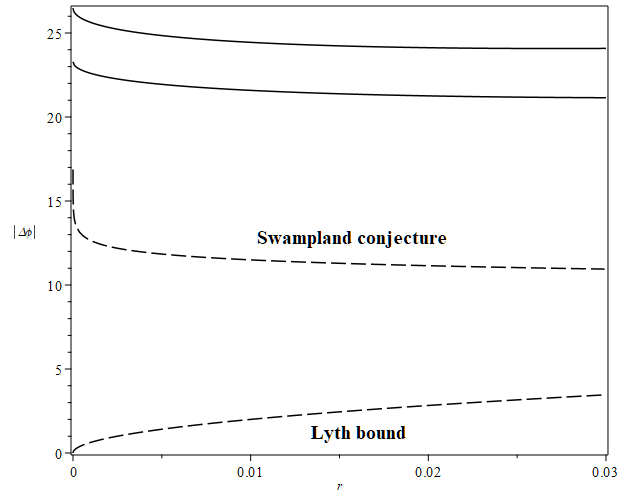}
	\caption{The dependence $|\Delta\phi|=|\Delta\phi|(r)$ for the spectral index of the scalar perturbations $n_S=0.974\pm0.0030$ (ACT). The values of $|\Delta\phi|$ lie within the region bounded by the solid curves.}
	\label{FIG1FACT}
\end{figure}

In Fig. \ref{FIG1} the dependence $\Delta N=\Delta N(r)$  is presented for the spectral index of the scalar perturbations $n_S=0.9649\pm 0.0042$ according to constraint (\ref{NS}) for relation (\ref{V}) between the slow-roll parameters.
Observational constraint on the tensor-to-scalar ratio (\ref{R}) can be satisfied for $\Delta N>88$. This value of the e-folds number does not meet constraint (\ref{EFOLDSOBS}).

In Fig. \ref{FIG1A} the dependence $\Delta N=\Delta N(r)$  is presented for the spectral index of the scalar perturbations $n_S=0.974\pm0.0030$ according to constraint (\ref{NSACT}) for relation (\ref{V}) between the slow-roll parameters.
Observational constraint on the tensor-to-scalar ratio (\ref{R}) can be satisfied for $\Delta N>112$. This value of the e-folds number does not meet constraint (\ref{EFOLDSOBS}).

Also, equations (\ref{C4}) and (\ref{VER5DEL}) lead to the following expression for the changing of a scalar field between the end of inflation and the crossing of the Hubble radius
\begin{eqnarray}
\label{VER6C}
\nonumber
&&|\Delta\phi|=2\sqrt{\frac{2}{\mu_{0}}}{\rm arctanh}\left(\frac{1}{4}\sqrt{\frac{r}{\mu_{0}}}\right)\\
&&-2\sqrt{\frac{2}{\mu_{0}}}{\rm arctanh}\left(\frac{1}{\sqrt{\mu_{0}}}\right),
\end{eqnarray}
where $\mu_{0}$ is defined by expression (\ref{VER5A}).

In Fig. \ref{FIG1F} the dependence $|\Delta\phi|=|\Delta\phi|(r)$  is presented for the spectral index of scalar perturbations $n_S=0.9649\pm 0.0042$ according to constraint (\ref{NS}). In this case, the change of a scalar field between the crossing of the Hubble radius and the end of inflation $|\Delta\phi|$ does not correspond to constraint (\ref{C6}).

In Fig. \ref{FIG1FACT} the dependence $|\Delta\phi|=|\Delta\phi|(r)$  is presented for the spectral index of scalar perturbations $n_S=0.974\pm0.0030$ according to constraint (\ref{NSACT}). In this case, the change in a scalar field between the crossing of the Hubble radius and the end of inflation $|\Delta\phi|$ does not correspond to constraint (\ref{C6}) as well.

$\text{2}.$ For the model with slow-roll parameters (\ref{III1})--(\ref{III2}) from (\ref{PERT})--(\ref{PERT2}) we obtain
\begin{eqnarray}
\label{VER6RS}
&&r=4\left(1-n_{S}+2\mu_{0}\right),
\end{eqnarray}
where $|\mu_{0}|\ll1$.

From (\ref{III2}) and (\ref{C2}) we obtain
\begin{eqnarray}
\label{VER6}
&&\Delta N=\frac{1}{2\mu_{0}}\ln\left[\frac{\epsilon_{\ast}(1-\mu_{0})}{\epsilon_{\ast}-\mu_{0}}\right].
\end{eqnarray}

Also, from (\ref{PERT2}), (\ref{VER6RS}) and (\ref{VER6}) we get
\begin{eqnarray}
\label{VER6R}
&&r=\frac{16\mu_{0}\exp(2\mu_{0}\Delta N)}{\exp(\mu_{0}\Delta N)+\mu_{0}-1},\\
\label{VER6NS}
&&1-n_{S}=\frac{2\mu_{0}\left(\exp(2\mu_{0}\Delta N)-\mu_{0}+1\right)}{\exp(2\mu_{0}\Delta N)+\mu_{0}-1}.
\end{eqnarray}

Now, we consider the dependence $\Delta N=\Delta N(r)$ with the values of the spectral index of the scalar perturbations restricted by observational constraints (\ref{NS}).

Taking into account that from condition (\ref{III2}) at the crossing of the Hubble radius we get
\begin{eqnarray}
\label{VER6A}
&&\mu_{0}=\delta_{\ast}=\frac{1}{2}\left(-1+n_{S}+\frac{r}{4}\right).
\end{eqnarray}

From (\ref{EPSHK}), (\ref{VER6}) and (\ref{VER6A}) we obtain
\begin{eqnarray}
\label{VER6B}
&&\Delta N=\frac{-12\ln\left(2\right)+4\ln\left(\frac{r \left(-12+4n_{S}+r\right)}{r-8+8n_{S}}\right)}{-4+4n_{S}+r}.
\end{eqnarray}

\begin{figure}[ht]
	\centering
		\includegraphics[width=0.45\textwidth]{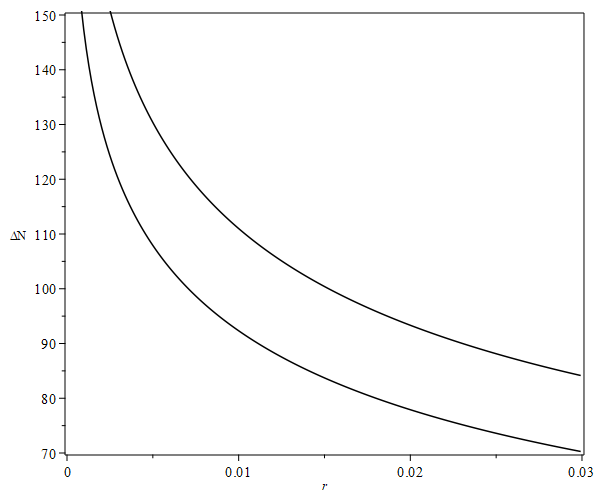}
	\caption{The dependence $\Delta N=\Delta N(r)$ for the spectral index of the scalar perturbations $n_S=0.9649\pm 0.0042$ (Planck). The values of $\Delta N$  lie within the region bounded by the curves.}
	\label{FIG2}
\end{figure}

\begin{figure}[ht]
	\centering
		\includegraphics[width=0.45\textwidth]{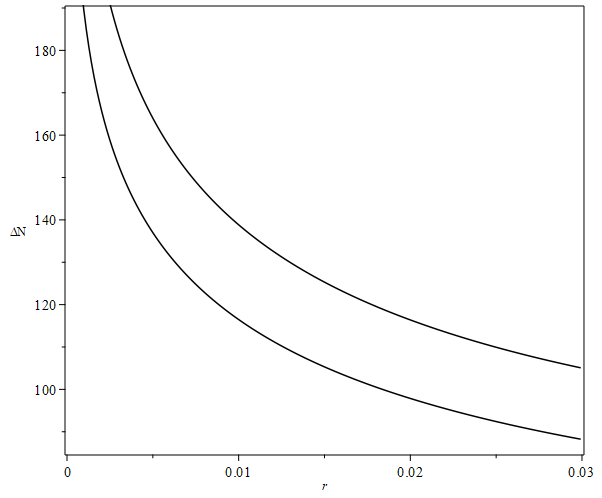}
	\caption{The dependence $\Delta N=\Delta N(r)$ for the spectral index of the scalar perturbations $n_S=0.974\pm0.0030$ (ACT). The values of $\Delta N$ lie within the region bounded by the curves.}
	\label{FIG2ACT}
\end{figure}

\begin{figure}[ht]
	\centering
		\includegraphics[width=0.45\textwidth]{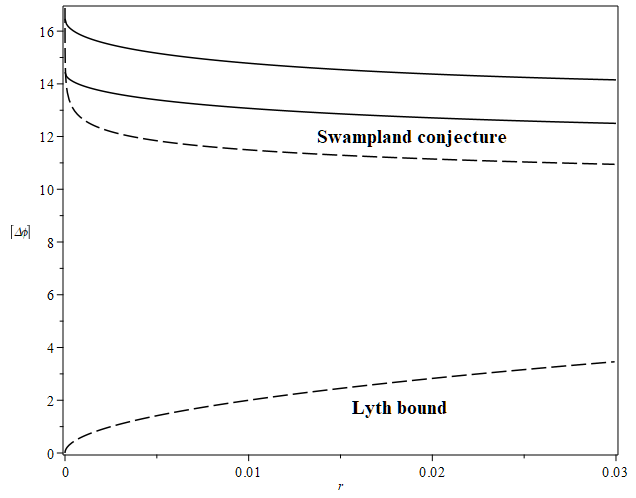}
	\caption{The dependence $|\Delta\phi|=|\Delta\phi|(r)$ for the spectral index of the scalar perturbations $n_S=0.9649\pm 0.0042$ (Planck). The values of $|\Delta\phi|$ lie within the region bounded by the solid curves.}
	\label{FIG2F}
\end{figure}

\begin{figure}[ht]
	\centering
		\includegraphics[width=0.45\textwidth]{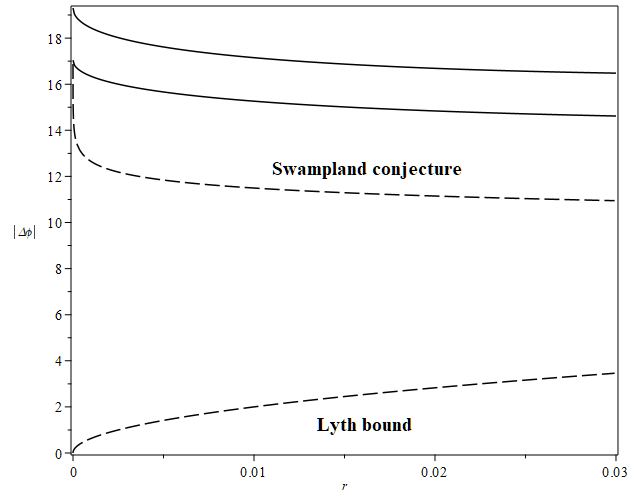}
	\caption{The dependence $|\Delta\phi|=|\Delta\phi|(r)$ for the spectral index of the scalar perturbations $n_S=0.974\pm0.0030$ (ACT). The values of $|\Delta\phi|$ lie within the region bounded by the solid curves.}
	\label{FIG2FACT}
\end{figure}

In Fig. \ref{FIG2} the dependence $\Delta N=\Delta N(r)$  is presented for the spectral index of the scalar perturbations $n_S=0.9649\pm 0.0042$ according to constraint (\ref{NS}) for slow-roll parameters (\ref{III1})--(\ref{III2}).
From expression (\ref{VER6B}) it follows that this type of inflationary model can satisfy the observational constraints (\ref{NS})--(\ref{R}) for e-folds number $\Delta N>72$, which does not correspond to condition (\ref{EFOLDSOBS}).

In Fig. \ref{FIG2ACT} the dependence $\Delta N=\Delta N(r)$  is presented for the spectral index of the scalar perturbations $n_S=0.974\pm0.0030$ according to constraint (\ref{NSACT}).
From expression (\ref{VER6B}) it follows that this type of inflationary model can satisfy the observational constraints (\ref{NS})--(\ref{R}) for e-folds number $\Delta N>90$, which does not correspond to condition (\ref{EFOLDSOBS}).

Also, from (\ref{C4}) and (\ref{III2}) we get
\begin{eqnarray}
\label{VER6CR}
\nonumber
&&|\Delta\phi|=\sqrt{\frac{2}{\mu_{0}}}{\rm arctanh}\left(\frac{1}{4}\sqrt{\frac{r}{\mu_{0}}}\right)\\
&&-\sqrt{\frac{2}{\mu_{0}}}{\rm arctanh}\left(\frac{1}{\sqrt{\mu_{0}}}\right),
\end{eqnarray}
where $\mu_{0}$ is defined by expression (\ref{VER6A}).

Fig. \ref{FIG2F} shows the dependence $|\Delta\phi|=|\Delta\phi|(r)$ for the spectral index of the scalar perturbations $n_S=0.9649\pm 0.0042$ according to constraint (\ref{NS}). The change of a scalar field between the crossing of the Hubble radius and the end of inflation $|\Delta\phi|$ does not correspond to constraint (\ref{C6}).

Fig. \ref{FIG2FACT} shows the dependence $|\Delta\phi|=|\Delta\phi|(r)$ for the spectral index of the scalar perturbations $n_S=0.974\pm0.0030$ according to constraint (\ref{NSACT}). The change of a scalar field between the crossing of the Hubble radius and the end of inflation $|\Delta\phi|$ does not correspond to constraint (\ref{C6}).

\subsection{Models without explicit dependence $r=r(1-n_{S})$}\label{SHIDTED}

Now, we consider relation (\ref{RELSRPAR}) in terms of the shifted slow-roll parameters
\begin{eqnarray}
\label{VER9}
&&\frac{\Delta^{2}}{\varepsilon}=\frac{\mu^{2}_{1}}{4\mu_{2}}=n_{1}=const,
\end{eqnarray}
where shifted slow-roll parameters are defined as
\begin{eqnarray}
\label{VER92}
&& \varepsilon=\epsilon+\left(\frac{\mu^{2}_{1}}{4\mu_{2}}-\mu_{0}\right)=\epsilon+n,\\
\label{VER93}
&&\Delta=\delta+\left(\frac{\mu^{2}_{1}}{4\mu_{2}}-\mu_{0}\right)=\delta+n,
\end{eqnarray}
and $\mu_{0}\neq0$, $\mu_{1}\neq0$, $\mu_{2}\neq0$.

Thus, in this case, the relation between the shifted slow-roll parameters $\left(\Delta^{2}/\varepsilon\right)=const$ is conserved.

Further, we rewrite expressions (\ref{C2}) and (\ref{C4}) on the basis of expressions (\ref{VER9})--(\ref{VER93}), (\ref{EPSHK}) and (\ref{RELMU1})--(\ref{RELMU2}) in terms of the shifted slow-roll parameters
\begin{eqnarray}
\label{VER95}
&&\Delta N=\frac{1}{2}\int^{\varepsilon_{e}}_{\varepsilon_{\ast}}
\frac{d\varepsilon}{(\varepsilon-n)(\varepsilon-\Delta(\varepsilon))},\\
\label{VER95F}
&&|\Delta\phi|=\frac{1}{\sqrt{2}}\int^{\varepsilon_{e}}_{\varepsilon_{\ast}}
\frac{d\varepsilon}{\sqrt{\varepsilon-n}(\varepsilon-\Delta(\varepsilon))},
\end{eqnarray}
where
\begin{eqnarray}
\label{VER95FDELTA}
&&\Delta^{2}(\varepsilon)=n_{1}\varepsilon,
\end{eqnarray}
and
\begin{eqnarray}
\label{VER96}
&&\Delta(\varepsilon)=-\sqrt{n_{1}\varepsilon},
\end{eqnarray}
for $\varepsilon_{e}>\varepsilon_{\ast}$.

The remaining parameters are
\begin{eqnarray}
\label{VER97}
&&n=\frac{\mu^{2}_{1}}{4\mu_{2}}-\mu_{0},~~~n_{1}=\frac{\mu^{2}_{1}}{4\mu_{2}},\\
\label{VER98}
&&\varepsilon_{\ast}=\epsilon_{\ast}+n,~~~~\varepsilon_{e}=1+n,\\
\label{VER99}
&&\epsilon_{\ast}=\frac{r}{16},~~~~~
\delta_{\ast}=\frac{1}{2}\left(-1+n_{S}+\frac{r}{4}\right).
\end{eqnarray}

Explicit expressions of integrals (\ref{VER95}) and (\ref{VER95F}) can be written as follows
\begin{eqnarray}
\label{INTN}
\nonumber
&&\Delta N=-\left(\frac{1}{n-n_{1}}\right)\ln\left[\frac{\sqrt{r}\left(\sqrt{n_{1}(1+n)}+n_{1}\right)}{\sqrt{n_{1}(1+16n)}+4n_{1}}\right]\\
&&-\frac{n_{1}\left[{\rm arctanh}\left(\sqrt{\frac{16n+r}{16n}}\right)-{\rm arctanh}\left(\sqrt{\frac{1+n}{n}}\right)\right]}{\sqrt{n_{1}n}(n-n_{1})},\\
\label{INPHI}
\nonumber
&&|\Delta\phi|=\frac{\sqrt{-2n^{2}_{1}(n-n_{1})}}{2(n-n_{1})n_{1}}\left(\ln g_{1}(r,n)-\ln g_{2}(n)\right)\\
&&+\frac{\sqrt{2}\left[\arctan\left(\frac{1}{\sqrt{n-n_{1}}}\right)
-\arctan\left(\frac{\sqrt{r}}{\sqrt{n-n_{1}}}\right)\right]}{\sqrt{n-n_{1}}},
\end{eqnarray}
where
\begin{eqnarray}
\label{G1}
\nonumber
&&g_{1}(r,n)=\frac{-16nn_{1}(n-n_{1})+rn_{1}(2-n)}{r+16n-16n_{1}}\\
&&+\frac{2\sqrt{n_{1}r(r+16n)}\sqrt{-n^{2}_{1}(n-n_{1})}}{r+16n-16n_{1}},\\
\label{G2}
\nonumber
&&g_{2}(n)=\frac{-n_{1}n^{2}+n^{2}_{1}n-n_{1}n+2n^{2}_{1}}{1+n-n_{1}}\\
&&+\frac{2\sqrt{n_{1}(n_{1}+1)}\sqrt{-n^{2}_{1}(n-n_{1})}}{1+n-n_{1}}.
\end{eqnarray}

\begin{figure}[ht]
	\centering
		\includegraphics[width=0.45\textwidth]{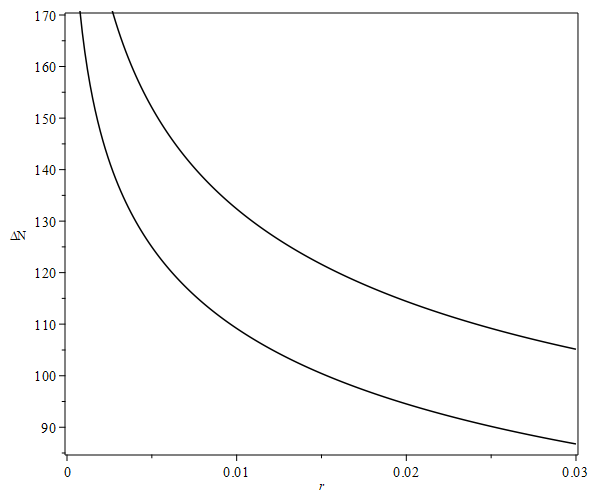}
	\caption{The dependence $\Delta N=\Delta N(r)$
for the spectral index of the scalar perturbations $n_S=0.9649\pm 0.0042$ (Planck).
The values of $\Delta N$ lie within the region bounded by the curves.}
	\label{FIG3G}
\end{figure}

\begin{figure}[ht]
	\centering
		\includegraphics[width=0.45\textwidth]{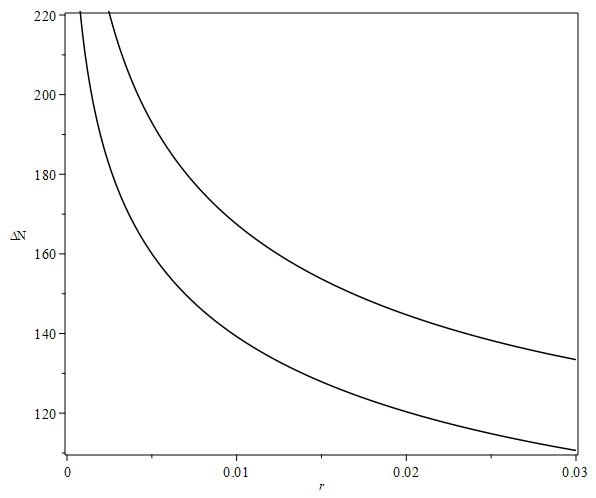}
	\caption{The dependence $\Delta N=\Delta N(r)$
for the spectral index of the scalar perturbations $n_S=0.974\pm0.0030$ (ACT).
The values of $\Delta N$ lie within the region bounded by the curves.}
	\label{FIG3GACT}
\end{figure}

\begin{figure}[ht]
	\centering
		\includegraphics[width=0.45\textwidth]{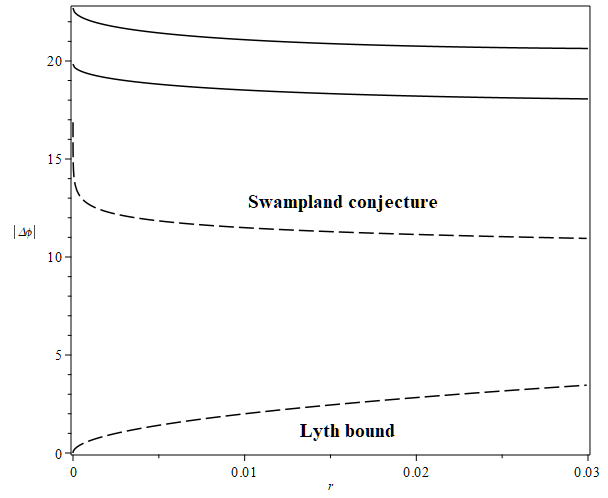}
	\caption{The dependence $|\Delta\phi|=|\Delta\phi|(r)$
for the spectral index of the scalar perturbations $n_S=0.9649\pm 0.0042$ (Planck).
The values of $|\Delta\phi|$ lie within the region bounded by the solid curves.}
	\label{FIG3F}
\end{figure}

\begin{figure}[ht]
	\centering
		\includegraphics[width=0.45\textwidth]{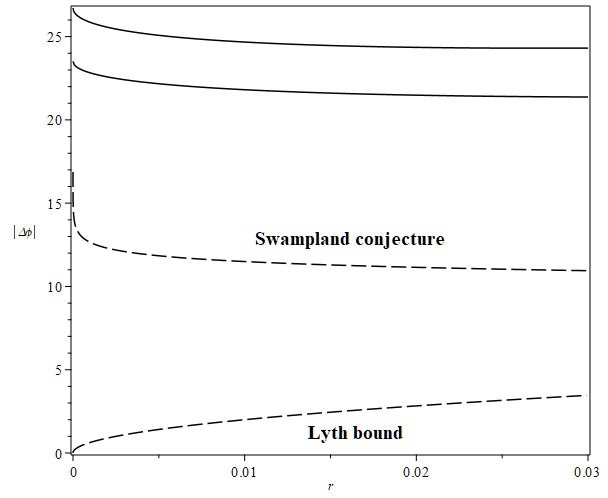}
	\caption{The dependence $|\Delta\phi|=|\Delta\phi|(r)$
for the spectral index of the scalar perturbations $n_S=0.974\pm0.0030$ (ACT).
The  values of $|\Delta\phi|$ lie within the region bounded by the solid curves.}
	\label{FIG3FACT}
\end{figure}

From expressions (\ref{GEN25}), (\ref{EXAMN3}) and (\ref{VER97})  we obtain
\begin{eqnarray}
\label{EXAMN4}
&&n=-3,\\
\label{EXAM5}
&&n_{1}=-3-\frac{m_{\ast}}{2}=-3\left(\frac{\lambda}{\lambda_{\ast}}\right)^{2}.
\end{eqnarray}

On the other hand, from (\ref{VER9}), (\ref{VER99}) and (\ref{EXAMN4})
we can define the constant parameter $n_{1}$ as \begin{eqnarray}
\label{EXAM6}
&&n_{1}=\frac{\left(\delta_{\ast}-3\right)^{2}}{\epsilon_{\ast}-3}
=-\frac{\left(7-n_{S}-\frac{r}{4}\right)^{2}}{12-\frac{r}{4}}.
\end{eqnarray}

In Fig. \ref{FIG3G} the dependence (\ref{INTN}) is presented for the spectral index of the scalar perturbations $n_S=0.9649\pm 0.0042$ according to constraint (\ref{NS}) for parameters (\ref{EXAMN4}) and (\ref{EXAM6}).
Observational constraint on the tensor-to-scalar ratio (\ref{R}) can be satisfied for $\Delta N>88$. This value of the e-folds number does not meet constraint (\ref{EFOLDSOBS}).

In Fig. \ref{FIG3GACT} the dependence (\ref{INTN}) is presented for the spectral index of the scalar perturbations $n_S=0.974\pm0.0030$ according to constraint (\ref{NSACT}) for parameters (\ref{EXAMN4}) and (\ref{EXAM6}).
Observational constraint on the tensor-to-scalar ratio (\ref{R}) can be satisfied for $\Delta N>112$. This value of the e-folds number does not meet constraint (\ref{EFOLDSOBS}).

Fig. \ref{FIG3F} shows the dependence $|\Delta\phi|=|\Delta\phi|(r)$ for the spectral index of the scalar perturbations $n_S=0.9649\pm 0.0042$ according to constraint (\ref{NS}). As one can see, the change of a scalar field between crossing of the Hubble radius and the end of inflation $|\Delta\phi|$ does not correspond to constraint (\ref{C6}).

Fig. \ref{FIG3FACT} shows the dependence $|\Delta\phi|=|\Delta\phi|(r)$ for the spectral index of the scalar perturbations $n_S=0.974\pm0.0030$ according to constraint (\ref{NSACT}). The change of a scalar field between crossing of the Hubble radius and the end of inflation $|\Delta\phi|$ does not correspond to constraint (\ref{C6}).

We also note that the values of the cosmological parameters presented in Figs. \ref{FIG3G}--\ref{FIG3FACT} coincide with the previously considered dependencies presented in Figs. \ref{FIG1}--\ref{FIG1FACT}. Thus, for these inflationary models, the dependence of the tensor-to-scalar ratio on the spectral index of scalar perturbations is closed to (\ref{VER5RS}).

\subsection{Models with quadratic dependence $r\sim(1-n_{S})^{2}$}\label{QUADR}

For the models with relation between the slow-roll parameters (\ref{VIII}) from (\ref{PERT})--(\ref{PERT2}) we obtain
\begin{eqnarray}
\label{VER8RS}
&&1-n_{S}=\frac{r}{4}+\frac{1}{2}\sqrt{\mu_{0}r},
\end{eqnarray}
where we consider $\delta_{\ast}<0$ and $\mu_{0}>0$.

Taking into account condition  $\delta_{\ast}<0$ and corresponding relation $\delta_{\ast}=-\sqrt{\mu_{0}\epsilon_{\ast}}$ we have the following expression for the second slow-roll parameter
\begin{eqnarray}
\label{VERAD}
&&\delta=-\sqrt{\mu_{0}\epsilon}.
\end{eqnarray}

After substituting (\ref{VERAD}) into (\ref{C2}) we obtain
\begin{eqnarray}
\label{VER8}
\nonumber
&&\Delta N=\frac{1}{\mu_{0}}\ln\left[\frac{\sqrt{\epsilon_{\ast}}(1+\sqrt{\mu_{0}})}{\sqrt{\epsilon_{\ast}}+\sqrt{\mu_{0}}}\right]\\
&&+\frac{1}{\sqrt{\mu_{0}}}\left(\frac{1}{\sqrt{\epsilon_{\ast}}}-1\right).
\end{eqnarray}

Taking into account the slow-roll condition $\epsilon_{\ast}\ll1$, for $\mu_{0}\sim1$ from (\ref{VER8}) one has
\begin{eqnarray}
\label{VER8A}
&&\Delta N\simeq\frac{1}{\sqrt{\epsilon_{\ast}\mu_{0}}}=\frac{4}{\sqrt{\mu_{0}r}},
\end{eqnarray}
and, therefore, the tensor-to-scalar ratio can be written as
\begin{eqnarray}
\label{VER8B}
&&r\simeq\frac{16}{\mu_{0}\left(\Delta N\right)^{2}}.
\end{eqnarray}

After substitution expression (\ref{VER8B}) into (\ref{VER8RS}) we obtain
\begin{eqnarray}
\label{VER8C}
&&1-n_{S}=\frac{4}{\mu_{0}\left(\Delta N\right)^{2}}+\frac{2}{\Delta N}\simeq\frac{2}{\Delta N}.
\end{eqnarray}

Also, from (\ref{VER8B})--(\ref{VER8C}) we obtain relation
\begin{eqnarray}
\label{VER8F}
&&r\simeq\frac{4}{\mu_{0}}(1-n_{S})^{2}.
\end{eqnarray}

Now, we can define $\mu_{0}$ and $\epsilon_{\ast}$ from expressions (\ref{VER8RS}) and (\ref{EPSHK}) as
\begin{eqnarray}
\label{VER8F7}
&&\mu_{0}=\frac{4}{r}\left(1-n_{S}-\frac{r}{4}\right)^{2},~~~~\epsilon_{\ast}=\frac{r}{16}.
\end{eqnarray}

From (\ref{EPSHK}), (\ref{C4}) and (\ref{VERAD}) we get
\begin{eqnarray}
\label{VERPHI}
\nonumber
&&|\Delta\phi|=\sqrt{\frac{2}{\mu_{0}}}{\rm arctanh}\left(\frac{1}{4}\sqrt{\frac{r}{\mu_{0}}}\right)\\
&&-\sqrt{\frac{2}{\mu_{0}}}{\rm arctanh}\left(\frac{1}{\sqrt{\mu_{0}}}\right)\\
&&-\frac{1}{\sqrt{2\mu_{0}}}\ln\left[\frac{r(1-\mu_{0})}{r-16\mu_{0}}\right].
\end{eqnarray}

\begin{figure}[ht]
	\centering
		\includegraphics[width=0.45\textwidth]{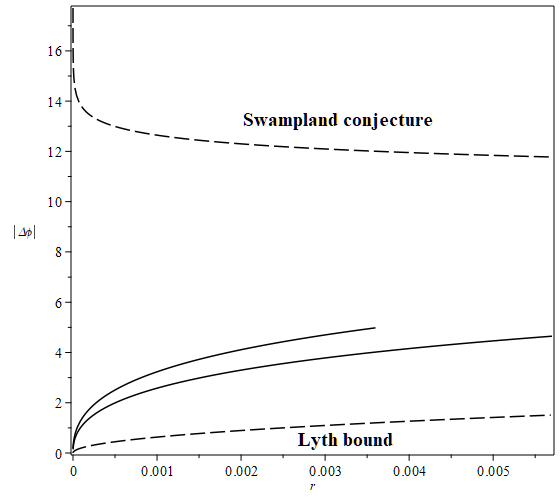}
	\caption{The dependence $|\Delta\phi|=|\Delta\phi|(r)$ for the spectral index of the scalar perturbations $n_S=0.9649\pm 0.0042$ (Planck). The values of $|\Delta\phi|$ correspond to the Lyth bound and Swampland conjecture. The possible values of the tensor-to-scalar ratio $0<r\leq0.0035$ and $0<r\leq0.0057$ are restricted by the real values of $|\Delta\phi|$ with taking into account constraint (\ref{R}).}
	\label{FIG4F}
\end{figure}

\begin{figure}[ht]
	\centering
		\includegraphics[width=0.4\textwidth]{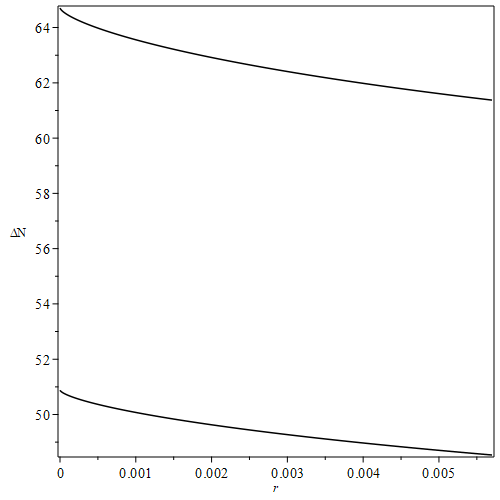}
	\caption{The dependence $\Delta N=\Delta N(r)$ for $0<r\leq0.0057$ and $n_S=0.9649\pm 0.0042$ (Planck). The values of $\Delta N$ lie within the region bounded by the curves.}
	\label{FIG4}
\end{figure}

\begin{figure}[ht]
	\centering
		\includegraphics[width=0.4\textwidth]{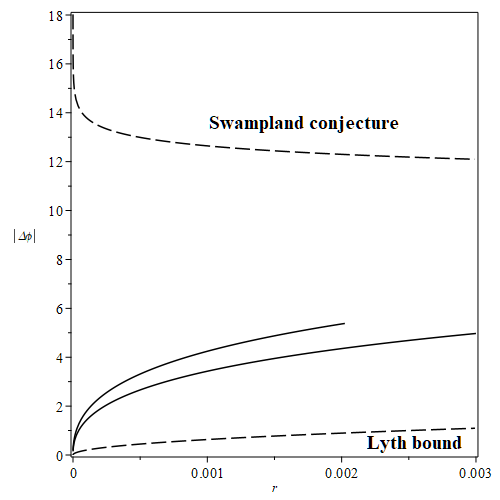}
	\caption{The dependence $|\Delta\phi|=|\Delta\phi|(r)$ for the spectral index of the scalar perturbations $n_S=0.974\pm0.0030$ (ACT). The values of $|\Delta\phi|$ correspond to the Lyth bound and Swampland conjecture. The possible values of the tensor-to-scalar ratio $0<r\leq0.002$ and $0<r\leq0.003$  are restricted by the real values of $|\Delta\phi|$ with taking into account constraint (\ref{R}).}
	\label{FIG4FACT}
\end{figure}

\begin{figure}[ht]
	\centering
		\includegraphics[width=0.4\textwidth]{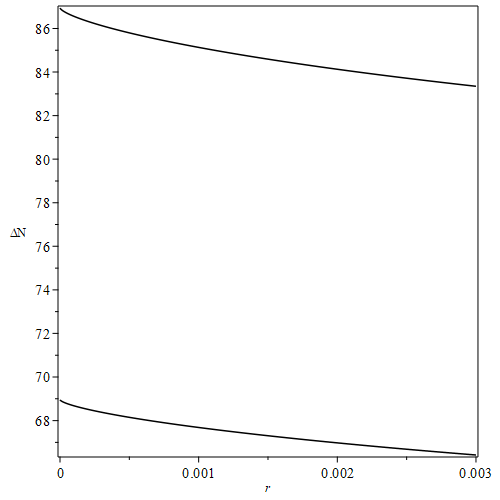}
	\caption{The dependence $\Delta N=\Delta N(r)$ for $0<r\leq0.003$ and $n_S=0.974\pm0.0030$ (ACT). The values of $\Delta N$ lie within the region bounded by the curves.}
	\label{FIG4ACT}
\end{figure}

In Fig. \ref{FIG4F} the dependence $|\Delta\phi|=|\Delta\phi|(r)$ is presented for the spectral index of scalar perturbations $n_S=0.9649\pm 0.0042$ according to constraint (\ref{NS}). From the condition of the real values of a change of a scalar field we have the following constraint on the tensor-to-scalar ratio $0<r\leq0.0057$ and from (\ref{VER8F}) it follows that $\mu_{0}\geq0.27$. Thus, for models with relation between slow-roll parameters (\ref{VIII}) the changing of a scalar field between crossing of the Hubble radius and the end of inflation satisfy
both constraints (\ref{C5}) and (\ref{C6}).

In Fig. \ref{FIG4} the dependence (\ref{VER8}) with parameters (\ref{VER8F7}) is presented for the spectral index of the scalar perturbations $n_S=0.9649\pm 0.0042$ according to the constraint (\ref{NS}). The e-folds number takes the values $\Delta N\geq49$ that correspond to constraint (\ref{EFOLDSOBS}).

In Fig. \ref{FIG4FACT} the dependence $|\Delta\phi|=|\Delta\phi|(r)$ is presented for $n_S=0.974\pm0.0030$ according to the constraint (\ref{NSACT}). From the condition of the real values of a change of a scalar field we have the following constraint on the tensor-to-scalar ratio $0<r\leq0.003$ and from (\ref{VER8F}) it follows that $\mu_{0}\geq0.28$. As one can see, the change of a scalar field  between the crossing of the Hubble radius and the end of inflation satisfy both constraints (\ref{C5}) and (\ref{C6}).

In Fig. \ref{FIG4ACT} the dependence (\ref{VER8}) with parameters (\ref{VER8F7}) is presented for $n_S=0.974\pm0.0030$ according to the constraint (\ref{NSACT}).
In this case, the e-folds number $\Delta N\ge66$ does not correspond to constraint (\ref{EFOLDSOBS}).

Also, from relation (\ref{VIIICOND}) and expressions (\ref{RELMU1})--(\ref{RELMU2}) we obtain the equation
\begin{eqnarray}
\label{VER8ALPHA1}
&&\frac{\left(r+8n_S-8\right)^{2}\left(-4+4n_S+r\right)^{2}}{4\left(r^{2}+\left(8n_S-8\right)r-32 \alpha_{S}\right)^{2}}=1.
\end{eqnarray}

The solutions of this equation are
\begin{eqnarray}
\label{VER8ALPHA2}
&&\alpha_{S}=-\frac{1}{2}\left(1-n_S\right)^{2}-\frac{r}{16}\left(1-n_S-\frac{r}{4}\right),\\
\label{VER8ALPHA3}
&&\alpha_{S}=\frac{1}{2}\left(1-n_S\right)^{2}-\frac{7r}{16}\left(1-n_S-\frac{3r}{28}\right).
\end{eqnarray}

Taking into account (\ref{VER8F}), from (\ref{VER8ALPHA2})--(\ref{VER8ALPHA3}) we obtain the expression for "running''
\begin{eqnarray}
\label{VER8ALPHA4}
&&|\alpha_{S}|\simeq\frac{1}{2}\left(1-n_S\right)^{2}\sim 10^{-4},
\end{eqnarray}
corresponding to the Planck constraint (\ref{RUNNING1}).

On the other hand, due to the ACT results (\ref{RUNNING1A}) the "running'' can be estimated as
\begin{eqnarray}
\label{RUNNING2ACT}
&&\alpha_{S}\sim10^{-2}-10^{-3}.
\end{eqnarray}

Thus, inflationary models with relation (\ref{VIII}) between slow-roll parameters correspond to the Planck constraints, and they do not correspond to the updated ACT constraints.

\subsection{Generalized dependence $r=r(1-n_{S})$}\label{GENRNS}

Now, we consider the dependence of the tensor-scalar ratio on the spectral index of scalar perturbations $r=r(1-n_{S})$. Since the value of the spectral index of scalar perturbations is $n_{S}\simeq0.97$ and $1-n_{S}\simeq0.03\ll1$, we can write the dependence $r=r(1-n_{S})$ as
\begin{eqnarray}
\label{pert13G}
&&r=\sum^{\infty}_{k=0}\beta_{k}(1-n_{S})^{k},
\end{eqnarray}
where $(1-n_{S})\ll1$ is the small expansion parameter and $\beta_{k}$ are the constant coefficients.
This representation of dependence $r=r(1-n_{S})$ was considered earlier in~\cite{Fomin:2024xzm}.

Based on the previously obtained results, we can write this dependence for the case of a single-field inflationary models for the case of Einstein gravity as
\begin{eqnarray}
\label{pert13}
&&r\simeq\beta_{0}+\beta_{1}(1-n_{S})+\beta_{2}(1-n_{S})^{2},
\end{eqnarray}
where the values of the constant coefficients $\beta_{0}$, $\beta_{1}$ and $\beta_{2}$ depend on the type of inflationary model.

Thus, condition (\ref{QDSRCE}) leads to the breaking of series (\ref{pert13G}) in the second order.

\section{Connection between solutions for the extreme values of a scalar field}\label{EXACTSEC}

Now, we will consider the exact solutions of equations (\ref{E1m})--(\ref{E2m}) for cosmological dynamics corresponding to extreme values of a scalar field and transitions between these solutions.

\subsection{The first type of extreme values of a scalar field}

For the slow-roll parameters (\ref{I1}) and equation (\ref{QDSRCE}) we obtain the following Hubble parameter and scale factor
\begin{eqnarray}
\label{F1}
&&H(t)=-\xi_{0}t+\eta_{0},\\
\label{F2}
&&a(t)\sim\exp\left(\eta_{0}t-\frac{\xi_{0}}{2}t^{2}\right),
\end{eqnarray}
corresponding to the first extreme value of a scalar field.

For $\xi_{0}=m^{2}_{\varphi}/3$, exact solutions of cosmological dynamic equations (\ref{E1m})--(\ref{E2m}) with Hubble parameter (\ref{F1}) can be written as
\begin{eqnarray}
\label{F3}
&&\varphi(t)=\phi(t)-\phi_{0}=-\sqrt{\frac{2}{3}}m_{\varphi}t+\frac{\sqrt{6}}{m_{\varphi}}\eta_{0},\\
\label{F7}
&&V(\varphi)=\frac{m^{2}_{\varphi}\varphi^{2}}{2}+\Lambda,
\end{eqnarray}
where the second term in expression (\ref{F7}) is the negative cosmological constant
\begin{eqnarray}
\label{F3CC}
&&\Lambda=-\frac{m^{2}_{\varphi}}{3},
\end{eqnarray}
and $m_{\varphi}$ is the mass of the scalar field.

However, we can consider potential (\ref{F7}) as
\begin{eqnarray}
\label{F8}
&&V(\varphi)=\frac{m^{2}_{\varphi}\varphi^{2}}{2},
\end{eqnarray}
taking into account the possibility of a redefinition of the potential up to a constant $V\rightarrow V+\Lambda$.

In Sec. \ref{wac} it was shown that this inflationary model does not correspond to modern observational constraints.
Nevertheless, potential (\ref{F8}) implies the solutions of dynamic equations (\ref{DE1})--(\ref{DE3}) corresponding to the post-inflationary reheating stage which differ from (\ref{F1})--(\ref{F3}). These solutions can be written as follows~\cite{Turner:1983he,Kofman:1994rk,Shtanov:1994ce,
Kofman:1997yn,Bassett:2005xm,Dai:2014jja,Martin:2014nya,Kaur:2023wos}
\begin{eqnarray}
\label{F9}
&&H(t)\simeq\frac{2}{3t}\left(1+\frac{\sin\left(2m_{\varphi}t\right)}{2m_{\varphi}t}\right),\\
\label{F10}
&&\phi(t)\simeq \frac{2\sqrt{2}}{\sqrt{3}m_{\varphi}t}\cos(m_{\varphi}t),
\end{eqnarray}
where time averaged Hubble parameter $H(t)\simeq\frac{2}{3t}$ corresponds to the transition from reheating stage to the radiation domination stage.

The corresponding slow-roll parameters are~\cite{Kaur:2023wos}
\begin{eqnarray}
\label{F9SR}
&&\epsilon=\epsilon_{1}=3\sin^{2}\left(m_{\varphi}t\right),\\
\label{F10SR1}
&&\epsilon_{2}=3m_{\varphi}t\cot\left(m_{\varphi}t\right),\\
\label{F10SR}
&&\delta=\epsilon_{1}-\frac{\epsilon_{2}}{2}=3\sin^{2}\left(m_{\varphi}t\right)-\frac{3}{2}m_{\varphi}t\cot\left(m_{\varphi}t\right),
\end{eqnarray}
where at the beginning of the reheating stage $\epsilon=\delta=3$ and $\epsilon_{2}=0$.

From equations (\ref{SLOWROLLP1})--(\ref{SLOWROLLP2}) it follows that the beginning of the reheating stage $\epsilon=\delta=3$ corresponds to conditions $V=0$ and $V'_{\phi}=0$.

Also, at the beginning of the reheating stage, the equation of state parameter is
\begin{eqnarray}
\label{F10REHV1SP}
&&w_{beg(r)}=-1+\frac{2}{3}\epsilon=1,
\end{eqnarray}
whereas during the reheating stage, the equation of state parameter  averaged over all oscillations is equal to $\langle w\rangle=0$~\cite{Turner:1983he}.

The reheating stage corresponding to the inflationary model with potential (\ref{F8}) was considered in more detail, for example, in~\cite{Turner:1983he,Kofman:1994rk,Shtanov:1994ce,
Kofman:1997yn,Bassett:2005xm,Dai:2014jja,Martin:2014nya,Kaur:2023wos}.

\subsection{The second type of extreme values of a scalar field}

Since inflationary models with relation (\ref{VIII}) between the slow-roll parameters satisfy at least the Planck observational constraints, we consider exact solutions of cosmological dynamics equations (\ref{E1m})--(\ref{E2m}) for the corresponding Hubble parameter (\ref{EXPL}), namely for
\begin{eqnarray}
\label{EXPLG}
&&H(t)=\lambda+\frac{s}{t-t_{0}},
\end{eqnarray}
where $s=\frac{1}{\xi_{2}}=\frac{1}{\mu_{0}}$, and $0<s\leq3.7$ according to the previously obtained value $\mu_{0}\geq0.27$.

From definitions of the slow-roll parameters (\ref{epsilonex})--(\ref{deltanex}) for Hubble parameter (\ref{EXPLG}) we obtain
\begin{eqnarray}
\label{SRPH1G}
&&\epsilon=s\left[s+\lambda(t-t_{0})\right]^{-2},\\
\label{SRPH2H}
&&\delta=\left[s+\lambda(t-t_{0})\right]^{-1}.
\end{eqnarray}

Also, from (\ref{EXPLG}) and (\ref{deltanex}) we get
\begin{eqnarray}
\label{FOC33AG}
&&t-t_{0}=\frac{s}{H-\lambda},\\
\label{FOC34G}
&&\delta=\frac{1}{s}\left(1-\frac{\lambda}{H}\right).
\end{eqnarray}

From expressions (\ref{CROSSINGH}), (\ref{FOC34G}) and conditions $s>0$, $s\sim1$, $|\delta_{\ast}|\ll1$ at the crossing of the Hubble radius, we obtain
\begin{eqnarray}
\label{CROSSINGH2}
&&\lambda=H_{\ast}(1-s\delta_{\ast})\simeq\frac{\pi}{2}\sqrt{2A_{S}r}>0.
\end{eqnarray}

Taking into account expression (\ref{VER8F}) and $s=1/\mu_{0}$, from (\ref{CROSSINGH2}) one has
\begin{eqnarray}
\label{CROSSINGH3}
&&\lambda\simeq\pi(1-n_{S})\sqrt{2sA_{S}},
\end{eqnarray}
where $\lambda\simeq6\times10^{-6}$ for $s\simeq1$.

Also, taking into account that $H>0$, $s>0$ and $\delta<0$, from (\ref{FOC33AG})--(\ref{FOC34G}) one has
\begin{eqnarray}
\label{FOC35G}
&&\lambda>0,~~~~H<\lambda,~~~~t-t_{0}<0.
\end{eqnarray}


Now, from (\ref{E2m}) and (\ref{EXPLG}) we obtain the following type of evolution of the scalar field
\begin{eqnarray}
\label{EXPLG12}
&&\varphi(t)=\phi(t)-\phi_{0}=\sqrt{2s}\ln\left(q\lambda(t_{0}-t)\right),
\end{eqnarray}
where $q>0$ is a positive non-zero constant.

After substituting inverse dependence (\ref{EXPLG12}) into (\ref{EXPLG}) we obtain
\begin{eqnarray}
\label{EXPLG3}
&&H(\varphi)=\lambda-qs\exp\left(-\frac{\varphi}{\sqrt{2s}}\right).
\end{eqnarray}

Thus,  for Hubble parameter (\ref{EXPLG3}) from equation (\ref{E1mF}) we get the following exact expression of the scalar field potential
\begin{eqnarray}
\label{EXPLG4}
\nonumber
&&V(\varphi)=3\lambda^{2}+sq^{2}\lambda^{2}(3s-1)\exp\left(-\sqrt{\frac{2}{s}}\varphi\right)\\
&&-6sq\lambda^{2}\exp\left(-\frac{\varphi}{\sqrt{2s}}\right).
\end{eqnarray}

For the partial case $s=1/3$ potential (\ref{EXPLG4}) is reduced to the following form
\begin{eqnarray}
\label{EXPLG5}
&&V(\varphi)=3\lambda^{2}\left[1-\frac{2}{3}q\exp\left(-\sqrt{\frac{3}{2}}\varphi\right)\right].
\end{eqnarray}

Potential (\ref{EXPLG5}) corresponds to the exponential SUSY inflation (ESI)~\cite{Martin:2013tda}. This potential was previously considered in various cosmological models with different values of the constant $q$~\cite{Obukhov:1993fd,Stewart:1994ts,Dvali:1998pa,Cicoli:2008gp,Giudice:2010ka}.

On the other hand, for $s\neq1/3$ we can consider exact solutions of cosmological dynamic equations (\ref{E1m})--(\ref{E2m}) for Hubble parameter (\ref{EXPLG}) in the other form, that is, as
\begin{eqnarray}
\label{EXACTEPL1}
&&\varphi(t)=\sqrt{2s}\ln\left[\frac{3\lambda\mu}{3s-1}(t_{0}-t)\right],\\
\label{EXACTEPL2}
&&V(\varphi)=V_{0}\left[1-\mu\exp\left(-\frac{\varphi}{\sqrt{2s}}\right)\right]^{2}+\Lambda,\\
\label{EXACTEPL3}
&&V_{0}=\frac{9\lambda^{2}s}{3s-1},~~~~\Lambda=-\frac{3\lambda^{2}}{3s-1}=-\frac{V_{0}}{3s},
\end{eqnarray}
where $\mu>0$ is some positive constant.

Also, taking into account that $s>0$ and $V_{0}>0$, we obtain the following condition
\begin{eqnarray}
\label{EXACTEPL3S}
&&s>1/3.
\end{eqnarray}

For the case $\mu=1$ potential (\ref{EXACTEPL2}) corresponds to the superconformal $\alpha$-attractor A inflation (SAAI)~\cite{Martin:2013tda}.
For the partial case $\mu=1$ and $s=3/4$ potential (\ref{EXACTEPL2}) corresponds to the Starobinsky inflation~\cite{Mishra:2018dtg}.
Models of $\alpha$-attractors were considered earlier, for example, in~\cite{Kallosh:2013yoa,Ferrara:2016fwe,Kallosh:2017ced,
Gunaydin:2020ric,Galante:2014ifa,Kallosh:2022feu,Bhattacharya:2022akq,Iacconi:2023mnw}.

Also, for $\mu=1$ under the condition $\varphi\ll1$ potential (\ref{EXACTEPL2}) is reduced to the following form
\begin{eqnarray}
\label{REH1}
&&V(\varphi)\simeq\frac{9\lambda^{2}}{2(3s-1)}\varphi^{2}+\Lambda
=\frac{m^{2}_{\varphi}}{2}\varphi^{2}+\Lambda,
\end{eqnarray}
where the cosmological constant is defined as follows
\begin{eqnarray}
\label{COSMCONSTA1}
&&\Lambda=-\frac{m^{2}_{\varphi}}{3}=-\frac{3\lambda^{2}}{3s-1}.
\end{eqnarray}

Thus, the decreasing scalar field (\ref{EXPLG12}) leads to a transition between models for the first and second types of its extreme values.

After substituting $\lambda\simeq6\times10^{-6}$ and $s\simeq1$ into (\ref{COSMCONSTA1}) we obtain
\begin{eqnarray}
\label{COSMCONSTA}
&&|\Lambda|\simeq5\times10^{-11}\gg9\times10^{-121}.
\end{eqnarray}

Thus, an inflationary model with potential (\ref{EXACTEPL2}) implies an additional negative cosmological constant, which greatly (by $110$ orders) exceeds (\ref{COSMCONST2}).

Also, Hubble parameter (\ref{EXPLG}) is restricted by condition  $t<t_{0}$, since the case $t=t_{0}$ implies future singularity
$H\rightarrow\infty$. This type of future singularity for cosmological models with Hubble parameter (\ref{EXPLG}) was considered, for example, in~\cite{Bamba:2013iga}.
Thus, the additional problems of these inflationary models are a future singularity and the large negative cosmological constant.

Therefore, all considered inflationary scenarios constrained by the extreme values of a scalar field for the case of Einstein gravity contain difficulties when compared with observational data or when describing the dynamics of the expansion of the universe.

\section{Late time solutions with additional material fields}\label{LATESEC}

Now, we consider the universe filled with the scalar field and the additional material field as the ideal barotropic fluid. The pressure $p_{m}$ and density $\rho_{m}$  of the additional material field (in the chosen system of units the density is equal to the energy density) are related by the following equation of state $p_{m} =w_{m}\rho_{m}$.

Thus, for the total energy density $\rho_{tot} = \rho_{m} + \rho_{\phi}+\rho_{\Lambda}$ and pressure $p_{tot} =
p_{m} + p_{\phi}+p_{\Lambda}$ define the following equation of state parameter
\begin{equation}
\label{weff}
w_{tot}=\frac{p_{tot}}{\rho_{tot}}=\frac{p_{\phi}+p_{\Lambda}+p_{m}}{\rho_{\phi}+\rho_{\Lambda}+\rho_{m}},
\end{equation}
corresponding to the system of a scalar field, cosmological constant, and additional material field.

The action for these models is
\begin{eqnarray}
\label{actionLm}
\nonumber
&&S=\int d^4x\sqrt{-g}\left[\frac{1}{2}R - \frac{1}{2}g^{\mu\nu}\partial_{\mu}\phi \partial_{\nu}\phi-V(\phi)-\Lambda\right]\\
&&+S_{m},
\end{eqnarray}
where $S_{m}=\int d^4x\sqrt{-g}\mathcal{L}_{m}$ is the action corresponding to
the additional material field.

The cosmological dynamic equations corresponding to action (\ref{actionLm}) can be written as follows~\cite{Fomin:2018xhq}
\begin{eqnarray}
\label{beq2ahm}
&&3H^{2}=\frac{1}{2}\dot{\phi}^{2}+V(\phi)+\Lambda+\rho_{m},\\
\label{beq3ahm}
&&-3H^{2}-2\dot{H}=\frac{1}{2}\dot{\phi}^{2}-V(\phi)-\Lambda+p_{m},\\
\label{beq3ahmV}
&&\ddot{\phi} + 3 H\dot{\phi}+V'_{\phi}= 0,\\
\label{beq4ahm}
&&\dot{\rho}_m+3H(\rho_m+p_m)=0,
\end{eqnarray}
where $p_{\Lambda}=-\rho_{\Lambda}$, and $\Lambda=\rho_{\Lambda}$ in chosen system of units.

The constant equation of state parameter of the additional material field can be defined as
\begin{eqnarray}
\label{StateParameter}
&&w_{m}\equiv\frac{(m-3)}{3}=const.
\end{eqnarray}

Different types of material fields with constant equation of state parameter
correspond to different values of the parameter $m$, namely:
\begin{itemize}
\item $m=3$ and $w_{m}=0$ -- baryonic matter and cold dark matter (CDM);
\item $m=4$ and $w_{m}=1/3$ -- radiation;
\item $m=6$ and $w_{m}=1$ -- extremely stiff matter,
\end{itemize}
where we consider condition $m\neq0$, since for $m=0$ one has a cosmological constant with equation of state parameter $w_{\Lambda}=-1$.
In this case, the cosmological constant was already defined in dynamic equations (\ref{beq2ahm})--(\ref{beq3ahm}).

The evolution of the density and pressure of the material field can be defined as a result of Eq. \eqref{beq4ahm} as follows~\cite{Barrow:2016qkh}
\begin{eqnarray}
\label{MatterS}
&&\rho_{m}=\frac{\rho^{(\ast)}_{m}}{(a(t))^{m}},\\
\label{MatterS2}
&&p_{m}=\left(\frac{m-3}{3}\right)\frac{\rho^{(\ast)}_{m}}{(a(t))^{m}},
\end{eqnarray}
where constant $\rho^{(\ast)}_{m}\geq0$ determines the initial value of the density of the material field.

It should be noted that one can consider the multi-component material field with
the same equation of state parameter (\ref{StateParameter}), and following density and pressure
\begin{eqnarray}
\label{MatterS3}
&&\rho_{m}=\sum_{i}\frac{\rho^{(\ast)}_{m(i)}}{(a(t))^{m}},\\
\label{MatterS4}
&&p_{m}=\left(\frac{m-3}{3}\right)\sum_{i}\frac{\rho^{(\ast)}_{m(i)}}{(a(t))^{m}},
\end{eqnarray}
with different (in the general case) initial densities $\rho^{(\ast)}_{m(i)}$
instead of a single material field.

Also, we can consider only two independent dynamic equations in system (\ref{beq2ahm})--(\ref{beq4ahm})
with (\ref{MatterS3})--(\ref{MatterS4}), which can be represented as
\begin{eqnarray}
\label{DE1m}
\nonumber
&&V(\phi)+\Lambda=3H^{2}+\dot{H}-\frac{1}{2}\left(\rho_{m}-p_{m}\right)\\
&&=3H^{2}+\dot{H}-\left(1-\frac{m}{6}\right)\frac{\rho^{(\ast)}_{m}}{(a(t))^{m}},\\
\label{DE2m}
\nonumber
&&X=\frac{1}{2}\dot{\phi}^{2}=-\dot{H}-\frac{1}{2}\left(\rho_{m}+p_{m}\right)\\
&&=-\dot{H}-\frac{m\rho^{(\ast)}_{m}}{6(a(t))^{m}},
\end{eqnarray}
since after substituting expressions (\ref{MatterS})--(\ref{MatterS2}) and (\ref{DE1m})--(\ref{DE2m})
into equations (\ref{beq3ahmV})--(\ref{beq4ahm}) we obtain an identity.

Now, we consider the condition
\begin{eqnarray}
\label{DE2mCOND}
&&\frac{1}{2}\left(\rho_{m}+p_{m}\right)=\left(\frac{m\rho^{(\ast)}_{m}}{6 (a(t))^{m}}\right)
=const\times\dot{H}.
\end{eqnarray}

Under condition (\ref{DE2mCOND}) equation (\ref{DE2m}) reduces to one similar to equation (\ref{VAR1}).
This condition means that we consider the cosmological dynamics restricted by the same extreme values of a scalar field for the entire period of the universe's evolution.

The Hubble parameters and the scale factors corresponding to condition (\ref{DE2mCOND}) can be written as follows:

1. For $m=0$ the solutions of equations (\ref{QDSRCE}) and  (\ref{DE2mCOND}) is
\begin{eqnarray}
\label{DE2CCLHC}
&&H_{\Lambda}=const,\\
\label{DE2CCL}
&&a_{\Lambda}(t)\propto\exp\left(H_{\Lambda}\hat{t}\right),
\end{eqnarray}
where $\hat{t}=t-t_{m}$, $t_{m}$ is some characteristic time scale.

2. The first type of solutions of equations (\ref{QDSRCE}) and (\ref{DE2mCOND}) for $m\neq0$ is
\begin{eqnarray}
\label{DE2mPL}
&&a(t)\propto \hat{t}^{2/m},\\
\label{DE2mPL2}
&&H(t)=\frac{2}{m\hat{t}},
\end{eqnarray}
corresponding to the power-law expansion of the universe for $m>0$.

3. The second type of solutions of equations (\ref{QDSRCE}) and (\ref{DE2mCOND}) for $m\neq0$ can be written as
\begin{eqnarray}
\label{DE2mCOND1}
&&a(t)=\left(k_{1}e^{k_{3}m\hat{t}}-k_{2}e^{-k_{3}m\hat{t}}\right)^{2/m},\\
\label{DE2mCOND2}
&&H(t)=\frac{2k_{3}\left(k_{1}e^{k_{3}m\hat{t}}+k_{2}e^{-k_{3}m\hat{t}}\right)}{\left(k_{1}e^{k_{3}m\hat{t}}-k_{2}e^{-k_{3}m\hat{t}}\right)},
\end{eqnarray}
where $k_{1}$, $k_{2}$ and $k_{3}$ are some constants.

For Hubble parameter (\ref{DE2mCOND2}) from (\ref{epsilonex})--(\ref{deltanex}) for Hubble parameter (\ref{DE2mCOND2}) we obtain slow-roll parameters
\begin{eqnarray}
\label{DE2SRP1}
&&\epsilon=-\frac{\dot{H}}{H^{2}}=\frac{2mk_{1}k_{2}}{\left(k_{1}e^{k_{3}m\hat{t}}+k_{2}e^{-k_{3}m\hat{t}}\right)^{2}},\\
\label{DE2SRP2}
&&\delta=-\frac{\ddot{H}}{2H\dot{H}}=\frac{m}{2}=constant.
\end{eqnarray}

Also, from (\ref{DE2mCOND1})--(\ref{DE2SRP1}) we obtain the following relation between the scale factor, the Hubble parameter and the first slow-roll parameter
\begin{eqnarray}
\label{DE2SRP3M}
&&\epsilon H^{2}=-\dot{H}=\frac{8mk_{1}k_{2}k^{2}_{3}}{(a(t))^{m}}.
\end{eqnarray}

Now, we consider exact solutions of the cosmological dynamics equations (\ref{DE1m})--(\ref{DE2m})
for scale factor (\ref{DE2mCOND1}).

\subsection{Generalized exact solutions of cosmological dynamic equations}\label{GENSOLL}

For scale factor (\ref{DE2mCOND1}) and Hubble parameter (\ref{DE2mCOND2}) from (\ref{DE2m}) we can define the evolution of the scalar field
\begin{eqnarray}
\label{ME2G}
&&\varphi(t)=\phi(t)-\phi_{0}=\pm\beta{\rm arctanh}\left(\sqrt{\frac{k_{1}}{k_{2}}}e^{k_{3}m\hat{t}}\right),
\end{eqnarray}
with the following constants
\begin{eqnarray}
\label{BETAG}
&&\beta=4\sqrt{\frac{\eta}{m}},\\
\label{RCONSTG}
&&\eta=1-\frac{\rho^{(\ast)}_{m}}{\rho^{(c)}_{m}},\\
\label{RHOINITIALLG}
&&\rho^{(c)}_{m}=48k_{1}k_{2}k^{2}_{3},
\end{eqnarray}
where $\phi_{0}$ is the integration constant and $\rho^{(c)}_{m}$ is a certain critical density parameter.

Due to (\ref{DE1m})--(\ref{DE2m}) and (\ref{DE2mCOND1})--(\ref{DE2mCOND2})  potential and kinetic energy of the scalar field as functions of cosmic time and cosmological constant
can be written as
\begin{eqnarray}
\label{VGEN1M}
&&V(t)=\frac{(6-m)\left(\rho^{(c)}_{m}-\rho^{(\ast)}_{m}\right)}{6(a(t))^{m}},\\
\label{XGEN1M}
&&X(t)=\frac{m\left(\rho^{(c)}_{m}-\rho^{(\ast)}_{m}\right)}{6(a(t))^{m}},\\
\label{XGEN1ML}
&&\Lambda=12k^{2}_{3},
\end{eqnarray}
where the scale factor is defined by expression (\ref{DE2mCOND1}).

From (\ref{weff}), (\ref{beq2ahm})--(\ref{beq3ahm})  we can define the total equation of state parameter
\begin{eqnarray}
\label{weffGENM}
&&w_{tot}=\frac{X-V-\Lambda+p_{m}}{X+V+\Lambda+\rho_{m}}=-1+\frac{2}{3}\epsilon,
\end{eqnarray}
where the slow-roll parameter $\epsilon$ is defined by expression (\ref{DE2SRP1}).

Also, from (\ref{ME2G}) and (\ref{VGEN1M}) we obtain the following potential of the scalar field
\begin{eqnarray}
\label{VGEN2M}
&&V(\varphi)=M\sinh^{2}\left(\frac{2\varphi}{\beta}\right),
\end{eqnarray}
where
\begin{eqnarray}
\label{MM}
M=2k^{2}_{3}(6-m)\eta=\Lambda\left(1-\frac{m}{6}\right)\eta.
\end{eqnarray}

This potential corresponds to those considered earlier, for example, in~\cite{Rubano:2001xi,Guo:2006ab}.

Due to (\ref{VGEN1M})--(\ref{XGEN1M}) the energy density and pressure of the scalar field are
\begin{eqnarray}
\label{RHOPHIRM2}
&&\rho_{\phi}=X+V=\frac{\left(\rho^{(c)}_{m}-\rho^{(\ast)}_{m}\right)}{(a(t))^{m}},\\
\label{PPHIRM2}
&&p_{\phi}=X-V=\frac{(m-3)\left(\rho^{(c)}_{m}-\rho^{(\ast)}_{m}\right)}{3(a(t))^{m}}.
\end{eqnarray}

Expression (\ref{RHOPHIRM2}) implies the following properties of the scalar field: the case $\rho^{(\ast)}_{m}<\rho^{(c)}_{m}\, (\eta>0)$ corresponds to the canonical scalar field with positive non-zero energy density $\rho_{\phi}>0$.
For $\rho^{(\ast)}_{m}=\rho^{(c)}_{m}\, (\eta=0)$ one has $\rho_{\phi}=0$.
The case $\rho^{(\ast)}_{m}>\rho^{(c)}_{m}\, (\eta<0)$ corresponds to the phantom field with a negative energy density $\rho_{\phi}<0$, which violates the weak energy condition~\cite{Kontou:2020bta}.

From expressions (\ref{MatterS})--(\ref{MatterS2}) and (\ref{RHOPHIRM2})--(\ref{PPHIRM2}) it follows that the state parameters of the additional material field and scalar field
\begin{eqnarray}
\label{STATRPARFM}
&&w_{\phi}=\frac{p_{\phi}}{\rho_{\phi}}=w_{m}=\frac{p_{m}}{\rho_{m}}=-1+\frac{m}{3},
\end{eqnarray}
are equal to each other.

It should be noted that condition $w_{\phi}=const$ was considered in~\cite{Rubano:2001xi} as initially given in order to construct the exact cosmological solutions for the universe filled with a scalar field and additional material field.

Condition (\ref{STATRPARFM}) implies that the scalar field can effectively describe the material field with the same equation of state parameter. Therefore, one can consider the scalar field as a real field or as effective field.

Thus, for the case of cosmological dynamics (\ref{DE2mCOND1}) the system of a scalar field with (\ref{RHOPHIRM2})--(\ref{PPHIRM2}) and additional material field with (\ref{MatterS})--(\ref{MatterS2}) can be considered as two-components ideal barotropic cosmic fluid with following density and pressure
\begin{eqnarray}
\label{MatterS3M}
&&\rho_{m}=\sum\frac{\rho^{(\ast)}_{m(i)}}{(a(t))^{m}}=
\frac{\rho^{(c)}_{m}}{(a(t))^{m}},\\
\label{MatterS4M}
\nonumber
&&p_{m}=\left(\frac{m-3}{3}\right)\sum\frac{\rho^{(\ast)}_{m(i)}}{(a(t))^{m}}\\
&&=\left(\frac{m-3}{3}\right)\frac{\rho^{(c)}_{m}}{(a(t))^{m}}.
\end{eqnarray}

It should also be noted that property (\ref{STATRPARFM}) is typical for scaling solutions~\cite{Copeland:1997et,Billyard:1998hv,Haro:2019peq}.

Taking into account $\rho_{\Lambda}=\Lambda=12k^{2}_{3}$, from (\ref{RHOINITIALLG}), (\ref{RHOPHIRM2}) and (\ref{MatterS}) we obtain
\begin{eqnarray}
\label{ME9M}
&&\frac{\rho_{\Lambda}}{\rho_{\phi}+\rho_{m}}=\frac{(a(t))^{m}}{4k_{1}k_{2}}.
\end{eqnarray}

Thus, cosmological constant dominates over the scalar field and additional material field at large times.

Under a special initial condition $\rho^{(c)}_{m}=\rho^{(\ast)}_{0m}$ the scalar field can be eliminated from these solutions. For the case $\rho^{(\ast)}_{m}=0$, the additional material field can also be eliminated from these solutions.

\subsection{Cosmographic parameters}\label{cosmographic}

Based on scale factor (\ref{DE2mCOND1}) and Hubble parameter (\ref{DE2mCOND2}) we can write the following expressions for the cosmographic parameters~\cite{Visser:2004bf,Capozziello:2011hj,Aviles:2012ay,
Dunsby:2015ers,Mehrabi:2021cob}
\begin{eqnarray}
\label{CPHOM}
\nonumber
&&q(n=2),j(n=3),l(n=4),s(n=5)\\
&&=\left(\frac{1}{a}\frac{d^{n}a}{dt^{n}}\right)\frac{1}{H^{n}},
\end{eqnarray}
in terms of the first slow-roll parameter (\ref{DE2SRP1}) as
\begin{eqnarray}
\label{deceleration}
&&q=-1+\epsilon,\\
\label{jerk}
&&j=1+(m-3)\epsilon,\\
\label{snap}
&&s=1-(m^{2}-4m+6)\epsilon+(3-m)\epsilon^{2},\\
\label{lerk}
\nonumber
&&l=1+(m^{3}-5m^{2}+10m-10)\epsilon\\
&&+(4m^{2}-15m-15)\epsilon^{2}.
\end{eqnarray}

To obtain dependence of the Hubble parameter on redshift $H=H(z)$ we will use the following expressions~\cite{Visser:2004bf,Capozziello:2011hj,
Aviles:2012ay,Dunsby:2015ers,Mehrabi:2021cob}
\begin{eqnarray}
\label{z1}
&&a(t)=\frac{a_{0}}{1+z},\\
\label{z2}
&&\frac{dH}{dt}=-(1+z)H\frac{dH}{dz},
\end{eqnarray}
where $z$ is the redshift, $a_{0}=a(\hat{t}=t_{0})=a(z=0)$, where $t_{0}$ is the cosmic time corresponding to condition $z=0$.

From (\ref{DE2mCOND1})--(\ref{DE2mCOND2}), (\ref{DE2SRP3M}) and (\ref{z1})--(\ref{z2}) we obtain
\begin{eqnarray}
\label{z3}
\nonumber
&&\epsilon H^{2}=-\frac{dH}{dt}=\frac{m\rho^{(c)}_{m}}{6}\left(\frac{1+z}{a_{0}}\right)^{m}\\
&&=(1+z)H(z)\frac{dH(z)}{dz}.
\end{eqnarray}

The solution of equation (\ref{z3}) can be written as
\begin{eqnarray}
\label{z4}
&&H(z)=H_{0}\sqrt{\Omega_{0m}(1+z)^{m}+\Omega_{0\Lambda}},
\end{eqnarray}
where $\Omega_{0m}=\frac{\rho^{(c)}_{m}}{3H^{2}_{0}a^{m}_{0}}$ is the relative density
(energy density) of the scalar field and the additional material filed, and $\Omega_{0\Lambda}=\frac{\rho_{\Lambda}}{3H^{2}_{0}a^{m}_{0}}$ is the relative
energy density of the cosmological constant, which we consider as the constant of integration.

It should be noted that we can consider relative densities as $\Omega_{0m}=\frac{\rho^{(c)}_{m}}{\rho_{c}}$ and $\Omega_{0\Lambda}=\frac{\rho_{\Lambda}}{\rho_{c}}$ under condition $a_{0}=1$ or by means of a redefinition of the constants $k_{1}$ and $k_{2}$, where the critical density is $\rho_{c}=3H^{2}_{0}$.

Also, from (\ref{z3}) and (\ref{z4}) we obtain expression for the first slow-roll parameter in terms of the redshift
\begin{eqnarray}
\label{z5}
&&\epsilon(z)=\left(\frac{m}{2}\right)\frac{\Omega_{0m}(1+z)^{m}}{\Omega_{0m}(1+z)^{m}+\Omega_{0\Lambda}}.
\end{eqnarray}

Expressions (\ref{deceleration})--(\ref{z5}) allow one to obtain a cosmographic description of the expansion of the universe filled with a cosmological constant, scalar field, and a material field(s) with equation of state parameter (\ref{STATRPARFM}).

\subsection{The connection between different stages of the universe's evolution}\label{GENS}

In order to connect different stages of the evolution of the universe, we consider the following condition
\begin{eqnarray}
\label{CCST}
&&k_{2}=k_{1}.
\end{eqnarray}

Thus, Hubble parameter (\ref{DE2mCOND2}) can be written as
\begin{eqnarray}
\label{GS1H}
&&H_{ms}(t)=2B\coth\left(Bm\hat{t}\right),
\end{eqnarray}
with corresponding scale factor
\begin{eqnarray}
\label{GS2A}
\nonumber
&&a_{ms}(t)=\left(\frac{A_{m}}{2}e^{Bm\hat{t}}-\frac{A_{m}}{2}e^{-Bm\hat{t}}\right)^{2/m}\\
&&=A^{2/m}_{m}\sinh^{2/m}\left(Bm\hat{t}\right),
\end{eqnarray}
where
\begin{eqnarray}
\label{AA2}
&&k_{1}=k_{2}=\frac{A_{m}}{2},\\
\label{RCONST}
&&k_{3}=B,\\
\label{RHOINITIALL}
&&\rho^{(c)}_{m}=12A^{2}_{m}B^{2},
\end{eqnarray}
and $A_{m}>0$ is the dimensionless positive scaling constant, which takes a different value for different $m$.

In this case, the exact solutions of cosmological dynamic equations are the same as presented in Sec. \ref{GENSOLL} with taking into account expressions (\ref{AA2})--(\ref{RHOINITIALL}) and following replacement of the scale factor $a\rightarrow a_{ms}$.

At early times $Bm\hat{t}\ll1$ Hubble parameter (\ref{GS1H}) is reduced to (\ref{DE2mPL2}) corresponding to the post-inflationary stages of radiation and matter domination, while at late times $Bm\hat{t}\gg1$ Hubble parameter (\ref{GS1H}) is reduced to (\ref{DE2CCLHC}) corresponding to cosmological constant domination with negligible energy density of the material fields.

\subsection{Post-inflationary stages of the universe's evolution}\label{post}

Now, we consider exact solutions (\ref{ME2G})--(\ref{XGEN1ML}) with Hubble parameter and scale factor (\ref{GS1H})--(\ref{GS2A}) at the small times  $Bm\hat{t}\ll1$.

We can write the main terms in the expansions of expressions (\ref{ME2G})--(\ref{XGEN1ML}) and (\ref{GS1H})--(\ref{GS2A}) in a series as
\begin{eqnarray}
\label{POST1}
&&H_{ms}(t)=\frac{2}{m\hat{t}},\\
\label{POST2}
&&a_{ms}(t)=\left(mA_{m}B\hat{t}\right)^{2/m},\\
\label{POST3}
&&\varphi(t)=2\sqrt{\frac{\eta}{m}}\,\ln\left(\frac{1}{2}Bm\hat{t}\right),\\
\label{POST4}
&&V(t)=\frac{2(6-m)\eta}{m^{2}\hat{t}^{2}},\\
\label{POST5}
&&X(t)=\frac{2\eta}{m\hat{t}^{2}},
\end{eqnarray}
where the cosmological constant can be neglected as in the inflationary stage.

Also, from (\ref{POST3}) and (\ref{POST4}) we obtain
\begin{eqnarray}
\label{POST6}
&&V(\varphi)= V_{0m}\exp\left(-\sqrt{\frac{m}{\eta}}\varphi\right),
\end{eqnarray}
where
\begin{eqnarray}
\label{POST7}
&&V_{0m}=\frac{(6-m)\left(\rho^{(\Lambda)}_{m}-\rho^{(\ast)}_{m}\right)}{24A^{2}_{m}}.
\end{eqnarray}

The equal signs in expressions (\ref{POST1})--(\ref{POST7}) are written here because
the main terms of the expansion of exact solutions (\ref{ME2G})--(\ref{XGEN1ML})
and (\ref{GS1H})--(\ref{GS2A}) in a series are themselves exact solutions of
cosmological dynamic equations (\ref{DE1m})--(\ref{DE2m}).
These solutions were considered earlier, for example, in~\cite{Hawkins:2001zx,Fomin:2018xhq}.

For this case, both slow-roll parameters are
\begin{eqnarray}
\label{POST9SR}
&&\epsilon=\delta=\frac{m}{2}=const.
\end{eqnarray}

From (\ref{weffGENM}) and (\ref{POST9SR}) we obtain the following equation of state parameter
\begin{eqnarray}
\label{weffGENMEFF}
&&w_{tot}=w_{m}=-1+\frac{m}{3}.
\end{eqnarray}

For the case $m=6$ from (\ref{POST7}), (\ref{POST9SR}) and (\ref{weffGENMEFF}) we obtain
\begin{eqnarray}
\label{REHEFF}
&&\epsilon=\delta=3,~~~w_{m}=1,~~~V=0,
\end{eqnarray}
where the first slow-roll parameter $\epsilon=\epsilon_{m}=3$ takes the maximal value.

Parameters (\ref{REHEFF}) correspond to the beginning of the reheating stage with subsequent transition to the radiation domination stage~\cite{Turner:1983he,Kofman:1994rk,Shtanov:1994ce,Kofman:1997yn,
Bassett:2005xm,Dai:2014jja,Martin:2014nya,Kaur:2023wos}.

For the cases $m=4$ and $m=3$ expressions (\ref{POST1})--(\ref{weffGENMEFF}) correspond to the post-inflationary stages of radiation domination (RD) with $w_{m}=1/3$ and matter domination (MD) with $w_{m}=0$. At these stages, for the case $\eta\neq0$, the scalar field behaves like radiation and matter accordingly.

Thus, the obtained solutions give the connection between different post-inflationary stages of the universe's evolution.

\subsection{$\Lambda$CDM-model}

Now, we can consider the solutions obtained in Sec.\ref{GENS} with $m=3$.
In this case, from (\ref{DE2SRP1}), (\ref{weffGENM}) and (\ref{GS1H})--(\ref{GS2A}) we have
\begin{eqnarray}
\label{LCDMH}
&&H(t)=2B\coth\left(3B\hat{t}\right),\\
\label{LCDM}
&&a(t)=A^{2/3}_{3}\sinh^{2/3}\left(3B\hat{t}\right),\\
\label{LCDM2}
&&w_{tot}=-\left(\frac{1-e^{6B\hat{t}}}{1+e^{6B\hat{t}}}\right)^{2}=-\tanh^{2}\left(3B\hat{t}\right),\\
\label{LCDM3}
&&w_{\phi}=w_{m}=0,~~~~~j=1.
\end{eqnarray}

In addition, we can estimate the constant $B$ from (\ref{COSMCONST2}) and the expression $\Lambda=12B^{2}$ as
\begin{eqnarray}
\label{B}
&&B=\frac{1}{2}\sqrt{\frac{\Lambda}{3}}\simeq2.7\times10^{-61}.
\end{eqnarray}

The corresponding scalar field evolution (\ref{ME2G}) and potential (\ref{VGEN2M}) with the cosmological constant for $m=3$ can be written as
\begin{eqnarray}
\label{ME2GLCDM}
&&\varphi(t)=\pm4\sqrt{\frac{\eta}{3}}{\rm arctanh}
\left[\exp\left(\frac{1}{2}\sqrt{3\Lambda}\,\hat{t}\right)\right],\\
\label{VGEN2LCDM}
&&V(\varphi)+\Lambda=\frac{\Lambda}{2}\eta\sinh^{2}\left(\sqrt{\frac{3}{4}}\varphi\right)+\Lambda.
\end{eqnarray}

These solutions correspond to the $\Lambda$CDM-model with the scalar
field~\cite{Akarsu:2013xha,Chakraborty:2022evc}
which is cosmographically equivalent to the $\Lambda$CDM-model without scalar
field due to the same dynamic of the universe's expansion (\ref{LCDM}).

Thus, for $\rho^{(\ast)}_{m}\neq\rho^{(c)}_{0m}$ ($\eta\neq0$) scalar field (\ref{ME2GLCDM})
can be considered as the real or effective source of the Cold Dark Matter (CDM).
For the case $\rho^{(\ast)}_{m}=\rho^{(c)}_{0m}$ ($\eta=0$) the scalar field is absent,
and the Cold Dark Matter can be described by some material field with state parameter $w_{m}=0$
due to expressions (\ref{MatterS3})--(\ref{MatterS4}).

Thus, unification of the analysis of cosmological dynamics
restricted by the extreme values of a scalar field for arbitrary cosmic time in the case of Einstein gravity leads to the $\Lambda$CDM-model for the description of the modern stage of the universe's evolution.

For the case of negligible spatial curvature $\Omega_{K}\simeq0$ in the modern stage of the evolution of the universe, the following relation is fulfilled
\begin{eqnarray}
\label{LCDM4}
&&\Omega_{0\Lambda}+\Omega_{0DM}+\Omega_{0b}+\Omega_{0r}=1.
\end{eqnarray}

Also, in the modern stage, the radiation density is $\Omega_{0r}\simeq5\times10^{-5}$, the baryonic matter density is $\Omega_{0b}\simeq0.04$, the Dark Matter density is $\Omega_{0DM}\simeq0.26$, and the cosmological constant energy density
is $\Omega_{0\Lambda}\simeq0.7$~\cite{Planck:2018vyg,ACT:2025fju,ACT:2025tim}.

\begin{figure}[ht]
	\centering
		\includegraphics[width=0.4\textwidth]{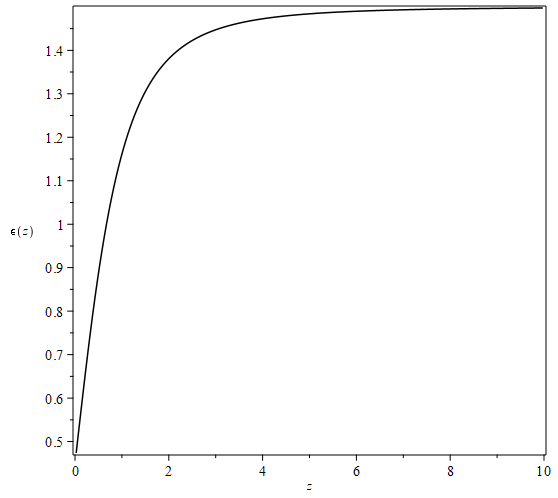}
	\caption{The dependence $\epsilon=\epsilon(z)$ for $m=3$, $\Omega_{0m}=0.3$
and $\Omega_{0\Lambda}=0.7$. The transition from the decelerated expansion of the universe
to the accelerated expansion $\epsilon(z=z_{c})=1$ corresponds to redshift $z_{c}\simeq0.672$.}
	\label{FIG5}
\end{figure}

Thus, we can consider
\begin{eqnarray}
\label{LCDM5}
&&\Omega_{0\Lambda}+\Omega_{0m}\simeq1,
\end{eqnarray}
where $\Omega_{0m}=\Omega_{0DM}+\Omega_{0b}\simeq0.3$.

From (\ref{z4}) and (\ref{LCDM5}) for $m=3$ we obtain
\begin{eqnarray}
\label{z4Lambda}
&&H(z)=H_{0}\sqrt{\Omega_{0m}(1+z)^{3}+(1-\Omega_{0m})}.
\end{eqnarray}

From expressions (\ref{deceleration}), (\ref{z5}) and (\ref{LCDM5}) for $m=3$ and $z=0$ we get
\begin{eqnarray}
\label{z0}
&&q(z=0)=-1+\epsilon(z=0)=-1+\frac{3}{2}\Omega_{0m}.
\end{eqnarray}

Thus, for $\Omega_{0m}\simeq0.3$ from (\ref{z0}) we get
\begin{eqnarray}
\label{z01}
&&\epsilon(z=0)\simeq0.45<1,~~~q(z=0)\simeq-0.55,
\end{eqnarray}
corresponding to accelerated expansion of the universe.

Also, for $m=3$ and $z\rightarrow\infty$ from (\ref{z5}) we get
\begin{eqnarray}
\label{zm}
&&\epsilon(z\rightarrow\infty)=\frac{3}{2}>1,
\end{eqnarray}
corresponding to a decelerated expansion of the early universe.

Thus, the early time asymptotic of the $\Lambda$CDM-model
is the matter dominance stage described by solutions (\ref{POST1})--(\ref{weffGENMEFF}) with $m=3$.
Fig. \ref{FIG5} demonstrate the transition from decelerated to accelerated expansion of the universe
in accordance with the $\Lambda$CDM-model.

Also, one can consider cosmographical description of the $\Lambda$CDM-model for $m=3$ by expressions presented in Sec.\ref{cosmographic} in more detail.
This cosmographical description of the $\Lambda$CDM-model based on modern observational data was presented, for example, in~\cite{Pourojaghi:2024bxa}.

Despite the fact that the $\Lambda$CDM-model successfully satisfies the most astronomical observations
~\cite{BOSS:2016wmc,Pan-STARRS1:2017jku,Benisty:2020otr,Dainotti:2022wli,
Cao:2021cix,Liu:2022inf,Pourojaghi:2022zrh}, it should be noted that there are some tensions between modern observational data and the $\Lambda$CDM-model, see, for example, in~\cite{Perivolaropoulos:2021jda,DiValentino:2021izs,Hu:2023jqc}.

One of the most significant problems is the Hubble tension ($4-6\sigma$), i.e. the difference between the Hubble parameter value obtained from the local measurements $H^{(Loc)}_{0}=73.04\pm1.04$ km/s/Mpc~\cite{Riess:2021jrx,Breuval:2024lsv} and by the CMB observations $H^{(CMB)}_{0}=68.43\pm0.27$ km/s/Mpc corresponding to the $\Lambda$CDM-model~\cite{ACT:2025fju,ACT:2025tim}.
At the moment there is no unambiguous interpretation of the Hubble tension problem. There are various options to solve this problem within the framework of the $\Lambda$CDM-model, as well as interpretations associated with the need to go beyond this model~\cite{Perivolaropoulos:2021jda,DiValentino:2021izs,Hu:2023jqc}.

Thus, single-scalar field cosmological models based on Einstein gravity with dynamics restricted by condition (\ref{QDSRCE}) imply difficulties in verifying them by observational data both at the early inflationary stage and at the second late stage of accelerated expansion of the universe.
Therefore, the result of the proposed analysis is the need to modify Einstein gravity theory in cosmological models with one canonical scalar field.

\section{Conclusion}\label{SEC7}
In this paper, the description of the evolution of the universe in cosmological models with a canonical scalar field based on Einstein gravity was considered. Possible types of cosmological models were restricted by extreme values of the scalar field. In order to determine the possible types of cosmological dynamics between extreme values of the scalar field, the expansion of dependence $\dot{H}=\dot{H}(H)$ was considered taking into account the non-stationarity of the universe and the stages of accelerated expansion of the universe.
The first extreme value of the scalar field was shown to correspond to the zeroth order of the expansion $\dot{H}=\dot{H}(H)$, and the second extreme value of the scalar field corresponds to the second order of this expansion.
This fact allowed us to consider equation (\ref{QDSRCE}) instead of expansion (\ref{QDSR2}). The solutions of equation (\ref{QDSRCE}) correspond to the possible types of cosmological dynamics and the possible relations between the slow-roll parameters as well.
It was shown that all the relations obtained between the slow-roll parameters correspond to the different well-known models of cosmological inflation. Thus, the proposed approach allows, on the one hand, to relate various inflationary models and, on the other hand, it limits their number.
Furthermore, it was shown that all the resulting inflationary models do not satisfy modern observational constraints on the values of the parameters of cosmological perturbations.
Exponential SUSY inflation (ESI) and superconformal $\alpha$-attractor inflationary model (SAAI) with Hubble parameter (\ref{EXPLG}) were examined in more detail. The additional problems of these models are the future singularity and the presence of a large negative cosmological constant $\left(-\frac{\Lambda}{\Lambda_{obs}}\right)\sim10^{110}$ (for SAAI).

Based on the fact that the scalar field is restricted by the same extreme values throughout the entire evolution of the universe, it was shown that the $\Lambda$-CDM model is the only possible model to describe the second accelerated expansion within the framework of the proposed approach.
It was also shown that the presence or absence of a scalar field in the post-inflationary eras is determined by the initial density of the additional material field(s).
Finally, based on the proposed analysis, it can be concluded that cosmological models with one canonical scalar field must be considered on the basis of modified gravity theories instead of Einstein gravity.

When considering single-field cosmological models based on modified theories of gravity, there are additional possibilities for their verification by observational constraints. For example, in~\cite{Fomin:2017vae,Fomin:2017qta,Fomin:2018typ,Fomin:2020hfh,Fomin:2017sbt,Fomin:2018blx,Fomin:2020woj,Fomin:2022ozv,Chervon:2023gio,Fomin:2024xzm}, a model-independent approach was proposed to construct verifiable single-field inflationary models based on Einstein-Gauss-Bonnet gravity~\cite{Fomin:2017vae,Fomin:2017qta,Fomin:2018typ,Fomin:2020hfh}, scalar-tensor gravity~\cite{Fomin:2017sbt,Fomin:2018blx,Fomin:2020woj,Fomin:2022ozv} and scalar-torsion gravity~\cite{Chervon:2023gio,Fomin:2024xzm}. 
It is also necessary to note the influence of the non-minimal coupling of the scalar field and curvature on the cosmological parameters, for example, for the case of non-minimal Higgs inflation~\cite{Mishra:2018dtg,Bezrukov:2007ep,Bezrukov:2013fka,Mohammedi:2022qqj}.
The other approaches to constructing and analyzing cosmological models based on these modified theories of gravity can be found, for example, in~\cite{Fernandes:2022zrq,vandeBruck:2015gjd,Chakraborty:2018scm,Odintsov:2018zhw,Odintsov:2020sqy,Odintsov:2025kyw,Odintsov:2025bmp,Pozdeeva:2021iwc,Pozdeeva:2024ihc,
Fujii:2003pa,Belinchon:2016lwr,Motohashi:2019tyj,Gonzalez-Espinoza:2019ajd,Gonzalez-Espinoza:2020azh,Leon:2022oyy}.
At the same time, there are a huge number of cosmological models with the other modified gravity theories (see, for example, in~\cite{Sotiriou:2008rp,Nojiri:2010wj,DeFelice:2010aj,DeFelice:2011bh,DeFelice:2011zh,DeFelice:2011jm,
Capozziello:2011et,Nojiri:2017ncd,Ishak:2018his,CANTATA:2021asi,Bahamonde:2021gfp,Odintsov:2023weg}).

Several cases following from expansion (\ref{QDSR2}) can be considered.
The first case corresponds to GR-like dynamics, i.e. cosmological dynamics defined by expression (\ref{QDSRCE}). In this case, modifications of Einstein gravity influence the values of the coefficients in dependence of the tensor-to-scalar ratio from the spectral index of the scalar perturbations (\ref{pert13}) under certain relationships between the model parameters. The details of constructing such cosmological models were considered in~\cite{Fomin:2020hfh,Fomin:2022ozv,Fomin:2024xzm}.
Thus, the cosmological models obtained due to expression (\ref{QDSRCE}) can be considered as some seed models. These models can be modified by using different possible types of modified gravity. The investigation of modifications of the seed models is interesting because it allows one to explicitly parameterize the influence of the modifications of Einstein gravity theory on the cosmological parameters.

As the second particular case, we can consider the Hubble parameter (\ref{EXPL}) corresponding to the combination of the de Sitter and Friedman solutions. This case is interesting because it separates GR-like cosmological dynamics from other possible types of dynamics in expansion (\ref{QDSR2}).
Cosmological models based on generalized scalar-tensor gravity theory with a single canonical scalar field and Hubble parameter (\ref{EXPL}) were considered in~\cite{Fomin:2020woj}. In this work it was shown that, in contrast to the case of Einstein gravity theory, the constant parameter $\lambda$ can be negative which solves the problem of a future singularity.
The other types of cosmological models with dynamics defined by Hubble parameter (\ref{EXPL}) were considered, for example, in~\cite{Akarsu:2013xha,Yadav:2014uoa,Mishra:2015jja,Zia:2018tss,Tripathy:2021vjt,Alhallak:2022szt,Varshney:2021mvx}.

The third case is the breaking of expansion (\ref{QDSR2}) in some $k$-order for $k>2$ or the absence of the breaking of expansion (\ref{QDSR2}). In this case, it is impossible to reconstruct the exact analytical form of all types of cosmological dynamics. Dynamical system analysis for selected parameters of cosmological models is a fairly effective method to consider these types of models~\cite{Copeland:2006wr,Bahamonde:2017ize,Boehmer:2022wln,Alho:2023pkl,
Chakraborty:2024zga,Duchaniya:2024vvc}.
However, it should be noted that if the expansion (\ref{QDSR2}) can be interrupted at some order $k>2$, then one type of Hubble parameter (\ref{QDSRB}) is explicitly defined. Cosmological models with this type of dynamics based on modified gravity theories were considered in~\cite{Bamba:2013iga}.
It should be noted that the requirement that the expansion (\ref{QDSR2}) terminates in some $k$-order leads to additional restrictions on the parameters of modified gravity theories.
Also, according to modern observations, the speed of gravitational waves is close to the speed of light in a vacuum with high accuracy~\cite{LIGOScientific:2017ync}, which limits the types and parameters of modified theories of gravity~\cite{LIGOScientific:2017ync,Ezquiaga:2017ekz,Odintsov:2019clh,Odintsov:2020zkl,Shoom:2021mdj}.

\bibliographystyle{utphys}
\bibliography{SM}

\providecommand{\href}[2]{#2}\begingroup\raggedright\begin{thebibliography}{100}

\bibitem{Starobinsky:1980te}
A.~A. Starobinsky, ``{A New Type of Isotropic Cosmological Models Without
  Singularity},'' \href{http://dx.doi.org/10.1016/0370-2693(80)90670-X}{{\em
  Phys. Lett. B} {\bf 91} (1980)  99--102}.

\bibitem{Guth:1980zm}
A.~H. Guth, ``{The Inflationary Universe: A Possible Solution to the Horizon
  and Flatness Problems},''
  \href{http://dx.doi.org/10.1103/PhysRevD.23.347}{{\em Phys. Rev. D} {\bf 23}
  (1981)  347--356}.

\bibitem{Linde:1981mu}
A.~D. Linde, ``{A New Inflationary Universe Scenario: A Possible Solution of
  the Horizon, Flatness, Homogeneity, Isotropy and Primordial Monopole
  Problems},'' \href{http://dx.doi.org/10.1016/0370-2693(82)91219-9}{{\em Phys.
  Lett. B} {\bf 108} (1982)  389--393}.

\bibitem{Albrecht:1982wi}
A.~Albrecht and P.~J. Steinhardt, ``{Cosmology for Grand Unified Theories with
  Radiatively Induced Symmetry Breaking},''
  \href{http://dx.doi.org/10.1103/PhysRevLett.48.1220}{{\em Phys. Rev. Lett.}
  {\bf 48} (1982)  1220--1223}.

\bibitem{Linde:1983gd}
A.~D. Linde, ``{Chaotic Inflation},''
  \href{http://dx.doi.org/10.1016/0370-2693(83)90837-7}{{\em Phys. Lett. B}
  {\bf 129} (1983)  177--181}.

\bibitem{Linde:1984ir}
A.~D. Linde, ``{The Inflationary Universe},''
  \href{http://dx.doi.org/10.1088/0034-4885/47/8/002}{{\em Rept. Prog. Phys.}
  {\bf 47} (1984)  925--986}.

\bibitem{Belinsky:1985zd}
V.~A. Belinsky, I.~M. Khalatnikov, L.~P. Grishchuk, and Y.~B. Zeldovich,
  ``{Inflationary stages in cosmological models with a scalar field},''
  \href{http://dx.doi.org/10.1016/0370-2693(85)90644-6}{{\em Phys. Lett. B}
  {\bf 155} (1985)  232--236}.

\bibitem{Piran:1986dh}
T.~Piran, ``{On General Conditions for Inflation},''
  \href{http://dx.doi.org/10.1016/0370-2693(86)90039-0}{{\em Phys. Lett. B}
  {\bf 181} (1986)  238--243}.

\bibitem{Baumann:2014nda}
D.~Baumann and L.~McAllister,
  \href{http://dx.doi.org/10.1017/CBO9781316105733}{{\em {Inflation and String
  Theory}}}.
\newblock Cambridge Monographs on Mathematical Physics. Cambridge University
  Press, 5, 2015.
\newblock \href{http://arxiv.org/abs/1404.2601}{{\tt arXiv:1404.2601
  [hep-th]}}.

\bibitem{Chervon:2019sey}
S.~Chervon, I.~Fomin, V.~Yurov, and A.~Yurov,
  \href{http://dx.doi.org/10.1142/11405}{{\em {Scalar Field Cosmology}}},
  vol.~13 of {\em Series on the Foundations of Natural Science and Technology}.
\newblock WSP, Singapur, 2019.

\bibitem{Mukhanov:1990me}
V.~F. Mukhanov, H.~A. Feldman, and R.~H. Brandenberger, ``{Theory of
  cosmological perturbations. Part 1. Classical perturbations. Part 2. Quantum
  theory of perturbations. Part 3. Extensions},''
  \href{http://dx.doi.org/10.1016/0370-1573(92)90044-Z}{{\em Phys. Rept.} {\bf
  215} (1992)  203--333}.

\bibitem{Riotto:2002yw}
A.~Riotto, ``{Inflation and the theory of cosmological perturbations},'' {\em
  ICTP Lect. Notes Ser.} {\bf 14} (2003)  317--413,
  \href{http://arxiv.org/abs/hep-ph/0210162}{{\tt arXiv:hep-ph/0210162}}.

\bibitem{Brandenberger:2003vk}
R.~H. Brandenberger, ``{Lectures on the theory of cosmological
  perturbations},'' \href{http://dx.doi.org/10.1007/978-3-540-40918-2_5}{{\em
  Lect. Notes Phys.} {\bf 646} (2004)  127--167},
  \href{http://arxiv.org/abs/hep-th/0306071}{{\tt arXiv:hep-th/0306071}}.

\bibitem{Straumann:2005mz}
N.~Straumann, ``{From primordial quantum fluctuations to the anisotropies of
  the cosmic microwave background radiation},''
  \href{http://dx.doi.org/10.1002/andp.200610212}{{\em Annalen Phys.} {\bf 15}
  (2006)  701--847}, \href{http://arxiv.org/abs/hep-ph/0505249}{{\tt
  arXiv:hep-ph/0505249}}.

\bibitem{Martin:2013tda}
J.~Martin, C.~Ringeval, and V.~Vennin, ``{Encyclop\ae{}dia Inflationaris}:
  {Opiparous Edition},''
  \href{http://dx.doi.org/10.1016/j.dark.2024.101653}{{\em Phys. Dark Univ.}
  {\bf 5-6} (2014)  75--235}, \href{http://arxiv.org/abs/1303.3787}{{\tt
  arXiv:1303.3787 [astro-ph.CO]}}.

\bibitem{Gron:2018rtj}
{\O}.~Gr{\o}n, ``{Predictions of Spectral Parameters by Several Inflationary
  Universe Models in Light of the Planck Results},''
  \href{http://dx.doi.org/10.3390/universe4020015}{{\em Universe} {\bf 4}
  (2018) no.~2, 15}.

\bibitem{Dimopoulos:2001ix}
K.~Dimopoulos and J.~W.~F. Valle, ``{Modeling quintessential inflation},''
  \href{http://dx.doi.org/10.1016/S0927-6505(02)00115-9}{{\em Astropart. Phys.}
  {\bf 18} (2002)  287--306}, \href{http://arxiv.org/abs/astro-ph/0111417}{{\tt
  arXiv:astro-ph/0111417}}.

\bibitem{Tsujikawa:2013fta}
S.~Tsujikawa, ``{Quintessence: A Review},''
  \href{http://dx.doi.org/10.1088/0264-9381/30/21/214003}{{\em Class. Quant.
  Grav.} {\bf 30} (2013)  214003}, \href{http://arxiv.org/abs/1304.1961}{{\tt
  arXiv:1304.1961 [gr-qc]}}.

\bibitem{deHaro:2021swo}
J.~de~Haro and L.~A. Sal{\'o}, ``{A Review of Quintessential Inflation},''
  \href{http://dx.doi.org/10.3390/galaxies9040073}{{\em Galaxies} {\bf 9}
  (2021) no.~4, 73}, \href{http://arxiv.org/abs/2108.11144}{{\tt
  arXiv:2108.11144 [gr-qc]}}.

\bibitem{Johri:2003rh}
V.~B. Johri, ``{Phantom cosmologies},''
  \href{http://dx.doi.org/10.1103/PhysRevD.70.041303}{{\em Phys. Rev. D} {\bf
  70} (2004)  041303}, \href{http://arxiv.org/abs/astro-ph/0311293}{{\tt
  arXiv:astro-ph/0311293}}.

\bibitem{Singh:2003vx}
P.~Singh, M.~Sami, and N.~Dadhich, ``{Cosmological dynamics of phantom
  field},'' \href{http://dx.doi.org/10.1103/PhysRevD.68.023522}{{\em Phys. Rev.
  D} {\bf 68} (2003)  023522}, \href{http://arxiv.org/abs/hep-th/0305110}{{\tt
  arXiv:hep-th/0305110}}.

\bibitem{Astashenok:2012tv}
A.~V. Astashenok, S.~Nojiri, S.~D. Odintsov, and A.~V. Yurov, ``{Phantom
  Cosmology without Big Rip Singularity},''
  \href{http://dx.doi.org/10.1016/j.physletb.2012.02.039}{{\em Phys. Lett. B}
  {\bf 709} (2012)  396--403}, \href{http://arxiv.org/abs/1201.4056}{{\tt
  arXiv:1201.4056 [gr-qc]}}.

\bibitem{Gomez-Valent:2024tdb}
A.~Gomez-Valent and J.~Sol{\`a}~Peracaula, ``{Phantom Matter: A Challenging
  Solution to the Cosmological Tensions},''
  \href{http://dx.doi.org/10.3847/1538-4357/ad7a62}{{\em Astrophys. J.} {\bf
  975} (2024) no.~1, 64}, \href{http://arxiv.org/abs/2404.18845}{{\tt
  arXiv:2404.18845 [astro-ph.CO]}}.

\bibitem{Cai:2009zp}
Y.-F. Cai, E.~N. Saridakis, M.~R. Setare, and J.-Q. Xia, ``{Quintom Cosmology:
  Theoretical implications and observations},''
  \href{http://dx.doi.org/10.1016/j.physrep.2010.04.001}{{\em Phys. Rept.} {\bf
  493} (2010)  1--60}, \href{http://arxiv.org/abs/0909.2776}{{\tt
  arXiv:0909.2776 [hep-th]}}.

\bibitem{Chervon:2013btx}
S.~V. Chervon, ``{Chiral Cosmological Models: Dark Sector Fields
  Description},'' {\em Quant. Matt.} {\bf 2} (2013)  71--82,
  \href{http://arxiv.org/abs/1403.7452}{{\tt arXiv:1403.7452 [gr-qc]}}.

\bibitem{Chervon:2019nwq}
S.~V. Chervon, I.~V. Fomin, E.~O. Pozdeeva, M.~Sami, and S.~Y. Vernov,
  ``{Superpotential method for chiral cosmological models connected with
  modified gravity},''
  \href{http://dx.doi.org/10.1103/PhysRevD.100.063522}{{\em Phys. Rev. D} {\bf
  100} (2019) no.~6, 063522}, \href{http://arxiv.org/abs/1904.11264}{{\tt
  arXiv:1904.11264 [gr-qc]}}.

\bibitem{Paliathanasis:2018vru}
A.~Paliathanasis, G.~Leon, and S.~Pan, ``{Exact Solutions in Chiral
  Cosmology},'' \href{http://dx.doi.org/10.1007/s10714-019-2594-2}{{\em Gen.
  Rel. Grav.} {\bf 51} (2019) no.~9, 106},
  \href{http://arxiv.org/abs/1811.10038}{{\tt arXiv:1811.10038 [gr-qc]}}.

\bibitem{Fomin:2021snm}
I.~V. Fomin and S.~V. Chervon, ``{New method of exponential potentials
  reconstruction based on given scale factor in phantonical two-field
  models},'' \href{http://dx.doi.org/10.1088/1475-7516/2022/04/025}{{\em JCAP}
  {\bf 04} (2022) no.~04, 025}, \href{http://arxiv.org/abs/2112.09359}{{\tt
  arXiv:2112.09359 [gr-qc]}}.

\bibitem{Gibbons:2002md}
G.~W. Gibbons, ``{Cosmological evolution of the rolling tachyon},''
  \href{http://dx.doi.org/10.1016/S0370-2693(02)01881-6}{{\em Phys. Lett. B}
  {\bf 537} (2002)  1--4}, \href{http://arxiv.org/abs/hep-th/0204008}{{\tt
  arXiv:hep-th/0204008}}.

\bibitem{Barbosa-Cendejas:2017pbo}
N.~Barbosa-Cendejas, J.~De-Santiago, G.~German, J.~C. Hidalgo, and R.~R.
  Mora-Luna, ``{Theoretical and observational constraints on Tachyon
  Inflation},'' \href{http://dx.doi.org/10.1088/1475-7516/2018/03/015}{{\em
  JCAP} {\bf 03} (2018)  015}, \href{http://arxiv.org/abs/1711.06693}{{\tt
  arXiv:1711.06693 [astro-ph.CO]}}.

\bibitem{Nautiyal:2018lyq}
A.~Nautiyal, ``{Reheating constraints on Tachyon Inflation},''
  \href{http://dx.doi.org/10.1103/PhysRevD.98.103531}{{\em Phys. Rev. D} {\bf
  98} (2018) no.~10, 103531}, \href{http://arxiv.org/abs/1806.03081}{{\tt
  arXiv:1806.03081 [astro-ph.CO]}}.

\bibitem{Aguirregabiria:2004rc}
J.~M. Aguirregabiria, L.~P. Chimento, and R.~Lazkoz, ``{Quintessence as
  k-essence},'' \href{http://dx.doi.org/10.1016/j.physletb.2005.10.011}{{\em
  Phys. Lett. B} {\bf 631} (2005)  93--99},
  \href{http://arxiv.org/abs/astro-ph/0411258}{{\tt arXiv:astro-ph/0411258}}.

\bibitem{Babichev:2007dw}
E.~Babichev, V.~Mukhanov, and A.~Vikman, ``{k-Essence, superluminal
  propagation, causality and emergent geometry},''
  \href{http://dx.doi.org/10.1088/1126-6708/2008/02/101}{{\em JHEP} {\bf 02}
  (2008)  101}, \href{http://arxiv.org/abs/0708.0561}{{\tt arXiv:0708.0561
  [hep-th]}}.

\bibitem{Bose:2008ew}
N.~Bose and A.~S. Majumdar, ``{A k-essence Model Of Inflation, Dark Matter and
  Dark Energy},'' \href{http://dx.doi.org/10.1103/PhysRevD.79.103517}{{\em
  Phys. Rev. D} {\bf 79} (2009)  103517},
  \href{http://arxiv.org/abs/0812.4131}{{\tt arXiv:0812.4131 [astro-ph]}}.

\bibitem{Dinda:2023mad}
B.~R. Dinda and N.~Banerjee, ``{Constraints on the speed of sound in the
  k-essence model of dark energy},''
  \href{http://dx.doi.org/10.1140/epjc/s10052-024-12547-6}{{\em Eur. Phys. J.
  C} {\bf 84} (2024) no.~2, 177}, \href{http://arxiv.org/abs/2309.10538}{{\tt
  arXiv:2309.10538 [astro-ph.CO]}}.

\bibitem{FerreiraJunior:2023qxi}
A.~L. Ferreira~Junior, N.~Pinto-Neto, and J.~Zanelli, ``{Inflation and
  late-time accelerated expansion driven by k-essence degenerate dynamics},''
  \href{http://dx.doi.org/10.1103/PhysRevD.109.023515}{{\em Phys. Rev. D} {\bf
  109} (2024) no.~2, 023515}, \href{http://arxiv.org/abs/2311.01456}{{\tt
  arXiv:2311.01456 [hep-th]}}.

\bibitem{Sotiriou:2008rp}
T.~P. Sotiriou and V.~Faraoni, ``{f(R) Theories Of Gravity},''
  \href{http://dx.doi.org/10.1103/RevModPhys.82.451}{{\em Rev. Mod. Phys.} {\bf
  82} (2010)  451--497}, \href{http://arxiv.org/abs/0805.1726}{{\tt
  arXiv:0805.1726 [gr-qc]}}.

\bibitem{Nojiri:2010wj}
S.~Nojiri and S.~D. Odintsov, ``{Unified cosmic history in modified gravity:
  from F(R) theory to Lorentz non-invariant models},''
  \href{http://dx.doi.org/10.1016/j.physrep.2011.04.001}{{\em Phys. Rept.} {\bf
  505} (2011)  59--144}, \href{http://arxiv.org/abs/1011.0544}{{\tt
  arXiv:1011.0544 [gr-qc]}}.

\bibitem{DeFelice:2010aj}
A.~De~Felice and S.~Tsujikawa, ``{f(R) theories},''
  \href{http://dx.doi.org/10.12942/lrr-2010-3}{{\em Living Rev. Rel.} {\bf 13}
  (2010)  3}, \href{http://arxiv.org/abs/1002.4928}{{\tt arXiv:1002.4928
  [gr-qc]}}.

\bibitem{DeFelice:2011bh}
A.~De~Felice and S.~Tsujikawa, ``{Conditions for the cosmological viability of
  the most general scalar-tensor theories and their applications to extended
  Galileon dark energy models},''
  \href{http://dx.doi.org/10.1088/1475-7516/2012/02/007}{{\em JCAP} {\bf 02}
  (2012)  007}, \href{http://arxiv.org/abs/1110.3878}{{\tt arXiv:1110.3878
  [gr-qc]}}.

\bibitem{DeFelice:2011zh}
A.~De~Felice and S.~Tsujikawa, ``{Primordial non-Gaussianities in general
  modified gravitational models of inflation},''
  \href{http://dx.doi.org/10.1088/1475-7516/2011/04/029}{{\em JCAP} {\bf 04}
  (2011)  029}, \href{http://arxiv.org/abs/1103.1172}{{\tt arXiv:1103.1172
  [astro-ph.CO]}}.

\bibitem{DeFelice:2011jm}
A.~De~Felice, S.~Tsujikawa, J.~Elliston, and R.~Tavakol, ``{Chaotic inflation
  in modified gravitational theories},''
  \href{http://dx.doi.org/10.1088/1475-7516/2011/08/021}{{\em JCAP} {\bf 08}
  (2011)  021}, \href{http://arxiv.org/abs/1105.4685}{{\tt arXiv:1105.4685
  [astro-ph.CO]}}.

\bibitem{Capozziello:2011et}
S.~Capozziello and M.~De~Laurentis, ``{Extended Theories of Gravity},''
  \href{http://dx.doi.org/10.1016/j.physrep.2011.09.003}{{\em Phys. Rept.} {\bf
  509} (2011)  167--321}, \href{http://arxiv.org/abs/1108.6266}{{\tt
  arXiv:1108.6266 [gr-qc]}}.

\bibitem{Nojiri:2017ncd}
S.~Nojiri, S.~D. Odintsov, and V.~K. Oikonomou, ``{Modified Gravity Theories on
  a Nutshell: Inflation, Bounce and Late-time Evolution},''
  \href{http://dx.doi.org/10.1016/j.physrep.2017.06.001}{{\em Phys. Rept.} {\bf
  692} (2017)  1--104}, \href{http://arxiv.org/abs/1705.11098}{{\tt
  arXiv:1705.11098 [gr-qc]}}.

\bibitem{Ishak:2018his}
M.~Ishak, ``{Testing General Relativity in Cosmology},''
  \href{http://dx.doi.org/10.1007/s41114-018-0017-4}{{\em Living Rev. Rel.}
  {\bf 22} (2019) no.~1, 1}, \href{http://arxiv.org/abs/1806.10122}{{\tt
  arXiv:1806.10122 [astro-ph.CO]}}.

\bibitem{CANTATA:2021asi}
{\bf CANTATA} Collaboration, Y.~Akrami {\em et al.},
  \href{http://dx.doi.org/10.1007/978-3-030-83715-0}{{\em {Modified Gravity and
  Cosmology. An Update by the CANTATA Network}}}.
\newblock Springer, 2021.
\newblock \href{http://arxiv.org/abs/2105.12582}{{\tt arXiv:2105.12582
  [gr-qc]}}.

\bibitem{Bahamonde:2021gfp}
S.~Bahamonde, K.~F. Dialektopoulos, C.~Escamilla-Rivera, G.~Farrugia, V.~Gakis,
  M.~Hendry, M.~Hohmann, J.~Levi~Said, J.~Mifsud, and E.~Di~Valentino,
  ``{Teleparallel gravity: from theory to cosmology},''
  \href{http://dx.doi.org/10.1088/1361-6633/ac9cef}{{\em Rept. Prog. Phys.}
  {\bf 86} (2023) no.~2, 026901}, \href{http://arxiv.org/abs/2106.13793}{{\tt
  arXiv:2106.13793 [gr-qc]}}.

\bibitem{Odintsov:2023weg}
S.~D. Odintsov, V.~K. Oikonomou, I.~Giannakoudi, F.~P. Fronimos, and E.~C.
  Lymperiadou, ``{Recent Advances in Inflation},''
  \href{http://dx.doi.org/10.3390/sym15091701}{{\em Symmetry} {\bf 15} (2023)
  no.~9, 1701}, \href{http://arxiv.org/abs/2307.16308}{{\tt arXiv:2307.16308
  [gr-qc]}}.

\bibitem{Copeland:2006wr}
E.~J. Copeland, M.~Sami, and S.~Tsujikawa, ``{Dynamics of dark energy},''
  \href{http://dx.doi.org/10.1142/S021827180600942X}{{\em Int. J. Mod. Phys. D}
  {\bf 15} (2006)  1753--1936}, \href{http://arxiv.org/abs/hep-th/0603057}{{\tt
  arXiv:hep-th/0603057}}.

\bibitem{Bahamonde:2017ize}
S.~Bahamonde, C.~G. B{\"o}hmer, S.~Carloni, E.~J. Copeland, W.~Fang, and
  N.~Tamanini, ``{Dynamical systems applied to cosmology: dark energy and
  modified gravity},''
  \href{http://dx.doi.org/10.1016/j.physrep.2018.09.001}{{\em Phys. Rept.} {\bf
  775-777} (2018)  1--122}, \href{http://arxiv.org/abs/1712.03107}{{\tt
  arXiv:1712.03107 [gr-qc]}}.

\bibitem{Boehmer:2022wln}
C.~G. Boehmer, E.~Jensko, and R.~Lazkoz, ``{Cosmological dynamical systems in
  modified gravity},''
  \href{http://dx.doi.org/10.1140/epjc/s10052-022-10412-y}{{\em Eur. Phys. J.
  C} {\bf 82} no.~6, 500}, \href{http://arxiv.org/abs/2201.09588}{{\tt
  arXiv:2201.09588 [gr-qc]}}.

\bibitem{Alho:2023pkl}
A.~Alho and C.~Uggla, ``{Quintessential {\ensuremath{\alpha}}-attractor
  inflation: a dynamical systems analysis},''
  \href{http://dx.doi.org/10.1088/1475-7516/2023/11/083}{{\em JCAP} {\bf 11}
  (2023)  083}, \href{http://arxiv.org/abs/2306.15326}{{\tt arXiv:2306.15326
  [gr-qc]}}.

\bibitem{Chakraborty:2024zga}
S.~Chakraborty, S.~Mishra, and S.~Chakraborty, ``{Dynamical system analysis of
  quintessence dark energy model},''
  \href{http://dx.doi.org/10.1142/S0219887824502505}{{\em Int. J. Geom. Meth.
  Mod. Phys.} {\bf 22} (2025) no.~01, 2450250},
  \href{http://arxiv.org/abs/2406.10692}{{\tt arXiv:2406.10692 [gr-qc]}}.

\bibitem{Duchaniya:2024vvc}
L.~K. Duchaniya, B.~Mishra, I.~V. Fomin, and S.~V. Chervon, ``{Dynamical system
  analysis in modified Galileon cosmology},''
  \href{http://dx.doi.org/10.1088/1361-6382/ad8a13}{{\em Class. Quant. Grav.}
  {\bf 41} (2024) no.~23, 235016}, \href{http://arxiv.org/abs/2407.11428}{{\tt
  arXiv:2407.11428 [gr-qc]}}.

\bibitem{deRitis:1990ba}
R.~de~Ritis, G.~Marmo, G.~Platania, C.~Rubano, P.~Scudellaro, and
  C.~Stornaiolo, ``{New approach to find exact solutions for cosmological
  models with a scalar field},''
  \href{http://dx.doi.org/10.1103/PhysRevD.42.1091}{{\em Phys. Rev. D} {\bf 42}
  (1990)  1091--1097}.

\bibitem{Capozziello:1996bi}
S.~Capozziello, R.~De~Ritis, C.~Rubano, and P.~Scudellaro, ``{Noether
  symmetries in cosmology},'' \href{http://dx.doi.org/10.1007/BF02742992}{{\em
  Riv. Nuovo Cim.} {\bf 19N4} (1996)  1--114}.

\bibitem{Chimento:2002gb}
L.~P. Chimento, ``{Symmetry and inflation},''
  \href{http://dx.doi.org/10.1103/PhysRevD.65.063517}{{\em Phys. Rev. D} {\bf
  65} (2002)  063517}.

\bibitem{Capozziello:2013bma}
S.~Capozziello and M.~Roshan, ``{Exact cosmological solutions from Hojman
  conservation quantities},''
  \href{http://dx.doi.org/10.1016/j.physletb.2013.08.047}{{\em Phys. Lett. B}
  {\bf 726} (2013)  471--480}, \href{http://arxiv.org/abs/1308.3910}{{\tt
  arXiv:1308.3910 [gr-qc]}}.

\bibitem{Tsamparlis:2018nyo}
M.~Tsamparlis and A.~Paliathanasis, ``{Symmetries of Differential Equations in
  Cosmology},'' \href{http://dx.doi.org/10.3390/sym10070233}{{\em Symmetry}
  {\bf 10} (2018) no.~7, 233}, \href{http://arxiv.org/abs/1806.05888}{{\tt
  arXiv:1806.05888 [gr-qc]}}.

\bibitem{Giacomini:2021xsx}
A.~Giacomini, E.~Gonz{\'a}lez, G.~Leon, and A.~Paliathanasis, ``{Variational
  symmetries and superintegrability in multifield cosmology},''
  \href{http://dx.doi.org/10.1103/PhysRevD.105.044010}{{\em Phys. Rev. D} {\bf
  105} (2022) no.~4, 044010}, \href{http://arxiv.org/abs/2104.13649}{{\tt
  arXiv:2104.13649 [gr-qc]}}.

\bibitem{Bajardi:2022ypn}
F.~Bajardi and S.~Capozziello,
  \href{http://dx.doi.org/10.1017/9781009208727}{{\em {Noether Symmetries in
  Theories of Gravity}}}.
\newblock Cambridge Monographs on Mathematical Physics. Cambridge University
  Press, 11, 2022.

\bibitem{Bhaumik:2022adi}
R.~Bhaumik, S.~Dutta, and S.~Chakraborty, ``{Noether symmetry analysis in
  chameleon field cosmology},''
  \href{http://dx.doi.org/10.1142/S0217751X2250018X}{{\em Int. J. Mod. Phys. A}
  {\bf 37} (2022) no.~05, 2250018}, \href{http://arxiv.org/abs/2302.14333}{{\tt
  arXiv:2302.14333 [gr-qc]}}.

\bibitem{Piedipalumbo:2023dzg}
E.~Piedipalumbo, S.~Vignolo, P.~Feola, and S.~Capozziello, ``{Interacting
  quintessence cosmology from Noether symmetries: Comparing theoretical
  predictions with observational data},''
  \href{http://dx.doi.org/10.1016/j.dark.2023.101274}{{\em Phys. Dark Univ.}
  {\bf 42} (2023)  101274}, \href{http://arxiv.org/abs/2307.02355}{{\tt
  arXiv:2307.02355 [gr-qc]}}.

\bibitem{Sousa:2023unz}
T.~Sousa, D.~J. Bartlett, H.~Desmond, and P.~G. Ferreira, ``{Optimal
  inflationary potentials},''
  \href{http://dx.doi.org/10.1103/PhysRevD.109.083524}{{\em Phys. Rev. D} {\bf
  109} (2024) no.~8, 083524}, \href{http://arxiv.org/abs/2310.16786}{{\tt
  arXiv:2310.16786 [astro-ph.CO]}}.

\bibitem{Planck:2018vyg}
{\bf Planck} Collaboration, N.~Aghanim {\em et al.}, ``{Planck 2018 results.
  VI. Cosmological parameters},''
  \href{http://dx.doi.org/10.1051/0004-6361/201833910}{{\em Astron. Astrophys.}
  {\bf 641} (2020)  A6}, \href{http://arxiv.org/abs/1807.06209}{{\tt
  arXiv:1807.06209 [astro-ph.CO]}}. [Erratum: Astron.Astrophys. 652, C4
  (2021)].

\bibitem{Galloni:2022mok}
G.~Galloni, N.~Bartolo, S.~Matarrese, M.~Migliaccio, A.~Ricciardone, and
  N.~Vittorio, ``{Updated constraints on amplitude and tilt of the tensor
  primordial spectrum},''
  \href{http://dx.doi.org/10.1088/1475-7516/2023/04/062}{{\em JCAP} {\bf 04}
  (2023)  062}, \href{http://arxiv.org/abs/2208.00188}{{\tt arXiv:2208.00188
  [astro-ph.CO]}}.

\bibitem{ACT:2025fju}
{\bf ACT} Collaboration, T.~Louis {\em et al.}, ``{The Atacama Cosmology
  Telescope: DR6 Power Spectra, Likelihoods and $\Lambda$CDM Parameters},''
  \href{http://arxiv.org/abs/2503.14452}{{\tt arXiv:2503.14452 [astro-ph.CO]}}.

\bibitem{ACT:2025tim}
{\bf ACT} Collaboration, E.~Calabrese {\em et al.}, ``{The Atacama Cosmology
  Telescope: DR6 Constraints on Extended Cosmological Models},''
  \href{http://arxiv.org/abs/2503.14454}{{\tt arXiv:2503.14454 [astro-ph.CO]}}.

\bibitem{Chervon:2017kgn}
S.~V. Chervon, I.~V. Fomin, and A.~Beesham, ``{The method of generating
  functions in exact scalar field inflationary cosmology},''
  \href{http://dx.doi.org/10.1140/epjc/s10052-018-5795-z}{{\em Eur. Phys. J. C}
  {\bf 78} (2018) no.~4, 301}, \href{http://arxiv.org/abs/1704.08712}{{\tt
  arXiv:1704.08712 [gr-qc]}}.

\bibitem{Rugh:2000ji}
S.~E. Rugh and H.~Zinkernagel, ``{The Quantum vacuum and the cosmological
  constant problem},''
  \href{http://dx.doi.org/10.1016/S1355-2198(02)00033-3}{{\em Stud. Hist. Phil.
  Sci. B} {\bf 33} (2002)  663--705},
  \href{http://arxiv.org/abs/hep-th/0012253}{{\tt arXiv:hep-th/0012253}}.

\bibitem{Padmanabhan:2002ji}
T.~Padmanabhan, ``{Cosmological constant: The Weight of the vacuum},''
  \href{http://dx.doi.org/10.1016/S0370-1573(03)00120-0}{{\em Phys. Rept.} {\bf
  380} (2003)  235--320}, \href{http://arxiv.org/abs/hep-th/0212290}{{\tt
  arXiv:hep-th/0212290}}.

\bibitem{Sahni:2002kh}
V.~Sahni, ``{The Cosmological constant problem and quintessence},''
  \href{http://dx.doi.org/10.1088/0264-9381/19/13/304}{{\em Class. Quant.
  Grav.} {\bf 19} (2002)  3435--3448},
  \href{http://arxiv.org/abs/astro-ph/0202076}{{\tt arXiv:astro-ph/0202076}}.

\bibitem{Nobbenhuis:2004wn}
S.~Nobbenhuis, ``{Categorizing different approaches to the cosmological
  constant problem},'' \href{http://dx.doi.org/10.1007/s10701-005-9042-8}{{\em
  Found. Phys.} {\bf 36} (2006)  613--680},
  \href{http://arxiv.org/abs/gr-qc/0411093}{{\tt arXiv:gr-qc/0411093}}.

\bibitem{Martin:2012bt}
J.~Martin, ``{Everything You Always Wanted To Know About The Cosmological
  Constant Problem (But Were Afraid To Ask)},''
  \href{http://dx.doi.org/10.1016/j.crhy.2012.04.008}{{\em Comptes Rendus
  Physique} {\bf 13} (2012)  566--665},
  \href{http://arxiv.org/abs/1205.3365}{{\tt arXiv:1205.3365 [astro-ph.CO]}}.

\bibitem{Sola:2013gha}
J.~Sola, ``{Cosmological constant and vacuum energy: old and new ideas},''
  \href{http://dx.doi.org/10.1088/1742-6596/453/1/012015}{{\em J. Phys. Conf.
  Ser.} {\bf 453} (2013)  012015}, \href{http://arxiv.org/abs/1306.1527}{{\tt
  arXiv:1306.1527 [gr-qc]}}.

\bibitem{Liddle:1994dx}
A.~R. Liddle, P.~Parsons, and J.~D. Barrow, ``{Formalizing the slow roll
  approximation in inflation},''
  \href{http://dx.doi.org/10.1103/PhysRevD.50.7222}{{\em Phys. Rev. D} {\bf 50}
  (1994)  7222--7232}, \href{http://arxiv.org/abs/astro-ph/9408015}{{\tt
  arXiv:astro-ph/9408015}}.

\bibitem{Schwarz:2001vv}
D.~J. Schwarz, C.~A. Terrero-Escalante, and A.~A. Garcia, ``{Higher order
  corrections to primordial spectra from cosmological inflation},''
  \href{http://dx.doi.org/10.1016/S0370-2693(01)01036-X}{{\em Phys. Lett. B}
  {\bf 517} (2001)  243--249},
  \href{http://arxiv.org/abs/astro-ph/0106020}{{\tt arXiv:astro-ph/0106020}}.

\bibitem{Leach:2002ar}
S.~M. Leach, A.~R. Liddle, J.~Martin, and D.~J. Schwarz, ``{Cosmological
  parameter estimation and the inflationary cosmology},''
  \href{http://dx.doi.org/10.1103/PhysRevD.66.023515}{{\em Phys. Rev. D} {\bf
  66} (2002)  023515}, \href{http://arxiv.org/abs/astro-ph/0202094}{{\tt
  arXiv:astro-ph/0202094}}.

\bibitem{Kinney:2005vj}
W.~H. Kinney, ``{Horizon crossing and inflation with large eta},''
  \href{http://dx.doi.org/10.1103/PhysRevD.72.023515}{{\em Phys. Rev. D} {\bf
  72} (2005)  023515}, \href{http://arxiv.org/abs/gr-qc/0503017}{{\tt
  arXiv:gr-qc/0503017}}.

\bibitem{Zhuravlev:1998ff}
V.~M. Zhuravlev, S.~V. Chervon, and V.~K. Shchigolev, ``{New classes of exact
  solutions in inflationary cosmology},''
  \href{http://dx.doi.org/10.1134/1.558649}{{\em J. Exp. Theor. Phys.} {\bf 87}
  (1998)  223--228}.

\bibitem{Pons:1988tj}
J.~M. Pons, ``{Ostrogradski Theorem for Higher Order Singular Lagrangians},''
  \href{http://dx.doi.org/10.1007/BF00401583}{{\em Lett. Math. Phys.} {\bf 17}
  (1989)  181}.

\bibitem{Baptista:2020adz}
R.~Baptista and O.~Bertolami, ``{An Ostrogradsky Instability Analysis of
  Non-minimally Coupled Weyl Connection Gravity Theories},''
  \href{http://dx.doi.org/10.1007/s10714-021-02784-5}{{\em Gen. Rel. Grav.}
  {\bf 53} (2021) no.~1, 12}, \href{http://arxiv.org/abs/2004.00544}{{\tt
  arXiv:2004.00544 [gr-qc]}}.

\bibitem{Abbott:1984fp}
L.~F. Abbott and M.~B. Wise, ``{Constraints on Generalized Inflationary
  Cosmologies},'' \href{http://dx.doi.org/10.1016/0550-3213(84)90329-8}{{\em
  Nucl. Phys. B} {\bf 244} (1984)  541--548}.

\bibitem{Lucchin:1984yf}
F.~Lucchin and S.~Matarrese, ``{Power Law Inflation},''
  \href{http://dx.doi.org/10.1103/PhysRevD.32.1316}{{\em Phys. Rev. D} {\bf 32}
  (1985)  1316}.

\bibitem{Sahni:1988zb}
V.~Sahni, ``{Scalar Field Fluctuations and Infrared Divergent States in
  Cosmological Models With Power Law Expansion},''
  \href{http://dx.doi.org/10.1088/0264-9381/5/7/002}{{\em Class. Quant. Grav.}
  {\bf 5} (1988)  L113}.

\bibitem{Motohashi:2014ppa}
H.~Motohashi, A.~A. Starobinsky, and J.~Yokoyama, ``{Inflation with a constant
  rate of roll},'' \href{http://dx.doi.org/10.1088/1475-7516/2015/09/018}{{\em
  JCAP} {\bf 09} (2015)  018}, \href{http://arxiv.org/abs/1411.5021}{{\tt
  arXiv:1411.5021 [astro-ph.CO]}}.

\bibitem{Motohashi:2017aob}
H.~Motohashi and A.~A. Starobinsky, ``{Constant-roll inflation: confrontation
  with recent observational data},''
  \href{http://dx.doi.org/10.1209/0295-5075/117/39001}{{\em EPL} {\bf 117}
  (2017) no.~3, 39001}, \href{http://arxiv.org/abs/1702.05847}{{\tt
  arXiv:1702.05847 [astro-ph.CO]}}.

\bibitem{Mishra:2018dtg}
S.~S. Mishra, V.~Sahni, and A.~V. Toporensky, ``{Initial conditions for
  Inflation in an FRW Universe},''
  \href{http://dx.doi.org/10.1103/PhysRevD.98.083538}{{\em Phys. Rev. D} {\bf
  98} (2018) no.~8, 083538}, \href{http://arxiv.org/abs/1801.04948}{{\tt
  arXiv:1801.04948 [gr-qc]}}.

\bibitem{Mazumdar:2010sa}
A.~Mazumdar and J.~Rocher, ``{Particle physics models of inflation and curvaton
  scenarios},'' \href{http://dx.doi.org/10.1016/j.physrep.2010.08.001}{{\em
  Phys. Rept.} {\bf 497} (2011)  85--215},
  \href{http://arxiv.org/abs/1001.0993}{{\tt arXiv:1001.0993 [hep-ph]}}.

\bibitem{Yamaguchi:2011kg}
M.~Yamaguchi, ``{Supergravity based inflation models: a review},''
  \href{http://dx.doi.org/10.1088/0264-9381/28/10/103001}{{\em Class. Quant.
  Grav.} {\bf 28} (2011)  103001}, \href{http://arxiv.org/abs/1101.2488}{{\tt
  arXiv:1101.2488 [astro-ph.CO]}}.

\bibitem{Basilakos:2015sza}
S.~Basilakos and J.~D. Barrow, ``{Hyperbolic Inflation in the Light of Planck
  2015 data},'' \href{http://dx.doi.org/10.1103/PhysRevD.91.103517}{{\em Phys.
  Rev. D} {\bf 91} (2015)  103517}, \href{http://arxiv.org/abs/1504.03469}{{\tt
  arXiv:1504.03469 [astro-ph.CO]}}.

\bibitem{Munoz:2014eqa}
J.~B. Munoz and M.~Kamionkowski, ``{Equation-of-State Parameter for
  Reheating},'' \href{http://dx.doi.org/10.1103/PhysRevD.91.043521}{{\em Phys.
  Rev. D} {\bf 91} (2015) no.~4, 043521},
  \href{http://arxiv.org/abs/1412.0656}{{\tt arXiv:1412.0656 [astro-ph.CO]}}.

\bibitem{Kitabayashi:2023vfe}
T.~Kitabayashi, ``{Generalized hybrid natural inflation},''
  \href{http://dx.doi.org/10.1103/PhysRevD.108.043514}{{\em Phys. Rev. D} {\bf
  108} (2023) no.~4, 043514}, \href{http://arxiv.org/abs/2305.03905}{{\tt
  arXiv:2305.03905 [hep-ph]}}.

\bibitem{Fomin:2020caa}
I.~Fomin and S.~Chervon, ``{Exact and Slow-Roll Solutions for Exponential
  Power-Law Inflation Connected with Modified Gravity and Observational
  Constraints},'' \href{http://dx.doi.org/10.3390/universe6110199}{{\em
  Universe} {\bf 6} (2020) no.~11, 199}.

\bibitem{Ketov:2021fww}
S.~V. Ketov, ``{Multi-Field versus Single-Field in the Supergravity Models of
  Inflation and Primordial Black Holes},''
  \href{http://dx.doi.org/10.3390/universe7050115}{{\em Universe} {\bf 7}
  (2021) no.~5, 115}.

\bibitem{Ivanov:2021chn}
V.~R. Ivanov, S.~V. Ketov, E.~O. Pozdeeva, and S.~Y. Vernov, ``{Analytic
  extensions of Starobinsky model of inflation},''
  \href{http://dx.doi.org/10.1088/1475-7516/2022/03/058}{{\em JCAP} {\bf 03}
  (2022) no.~03, 058}, \href{http://arxiv.org/abs/2111.09058}{{\tt
  arXiv:2111.09058 [gr-qc]}}.

\bibitem{Ketov:2024klm}
S.~V. Ketov, ``{Starobinsky inflation and swampland conjectures},''
  \href{http://dx.doi.org/10.1007/s11182-024-03318-7}{{\em Russ. Phys. J.} {\bf
  67} (2024) no.~11, 1819--1826}, \href{http://arxiv.org/abs/2406.06923}{{\tt
  arXiv:2406.06923 [hep-th]}}.

\bibitem{Pozo:2024fvo}
D.~Pozo, L.~Calvache, E.~Orozco, V.~A. Ar\'evalo, and C.~Rojas,
  ``{Observational predictions of some inflationary models},''
  \href{http://dx.doi.org/10.1016/j.nuclphysb.2024.116726}{{\em Nucl. Phys. B}
  {\bf 1009} (2024)  116726}, \href{http://arxiv.org/abs/2401.00835}{{\tt
  arXiv:2401.00835 [gr-qc]}}.

\bibitem{Liddle:2003as}
A.~R. Liddle and S.~M. Leach, ``{How long before the end of inflation were
  observable perturbations produced?},''
  \href{http://dx.doi.org/10.1103/PhysRevD.68.103503}{{\em Phys. Rev. D} {\bf
  68} (2003)  103503}, \href{http://arxiv.org/abs/astro-ph/0305263}{{\tt
  arXiv:astro-ph/0305263}}.

\bibitem{German:2022sjd}
G.~Germ{\'a}n, R.~G. Quaglia, and A.~M.~M. Colorado, ``{Model independent
  bounds for the number of e-folds during the evolution of the universe},''
  \href{http://dx.doi.org/10.1088/1475-7516/2023/03/004}{{\em JCAP} {\bf 03}
  (2023)  004}, \href{http://arxiv.org/abs/2212.03730}{{\tt arXiv:2212.03730
  [gr-qc]}}.

\bibitem{DiMarco:2024yzn}
A.~D. Di~Marco, E.~Orazi, and G.~Pradisi, ``{Introduction to the Number of
  e-Folds in Slow-Roll Inflation},''
  \href{http://dx.doi.org/10.3390/universe10070284}{{\em Universe} {\bf 10}
  (2024) no.~7, 284}, \href{http://arxiv.org/abs/2408.01854}{{\tt
  arXiv:2408.01854 [astro-ph.CO]}}.

\bibitem{Lyth:1996im}
D.~H. Lyth, ``{What would we learn by detecting a gravitational wave signal in
  the cosmic microwave background anisotropy?},''
  \href{http://dx.doi.org/10.1103/PhysRevLett.78.1861}{{\em Phys. Rev. Lett.}
  {\bf 78} (1997)  1861--1863}, \href{http://arxiv.org/abs/hep-ph/9606387}{{\tt
  arXiv:hep-ph/9606387}}.

\bibitem{Efstathiou:2005tq}
G.~Efstathiou and K.~J. Mack, ``{The Lyth bound revisited},''
  \href{http://dx.doi.org/10.1088/1475-7516/2005/05/008}{{\em JCAP} {\bf 05}
  (2005)  008}, \href{http://arxiv.org/abs/astro-ph/0503360}{{\tt
  arXiv:astro-ph/0503360}}.

\bibitem{DiMarco:2017ihz}
A.~Di~Marco, ``{Lyth Bound, eternal inflation and future cosmological
  missions},'' \href{http://dx.doi.org/10.1103/PhysRevD.96.023511}{{\em Phys.
  Rev. D} {\bf 96} (2017) no.~2, 023511},
  \href{http://arxiv.org/abs/1706.04144}{{\tt arXiv:1706.04144 [astro-ph.CO]}}.

\bibitem{Scalisi:2018eaz}
M.~Scalisi and I.~Valenzuela, ``{Swampland distance conjecture, inflation and
  $\alpha$-attractors},'' \href{http://dx.doi.org/10.1007/JHEP08(2019)160}{{\em
  JHEP} {\bf 08} (2019)  160}, \href{http://arxiv.org/abs/1812.07558}{{\tt
  arXiv:1812.07558 [hep-th]}}.

\bibitem{Palti:2019pca}
E.~Palti, ``{The Swampland: Introduction and Review},''
  \href{http://dx.doi.org/10.1002/prop.201900037}{{\em Fortsch. Phys.} {\bf 67}
  (2019) no.~6, 1900037}, \href{http://arxiv.org/abs/1903.06239}{{\tt
  arXiv:1903.06239 [hep-th]}}.

\bibitem{vanBeest:2021lhn}
M.~van Beest, J.~Calder{\'o}n-Infante, D.~Mirfendereski, and I.~Valenzuela,
  ``{Lectures on the Swampland Program in String Compactifications},''
  \href{http://dx.doi.org/10.1016/j.physrep.2022.09.002}{{\em Phys. Rept.} {\bf
  989} (2022)  1--50}, \href{http://arxiv.org/abs/2102.01111}{{\tt
  arXiv:2102.01111 [hep-th]}}.

\bibitem{Obied:2018sgi}
G.~Obied, H.~Ooguri, L.~Spodyneiko, and C.~Vafa, ``{De Sitter Space and the
  Swampland},'' \href{http://arxiv.org/abs/1806.08362}{{\tt arXiv:1806.08362
  [hep-th]}}.

\bibitem{Kehagias:2018uem}
A.~Kehagias and A.~Riotto, ``{A note on Inflation and the Swampland},''
  \href{http://dx.doi.org/10.1002/prop.201800052}{{\em Fortsch. Phys.} {\bf 66}
  (2018) no.~10, 1800052}, \href{http://arxiv.org/abs/1807.05445}{{\tt
  arXiv:1807.05445 [hep-th]}}.

\bibitem{Denef:2018etk}
F.~Denef, A.~Hebecker, and T.~Wrase, ``{de Sitter swampland conjecture and the
  Higgs potential},'' \href{http://dx.doi.org/10.1103/PhysRevD.98.086004}{{\em
  Phys. Rev. D} {\bf 98} (2018) no.~8, 086004},
  \href{http://arxiv.org/abs/1807.06581}{{\tt arXiv:1807.06581 [hep-th]}}.

\bibitem{Fomin:2024xzm}
I.~V. Fomin, S.~V. Chervon, L.~K. Duchaniya, and B.~Mishra, ``{The
  scalar-torsion gravity corrections in the first-order inflationary models},''
  \href{http://dx.doi.org/10.1016/j.dark.2025.101895}{{\em Phys. Dark Univ.}
  {\bf 48} (2025)  101895}, \href{http://arxiv.org/abs/2407.14542}{{\tt
  arXiv:2407.14542 [gr-qc]}}.

\bibitem{Turner:1983he}
M.~S. Turner, ``{Coherent Scalar Field Oscillations in an Expanding
  Universe},'' \href{http://dx.doi.org/10.1103/PhysRevD.28.1243}{{\em Phys.
  Rev. D} {\bf 28} (1983)  1243}.

\bibitem{Kofman:1994rk}
L.~Kofman, A.~D. Linde, and A.~A. Starobinsky, ``{Reheating after inflation},''
  \href{http://dx.doi.org/10.1103/PhysRevLett.73.3195}{{\em Phys. Rev. Lett.}
  {\bf 73} (1994)  3195--3198}, \href{http://arxiv.org/abs/hep-th/9405187}{{\tt
  arXiv:hep-th/9405187}}.

\bibitem{Shtanov:1994ce}
Y.~Shtanov, J.~H. Traschen, and R.~H. Brandenberger, ``{Universe reheating
  after inflation},'' \href{http://dx.doi.org/10.1103/PhysRevD.51.5438}{{\em
  Phys. Rev. D} {\bf 51} (1995)  5438--5455},
  \href{http://arxiv.org/abs/hep-ph/9407247}{{\tt arXiv:hep-ph/9407247}}.

\bibitem{Kofman:1997yn}
L.~Kofman, A.~D. Linde, and A.~A. Starobinsky, ``{Towards the theory of
  reheating after inflation},''
  \href{http://dx.doi.org/10.1103/PhysRevD.56.3258}{{\em Phys. Rev. D} {\bf 56}
  (1997)  3258--3295}, \href{http://arxiv.org/abs/hep-ph/9704452}{{\tt
  arXiv:hep-ph/9704452}}.

\bibitem{Bassett:2005xm}
B.~A. Bassett, S.~Tsujikawa, and D.~Wands, ``{Inflation dynamics and
  reheating},'' \href{http://dx.doi.org/10.1103/RevModPhys.78.537}{{\em Rev.
  Mod. Phys.} {\bf 78} (2006)  537--589},
  \href{http://arxiv.org/abs/astro-ph/0507632}{{\tt arXiv:astro-ph/0507632}}.

\bibitem{Dai:2014jja}
L.~Dai, M.~Kamionkowski, and J.~Wang, ``{Reheating constraints to inflationary
  models},'' \href{http://dx.doi.org/10.1103/PhysRevLett.113.041302}{{\em Phys.
  Rev. Lett.} {\bf 113} (2014)  041302},
  \href{http://arxiv.org/abs/1404.6704}{{\tt arXiv:1404.6704 [astro-ph.CO]}}.

\bibitem{Martin:2014nya}
J.~Martin, C.~Ringeval, and V.~Vennin, ``{Observing Inflationary Reheating},''
  \href{http://dx.doi.org/10.1103/PhysRevLett.114.081303}{{\em Phys. Rev.
  Lett.} {\bf 114} (2015) no.~8, 081303},
  \href{http://arxiv.org/abs/1410.7958}{{\tt arXiv:1410.7958 [astro-ph.CO]}}.

\bibitem{Kaur:2023wos}
M.~Kaur, D.~Nandi, and S.~R. B, ``{Unifying inflationary and reheating
  solution},'' \href{http://dx.doi.org/10.1088/1475-7516/2024/05/045}{{\em
  JCAP} {\bf 05} (2024)  045}, \href{http://arxiv.org/abs/2309.10570}{{\tt
  arXiv:2309.10570 [astro-ph.CO]}}.

\bibitem{Obukhov:1993fd}
Y.~N. Obukhov, ``{Spin driven inflation},''
  \href{http://dx.doi.org/10.1016/0375-9601(93)91059-E}{{\em Phys. Lett. A}
  {\bf 182} (1993)  214--216}, \href{http://arxiv.org/abs/gr-qc/0008015}{{\tt
  arXiv:gr-qc/0008015}}.

\bibitem{Stewart:1994ts}
E.~D. Stewart, ``{Inflation, supergravity and superstrings},''
  \href{http://dx.doi.org/10.1103/PhysRevD.51.6847}{{\em Phys. Rev. D} {\bf 51}
  (1995)  6847--6853}, \href{http://arxiv.org/abs/hep-ph/9405389}{{\tt
  arXiv:hep-ph/9405389}}.

\bibitem{Dvali:1998pa}
G.~R. Dvali and S.~H.~H. Tye, ``{Brane inflation},''
  \href{http://dx.doi.org/10.1016/S0370-2693(99)00132-X}{{\em Phys. Lett. B}
  {\bf 450} (1999)  72--82}, \href{http://arxiv.org/abs/hep-ph/9812483}{{\tt
  arXiv:hep-ph/9812483}}.

\bibitem{Cicoli:2008gp}
M.~Cicoli, C.~P. Burgess, and F.~Quevedo, ``{Fibre Inflation: Observable
  Gravity Waves from IIB String Compactifications},''
  \href{http://dx.doi.org/10.1088/1475-7516/2009/03/013}{{\em JCAP} {\bf 03}
  (2009)  013}, \href{http://arxiv.org/abs/0808.0691}{{\tt arXiv:0808.0691
  [hep-th]}}.

\bibitem{Giudice:2010ka}
G.~F. Giudice and H.~M. Lee, ``{Unitarizing Higgs Inflation},''
  \href{http://dx.doi.org/10.1016/j.physletb.2010.10.035}{{\em Phys. Lett. B}
  {\bf 694} (2011)  294--300}, \href{http://arxiv.org/abs/1010.1417}{{\tt
  arXiv:1010.1417 [hep-ph]}}.

\bibitem{Kallosh:2013yoa}
R.~Kallosh, A.~Linde, and D.~Roest, ``{Superconformal Inflationary
  $\alpha$-Attractors},'' \href{http://dx.doi.org/10.1007/JHEP11(2013)198}{{\em
  JHEP} {\bf 11} (2013)  198}, \href{http://arxiv.org/abs/1311.0472}{{\tt
  arXiv:1311.0472 [hep-th]}}.

\bibitem{Ferrara:2016fwe}
S.~Ferrara and R.~Kallosh, ``{Seven-disk manifold, $\alpha$-attractors, and $B$
  modes},'' \href{http://dx.doi.org/10.1103/PhysRevD.94.126015}{{\em Phys. Rev.
  D} {\bf 94} (2016) no.~12, 126015},
  \href{http://arxiv.org/abs/1610.04163}{{\tt arXiv:1610.04163 [hep-th]}}.

\bibitem{Kallosh:2017ced}
R.~Kallosh, A.~Linde, T.~Wrase, and Y.~Yamada, ``{Maximal Supersymmetry and
  B-Mode Targets},'' \href{http://dx.doi.org/10.1007/JHEP04(2017)144}{{\em
  JHEP} {\bf 04} (2017)  144}, \href{http://arxiv.org/abs/1704.04829}{{\tt
  arXiv:1704.04829 [hep-th]}}.

\bibitem{Gunaydin:2020ric}
M.~Gunaydin, R.~Kallosh, A.~Linde, and Y.~Yamada, ``{M-theory Cosmology,
  Octonions, Error Correcting Codes},''
  \href{http://dx.doi.org/10.1007/JHEP01(2021)160}{{\em JHEP} {\bf 01} (2021)
  160}, \href{http://arxiv.org/abs/2008.01494}{{\tt arXiv:2008.01494
  [hep-th]}}.

\bibitem{Galante:2014ifa}
M.~Galante, R.~Kallosh, A.~Linde, and D.~Roest, ``{Unity of Cosmological
  Inflation Attractors},''
  \href{http://dx.doi.org/10.1103/PhysRevLett.114.141302}{{\em Phys. Rev.
  Lett.} {\bf 114} (2015) no.~14, 141302},
  \href{http://arxiv.org/abs/1412.3797}{{\tt arXiv:1412.3797 [hep-th]}}.

\bibitem{Kallosh:2022feu}
R.~Kallosh and A.~Linde, ``{Polynomial \ensuremath{\alpha}-attractors},''
  \href{http://dx.doi.org/10.1088/1475-7516/2022/04/017}{{\em JCAP} {\bf 04}
  (2022) no.~04, 017}, \href{http://arxiv.org/abs/2202.06492}{{\tt
  arXiv:2202.06492 [astro-ph.CO]}}.

\bibitem{Bhattacharya:2022akq}
S.~Bhattacharya, K.~Dutta, M.~R. Gangopadhyay, and A.~Maharana,
  ``{\ensuremath{\alpha}-attractor inflation: Models and predictions},''
  \href{http://dx.doi.org/10.1103/PhysRevD.107.103530}{{\em Phys. Rev. D} {\bf
  107} (2023) no.~10, 103530}, \href{http://arxiv.org/abs/2212.13363}{{\tt
  arXiv:2212.13363 [astro-ph.CO]}}.

\bibitem{Iacconi:2023mnw}
L.~Iacconi, M.~Fasiello, J.~V\"aliviita, and D.~Wands, ``{Novel CMB constraints
  on the \ensuremath{\alpha} parameter in alpha-attractor models},''
  \href{http://dx.doi.org/10.1088/1475-7516/2023/10/015}{{\em JCAP} {\bf 10}
  (2023)  015}, \href{http://arxiv.org/abs/2306.00918}{{\tt arXiv:2306.00918
  [astro-ph.CO]}}.

\bibitem{Bamba:2013iga}
K.~Bamba, S.~Nojiri, and S.~D. Odintsov, ``{Modified gravity: walk through
  accelerating cosmology},'' in {\em {7th Mathematical Physics Meeting}:
  {Summer School and Conference on Modern Mathematical Physics}}, pp.~19--35.
\newblock 2, 2013.
\newblock \href{http://arxiv.org/abs/1302.4831}{{\tt arXiv:1302.4831 [gr-qc]}}.

\bibitem{Fomin:2018xhq}
I.~V. Fomin, ``{Generalized Exact Solutions in the Friedmann Cosmology},''
  \href{http://dx.doi.org/10.1007/s11182-018-1468-5}{{\em Russ. Phys. J.} {\bf
  61} (2018) no.~5, 843--851}.

\bibitem{Barrow:2016qkh}
J.~D. Barrow and A.~Paliathanasis, ``{Observational Constraints on New Exact
  Inflationary Scalar-field Solutions},''
  \href{http://dx.doi.org/10.1103/PhysRevD.94.083518}{{\em Phys. Rev. D} {\bf
  94} (2016) no.~8, 083518}, \href{http://arxiv.org/abs/1609.01126}{{\tt
  arXiv:1609.01126 [gr-qc]}}.

\bibitem{Rubano:2001xi}
C.~Rubano and J.~D. Barrow, ``{Scaling solutions and reconstruction of scalar
  field potentials},'' \href{http://dx.doi.org/10.1103/PhysRevD.64.127301}{{\em
  Phys. Rev. D} {\bf 64} (2001)  127301},
  \href{http://arxiv.org/abs/gr-qc/0105037}{{\tt arXiv:gr-qc/0105037}}.

\bibitem{Guo:2006ab}
Z.-K. Guo, N.~Ohta, and Y.-Z. Zhang, ``{Parametrizations of the dark energy
  density and scalar potentials},''
  \href{http://dx.doi.org/10.1142/S0217732307022839}{{\em Mod. Phys. Lett. A}
  {\bf 22} (2007)  883--890}, \href{http://arxiv.org/abs/astro-ph/0603109}{{\tt
  arXiv:astro-ph/0603109}}.

\bibitem{Kontou:2020bta}
E.-A. Kontou and K.~Sanders, ``{Energy conditions in general relativity and
  quantum field theory},''
  \href{http://dx.doi.org/10.1088/1361-6382/ab8fcf}{{\em Class. Quant. Grav.}
  {\bf 37} (2020) no.~19, 193001}, \href{http://arxiv.org/abs/2003.01815}{{\tt
  arXiv:2003.01815 [gr-qc]}}.

\bibitem{Copeland:1997et}
E.~J. Copeland, A.~R. Liddle, and D.~Wands, ``{Exponential potentials and
  cosmological scaling solutions},''
  \href{http://dx.doi.org/10.1103/PhysRevD.57.4686}{{\em Phys. Rev. D} {\bf 57}
  (1998)  4686--4690}, \href{http://arxiv.org/abs/gr-qc/9711068}{{\tt
  arXiv:gr-qc/9711068}}.

\bibitem{Billyard:1998hv}
A.~P. Billyard, A.~A. Coley, and R.~J. van~den Hoogen, ``{The Stability of
  cosmological scaling solutions},''
  \href{http://dx.doi.org/10.1103/PhysRevD.58.123501}{{\em Phys. Rev. D} {\bf
  58} (1998)  123501}, \href{http://arxiv.org/abs/gr-qc/9805085}{{\tt
  arXiv:gr-qc/9805085}}.

\bibitem{Haro:2019peq}
J.~Haro, J.~Amor{\'o}s, and S.~Pan, ``{Scaling solutions in quintessential
  inflation},'' \href{http://dx.doi.org/10.1140/epjc/s10052-020-7950-6}{{\em
  Eur. Phys. J. C} {\bf 80} (2020) no.~5, 404},
  \href{http://arxiv.org/abs/1908.01516}{{\tt arXiv:1908.01516 [gr-qc]}}.

\bibitem{Visser:2004bf}
M.~Visser, ``{Cosmography: Cosmology without the Einstein equations},''
  \href{http://dx.doi.org/10.1007/s10714-005-0134-8}{{\em Gen. Rel. Grav.} {\bf
  37} (2005)  1541--1548}, \href{http://arxiv.org/abs/gr-qc/0411131}{{\tt
  arXiv:gr-qc/0411131}}.

\bibitem{Capozziello:2011hj}
S.~Capozziello, V.~F. Cardone, H.~Farajollahi, and A.~Ravanpak, ``{Cosmography
  in f(T)-gravity},'' \href{http://dx.doi.org/10.1103/PhysRevD.84.043527}{{\em
  Phys. Rev. D} {\bf 84} (2011)  043527},
  \href{http://arxiv.org/abs/1108.2789}{{\tt arXiv:1108.2789 [astro-ph.CO]}}.

\bibitem{Aviles:2012ay}
A.~Aviles, C.~Gruber, O.~Luongo, and H.~Quevedo, ``{Cosmography and constraints
  on the equation of state of the Universe in various parametrizations},''
  \href{http://dx.doi.org/10.1103/PhysRevD.86.123516}{{\em Phys. Rev. D} {\bf
  86} (2012)  123516}, \href{http://arxiv.org/abs/1204.2007}{{\tt
  arXiv:1204.2007 [astro-ph.CO]}}.

\bibitem{Dunsby:2015ers}
P.~K.~S. Dunsby and O.~Luongo, ``{On the theory and applications of modern
  cosmography},'' \href{http://dx.doi.org/10.1142/S0219887816300026}{{\em Int.
  J. Geom. Meth. Mod. Phys.} {\bf 13} (2016) no.~03, 1630002},
  \href{http://arxiv.org/abs/1511.06532}{{\tt arXiv:1511.06532 [gr-qc]}}.

\bibitem{Mehrabi:2021cob}
A.~Mehrabi and M.~Rezaei, ``{Cosmographic Parameters in Model-independent
  Approaches},'' \href{http://dx.doi.org/10.3847/1538-4357/ac2fff}{{\em
  Astrophys. J.} {\bf 923} (2021) no.~2, 274},
  \href{http://arxiv.org/abs/2110.14950}{{\tt arXiv:2110.14950 [astro-ph.CO]}}.

\bibitem{Hawkins:2001zx}
R.~M. Hawkins and J.~E. Lidsey, ``{The Ermakov-Pinney equation in scalar field
  cosmologies},'' \href{http://dx.doi.org/10.1103/PhysRevD.66.023523}{{\em
  Phys. Rev. D} {\bf 66} (2002)  023523},
  \href{http://arxiv.org/abs/astro-ph/0112139}{{\tt arXiv:astro-ph/0112139}}.

\bibitem{Akarsu:2013xha}
O.~Akarsu, S.~Kumar, R.~Myrzakulov, M.~Sami, and L.~Xu, ``{Cosmology with
  hybrid expansion law: scalar field reconstruction of cosmic history and
  observational constraints},''
  \href{http://dx.doi.org/10.1088/1475-7516/2014/01/022}{{\em JCAP} {\bf 01}
  (2014)  022}, \href{http://arxiv.org/abs/1307.4911}{{\tt arXiv:1307.4911
  [gr-qc]}}.

\bibitem{Chakraborty:2022evc}
S.~Chakraborty, D.~Gregoris, and B.~Mishra, ``{On the uniqueness of
  {\ensuremath{\Lambda}}CDM-like evolution for homogeneous and isotropic
  cosmology in General Relativity},''
  \href{http://dx.doi.org/10.1016/j.physletb.2023.137962}{{\em Phys. Lett. B}
  {\bf 842} (2023)  137962}, \href{http://arxiv.org/abs/2208.04596}{{\tt
  arXiv:2208.04596 [gr-qc]}}.

\bibitem{Pourojaghi:2024bxa}
S.~Pourojaghi, M.~Malekjani, and Z.~Davari, ``{{\ensuremath{\Lambda}}CDM model
  against cosmography: a possible deviation after DESI 2024},''
  \href{http://dx.doi.org/10.1093/mnras/staf037}{{\em Mon. Not. Roy. Astron.
  Soc.} {\bf 537} (2025) no.~1, 436--447},
  \href{http://arxiv.org/abs/2408.10704}{{\tt arXiv:2408.10704 [astro-ph.CO]}}.

\bibitem{BOSS:2016wmc}
{\bf BOSS} Collaboration, S.~Alam {\em et al.}, ``{The clustering of galaxies
  in the completed SDSS-III Baryon Oscillation Spectroscopic Survey:
  cosmological analysis of the DR12 galaxy sample},''
  \href{http://dx.doi.org/10.1093/mnras/stx721}{{\em Mon. Not. Roy. Astron.
  Soc.} {\bf 470} (2017) no.~3, 2617--2652},
  \href{http://arxiv.org/abs/1607.03155}{{\tt arXiv:1607.03155 [astro-ph.CO]}}.

\bibitem{Pan-STARRS1:2017jku}
{\bf Pan-STARRS1} Collaboration, D.~M. Scolnic {\em et al.}, ``{The Complete
  Light-curve Sample of Spectroscopically Confirmed SNe Ia from Pan-STARRS1 and
  Cosmological Constraints from the Combined Pantheon Sample},''
  \href{http://dx.doi.org/10.3847/1538-4357/aab9bb}{{\em Astrophys. J.} {\bf
  859} (2018) no.~2, 101}, \href{http://arxiv.org/abs/1710.00845}{{\tt
  arXiv:1710.00845 [astro-ph.CO]}}.

\bibitem{Benisty:2020otr}
D.~Benisty and D.~Staicova, ``{Testing late-time cosmic acceleration with
  uncorrelated baryon acoustic oscillation dataset},''
  \href{http://dx.doi.org/10.1051/0004-6361/202039502}{{\em Astron. Astrophys.}
  {\bf 647} (2021)  A38}, \href{http://arxiv.org/abs/2009.10701}{{\tt
  arXiv:2009.10701 [astro-ph.CO]}}.

\bibitem{Dainotti:2022wli}
M.~G. Dainotti, V.~Nielson, G.~Sarracino, E.~Rinaldi, S.~Nagataki,
  S.~Capozziello, O.~Y. Gnedin, and G.~Bargiacchi, ``{Optical and X-ray GRB
  Fundamental Planes as cosmological distance indicators},''
  \href{http://dx.doi.org/10.1093/mnras/stac1141}{{\em Mon. Not. Roy. Astron.
  Soc.} {\bf 514} (2022) no.~2, 1828--1856},
  \href{http://arxiv.org/abs/2203.15538}{{\tt arXiv:2203.15538 [astro-ph.CO]}}.

\bibitem{Cao:2021cix}
S.~Cao, J.~Ryan, and B.~Ratra, ``{Cosmological constraints from H{\,}ii
  starburst galaxy, quasar angular size, and other measurements},''
  \href{http://dx.doi.org/10.1093/mnras/stab3304}{{\em Mon. Not. Roy. Astron.
  Soc.} {\bf 509} (2022)  4745--4757},
  \href{http://arxiv.org/abs/2109.01987}{{\tt arXiv:2109.01987 [astro-ph.CO]}}.

\bibitem{Liu:2022inf}
Y.~Liu, N.~Liang, X.~Xie, Z.~Yuan, H.~Yu, and P.~Wu, ``{Gamma-Ray Burst
  Constraints on Cosmological Models from the Improved Amati Correlation},''
  \href{http://dx.doi.org/10.3847/1538-4357/ac7de5}{{\em Astrophys. J.} {\bf
  935} (2022) no.~1, 7}, \href{http://arxiv.org/abs/2207.00455}{{\tt
  arXiv:2207.00455 [astro-ph.CO]}}.

\bibitem{Pourojaghi:2022zrh}
S.~Pourojaghi, N.~F. Zabihi, and M.~Malekjani, ``{Can high-redshift Hubble
  diagrams rule out the standard model of cosmology in the context of
  cosmography?},'' \href{http://dx.doi.org/10.1103/PhysRevD.106.123523}{{\em
  Phys. Rev. D} {\bf 106} (2022) no.~12, 123523},
  \href{http://arxiv.org/abs/2212.04118}{{\tt arXiv:2212.04118 [astro-ph.CO]}}.

\bibitem{Perivolaropoulos:2021jda}
L.~Perivolaropoulos and F.~Skara, ``{Challenges for {\ensuremath{\Lambda}}CDM:
  An update},'' \href{http://dx.doi.org/10.1016/j.newar.2022.101659}{{\em New
  Astron. Rev.} {\bf 95} (2022)  101659},
  \href{http://arxiv.org/abs/2105.05208}{{\tt arXiv:2105.05208 [astro-ph.CO]}}.

\bibitem{DiValentino:2021izs}
E.~Di~Valentino, O.~Mena, S.~Pan, L.~Visinelli, W.~Yang, A.~Melchiorri, D.~F.
  Mota, A.~G. Riess, and J.~Silk, ``{In the realm of the Hubble
  tension{\textemdash}a review of solutions},''
  \href{http://dx.doi.org/10.1088/1361-6382/ac086d}{{\em Class. Quant. Grav.}
  {\bf 38} (2021) no.~15, 153001}, \href{http://arxiv.org/abs/2103.01183}{{\tt
  arXiv:2103.01183 [astro-ph.CO]}}.

\bibitem{Hu:2023jqc}
J.-P. Hu and F.-Y. Wang, ``{Hubble Tension: The Evidence of New Physics},''
  \href{http://dx.doi.org/10.3390/universe9020094}{{\em Universe} {\bf 9}
  (2023) no.~2, 94}, \href{http://arxiv.org/abs/2302.05709}{{\tt
  arXiv:2302.05709 [astro-ph.CO]}}.

\bibitem{Riess:2021jrx}
A.~G. Riess {\em et al.}, ``{A Comprehensive Measurement of the Local Value of
  the Hubble Constant with 1 km s$^{-1}$ Mpc$^{-1}$ Uncertainty from the Hubble
  Space Telescope and the SH0ES Team},''
  \href{http://dx.doi.org/10.3847/2041-8213/ac5c5b}{{\em Astrophys. J. Lett.}
  {\bf 934} (2022) no.~1, L7}, \href{http://arxiv.org/abs/2112.04510}{{\tt
  arXiv:2112.04510 [astro-ph.CO]}}.

\bibitem{Breuval:2024lsv}
L.~Breuval, A.~G. Riess, S.~Casertano, W.~Yuan, L.~M. Macri, M.~Romaniello,
  Y.~S. Murakami, D.~Scolnic, G.~S. Anand, and I.~Soszy{\'n}ski, ``{Small
  Magellanic Cloud Cepheids Observed with the Hubble Space Telescope Provide a
  New Anchor for the SH0ES Distance Ladder},''
  \href{http://dx.doi.org/10.3847/1538-4357/ad630e}{{\em Astrophys. J.} {\bf
  973} (2024) no.~1, 30}, \href{http://arxiv.org/abs/2404.08038}{{\tt
  arXiv:2404.08038 [astro-ph.CO]}}.

\bibitem{Fomin:2017vae}
I.~V. Fomin and S.~V. Chervon, ``{Exact inflation in
  Einstein{\textendash}Gauss{\textendash}Bonnet gravity},''
  \href{http://dx.doi.org/10.1134/S0202289317040090}{{\em Grav. Cosmol.} {\bf
  23} (2017) no.~4, 367--374}, \href{http://arxiv.org/abs/1704.03634}{{\tt
  arXiv:1704.03634 [gr-qc]}}.

\bibitem{Fomin:2017qta}
I.~V. Fomin and S.~V. Chervon, ``{A new approach to exact solutions
  construction in scalar cosmology with a Gauss-Bonnet term},''
  \href{http://dx.doi.org/10.1142/S0217732317501292}{{\em Mod. Phys. Lett. A}
  {\bf 32} (2017) no.~25, 1750129}, \href{http://arxiv.org/abs/1704.07786}{{\tt
  arXiv:1704.07786 [gr-qc]}}.

\bibitem{Fomin:2018typ}
I.~V. Fomin, ``{Cosmological Inflation with
  Einstein{\textendash}Gauss{\textendash}Bonnet Gravity},''
  \href{http://dx.doi.org/10.1134/S1063779618040226}{{\em Phys. Part. Nucl.}
  {\bf 49} (2018) no.~4, 525--529}.

\bibitem{Fomin:2020hfh}
I.~Fomin, ``{Gauss-Bonnet term corrections in scalar field cosmology},''
  \href{http://dx.doi.org/10.1140/epjc/s10052-020-08718-w}{{\em Eur. Phys. J.
  C} {\bf 80} (2020) no.~12, 1145}, \href{http://arxiv.org/abs/2004.08065}{{\tt
  arXiv:2004.08065 [gr-qc]}}.

\bibitem{Fomin:2017sbt}
I.~V. Fomin and S.~V. Chervon, ``{Non-minimal coupling influence on the
  deviation from de Sitter cosmological expansion},''
  \href{http://dx.doi.org/10.1140/epjc/s10052-018-6409-5}{{\em Eur. Phys. J. C}
  {\bf 78} (2018) no.~11, 918}, \href{http://arxiv.org/abs/1711.06870}{{\tt
  arXiv:1711.06870 [gr-qc]}}.

\bibitem{Fomin:2018blx}
I.~Fomin and S.~Chervon, ``{Inflation with explicit parametric connection
  between general relativity and scalar{\textendash}tensor gravity},''
  \href{http://dx.doi.org/10.1142/S0217732318501614}{{\em Mod. Phys. Lett. A}
  {\bf 33} (2018) no.~28, 1850161}, \href{http://arxiv.org/abs/1802.10462}{{\tt
  arXiv:1802.10462 [gr-qc]}}.

\bibitem{Fomin:2020woj}
I.~V. Fomin, S.~V. Chervon, and A.~V. Tsyganov, ``{Generalized scalar-tensor
  theory of gravity reconstruction from physical potentials of a scalar
  field},'' \href{http://dx.doi.org/10.1140/epjc/s10052-020-7893-y}{{\em Eur.
  Phys. J. C} {\bf 80} (2020) no.~4, 350},
  \href{http://arxiv.org/abs/2004.08544}{{\tt arXiv:2004.08544 [gr-qc]}}.

\bibitem{Fomin:2022ozv}
I.~V. Fomin, S.~V. Chervon, A.~N. Morozov, and I.~S. Golyak, ``{Relic
  gravitational waves in verified inflationary models based on the generalized
  scalar{\textendash}tensor gravity},''
  \href{http://dx.doi.org/10.1140/epjc/s10052-022-10601-9}{{\em Eur. Phys. J.
  C} {\bf 82} (2022) no.~7, 642}.

\bibitem{Chervon:2023gio}
S.~V. Chervon and I.~V. Fomin, ``{Reconstruction of Scalar-Torsion Gravity
  Theories from the Physical Potential of a Scalar Field},''
  \href{http://dx.doi.org/10.3390/sym15020291}{{\em Symmetry} {\bf 15} (2023)
  no.~2, 291}.

\bibitem{Bezrukov:2007ep}
F.~L. Bezrukov and M.~Shaposhnikov, ``{The Standard Model Higgs boson as the
  inflaton},'' \href{http://dx.doi.org/10.1016/j.physletb.2007.11.072}{{\em
  Phys. Lett. B} {\bf 659} (2008)  703--706},
  \href{http://arxiv.org/abs/0710.3755}{{\tt arXiv:0710.3755 [hep-th]}}.

\bibitem{Bezrukov:2013fka}
F.~Bezrukov, ``{The Higgs field as an inflaton},''
  \href{http://dx.doi.org/10.1088/0264-9381/30/21/214001}{{\em Class. Quant.
  Grav.} {\bf 30} (2013)  214001}, \href{http://arxiv.org/abs/1307.0708}{{\tt
  arXiv:1307.0708 [hep-ph]}}.

\bibitem{Mohammedi:2022qqj}
N.~Mohammedi, ``{On Higgs inflation in non-minimally coupled models of
  gravity},'' \href{http://dx.doi.org/10.1016/j.physletb.2022.137180}{{\em
  Phys. Lett. B} {\bf 831} (2022)  137180},
  \href{http://arxiv.org/abs/2202.05696}{{\tt arXiv:2202.05696 [hep-th]}}.

\bibitem{Fernandes:2022zrq}
P.~G.~S. Fernandes, P.~Carrilho, T.~Clifton, and D.~J. Mulryne, ``{The 4D
  Einstein{\textendash}Gauss{\textendash}Bonnet theory of gravity: a review},''
  \href{http://dx.doi.org/10.1088/1361-6382/ac500a}{{\em Class. Quant. Grav.}
  {\bf 39} (2022) no.~6, 063001}, \href{http://arxiv.org/abs/2202.13908}{{\tt
  arXiv:2202.13908 [gr-qc]}}.

\bibitem{vandeBruck:2015gjd}
C.~van~de Bruck and C.~Longden, ``{Higgs Inflation with a Gauss-Bonnet term in
  the Jordan Frame},'' \href{http://dx.doi.org/10.1103/PhysRevD.93.063519}{{\em
  Phys. Rev. D} {\bf 93} (2016) no.~6, 063519},
  \href{http://arxiv.org/abs/1512.04768}{{\tt arXiv:1512.04768 [hep-ph]}}.

\bibitem{Chakraborty:2018scm}
S.~Chakraborty, T.~Paul, and S.~SenGupta, ``{Inflation driven by
  Einstein-Gauss-Bonnet gravity},''
  \href{http://dx.doi.org/10.1103/PhysRevD.98.083539}{{\em Phys. Rev. D} {\bf
  98} (2018) no.~8, 083539}, \href{http://arxiv.org/abs/1804.03004}{{\tt
  arXiv:1804.03004 [gr-qc]}}.

\bibitem{Odintsov:2018zhw}
S.~D. Odintsov and V.~K. Oikonomou, ``{Viable Inflation in Scalar-Gauss-Bonnet
  Gravity and Reconstruction from Observational Indices},''
  \href{http://dx.doi.org/10.1103/PhysRevD.98.044039}{{\em Phys. Rev. D} {\bf
  98} (2018) no.~4, 044039}, \href{http://arxiv.org/abs/1808.05045}{{\tt
  arXiv:1808.05045 [gr-qc]}}.

\bibitem{Odintsov:2020sqy}
S.~D. Odintsov, V.~K. Oikonomou, and F.~P. Fronimos, ``{Rectifying
  Einstein-Gauss-Bonnet Inflation in View of GW170817},''
  \href{http://dx.doi.org/10.1016/j.nuclphysb.2020.115135}{{\em Nucl. Phys. B}
  {\bf 958} (2020)  115135}, \href{http://arxiv.org/abs/2003.13724}{{\tt
  arXiv:2003.13724 [gr-qc]}}.

\bibitem{Odintsov:2025kyw}
S.~D. Odintsov, V.~K. Oikonomou, and G.~S. Sharov, ``{Einstein-Gauss-Bonnet
  cosmology confronted with observations},''
  \href{http://dx.doi.org/10.1016/j.jheap.2025.100398}{{\em JHEAp} {\bf 47}
  (2025)  100398}, \href{http://arxiv.org/abs/2503.17946}{{\tt arXiv:2503.17946
  [gr-qc]}}.

\bibitem{Odintsov:2025bmp}
S.~D. Odintsov and T.~Paul, ``{ACT inflation and its influence on reheating era
  in Einstein-Gauss-Bonnet gravity},''
  \href{http://arxiv.org/abs/2508.11377}{{\tt arXiv:2508.11377 [gr-qc]}}.

\bibitem{Pozdeeva:2021iwc}
E.~O. Pozdeeva and S.~Y. Vernov, ``{Construction of inflationary scenarios with
  the Gauss{\textendash}Bonnet term and nonminimal coupling},''
  \href{http://dx.doi.org/10.1140/epjc/s10052-021-09435-8}{{\em Eur. Phys. J.
  C} {\bf 81} (2021) no.~7, 633}, \href{http://arxiv.org/abs/2104.04995}{{\tt
  arXiv:2104.04995 [gr-qc]}}.

\bibitem{Pozdeeva:2024ihc}
E.~O. Pozdeeva, M.~A. Skugoreva, A.~V. Toporensky, and S.~Y. Vernov, ``{New
  slow-roll approximations for inflation in Einstein-Gauss-Bonnet gravity},''
  \href{http://dx.doi.org/10.1088/1475-7516/2024/09/050}{{\em JCAP} {\bf 09}
  (2024)  050}, \href{http://arxiv.org/abs/2403.06147}{{\tt arXiv:2403.06147
  [gr-qc]}}.

\bibitem{Fujii:2003pa}
Y.~Fujii and K.~Maeda, \href{http://dx.doi.org/10.1017/CBO9780511535093}{{\em
  {The scalar-tensor theory of gravitation}}}.
\newblock Cambridge Monographs on Mathematical Physics. Cambridge University
  Press, 7, 2007.

\bibitem{Belinchon:2016lwr}
J.~A. Belinch{\'o}n, T.~Harko, and M.~K. Mak, ``{Exact
  scalar{\textendash}tensor cosmological models},''
  \href{http://dx.doi.org/10.1142/S0218271817500730}{{\em Int. J. Mod. Phys. D}
  {\bf 26} (2017) no.~07, 1750073}, \href{http://arxiv.org/abs/1612.05446}{{\tt
  arXiv:1612.05446 [gr-qc]}}.

\bibitem{Motohashi:2019tyj}
H.~Motohashi and A.~A. Starobinsky, ``{Constant-roll inflation in scalar-tensor
  gravity},'' \href{http://dx.doi.org/10.1088/1475-7516/2019/11/025}{{\em JCAP}
  {\bf 11} (2019)  025}, \href{http://arxiv.org/abs/1909.10883}{{\tt
  arXiv:1909.10883 [gr-qc]}}.

\bibitem{Gonzalez-Espinoza:2019ajd}
M.~Gonzalez-Espinoza, G.~Otalora, N.~Videla, and J.~Saavedra, ``{Slow-roll
  inflation in generalized scalar-torsion gravity},''
  \href{http://dx.doi.org/10.1088/1475-7516/2019/08/029}{{\em JCAP} {\bf 08}
  (2019)  029}, \href{http://arxiv.org/abs/1904.08068}{{\tt arXiv:1904.08068
  [gr-qc]}}.

\bibitem{Gonzalez-Espinoza:2020azh}
M.~Gonzalez-Espinoza and G.~Otalora, ``{Generating primordial fluctuations from
  modified teleparallel gravity with local Lorentz-symmetry breaking},''
  \href{http://dx.doi.org/10.1016/j.physletb.2020.135696}{{\em Phys. Lett. B}
  {\bf 809} (2020)  135696}, \href{http://arxiv.org/abs/2005.03753}{{\tt
  arXiv:2005.03753 [gr-qc]}}.

\bibitem{Leon:2022oyy}
G.~Leon, A.~Paliathanasis, E.~N. Saridakis, and S.~Basilakos, ``{Unified dark
  sectors in scalar-torsion theories of gravity},''
  \href{http://dx.doi.org/10.1103/PhysRevD.106.024055}{{\em Phys. Rev. D} {\bf
  106} (2022) no.~2, 024055}, \href{http://arxiv.org/abs/2203.14866}{{\tt
  arXiv:2203.14866 [gr-qc]}}.

\bibitem{Yadav:2014uoa}
A.~K. Yadav, P.~K. Srivastava, and L.~Yadav, ``{Hybrid Expansion Law for Dark
  Energy Dominated Universe in f (R,T) Gravity},''
  \href{http://dx.doi.org/10.1007/s10773-014-2368-2}{{\em Int. J. Theor. Phys.}
  {\bf 54} (2015) no.~5, 1671--1679}.

\bibitem{Mishra:2015jja}
B.~Mishra and S.~K. Tripathy, ``{Anisotropic dark energy model with a hybrid
  scale factor},'' \href{http://dx.doi.org/10.1142/S0217732315501758}{{\em Mod.
  Phys. Lett. A} {\bf 30} (2015) no.~36, 1550175},
  \href{http://arxiv.org/abs/1507.03515}{{\tt arXiv:1507.03515
  [physics.gen-ph]}}.

\bibitem{Zia:2018tss}
R.~Zia, D.~Chandra~Maurya, and A.~Pradhan, ``{Transit dark energy string
  cosmological models with perfect fluid in $F(R,T)$-gravity},''
  \href{http://dx.doi.org/10.1142/S0219887818501682}{{\em Int. J. Geom. Meth.
  Mod. Phys.} {\bf 15} (2018) no.~10, 1850168}.

\bibitem{Tripathy:2021vjt}
S.~K. Tripathy, B.~Mishra, M.~Khlopov, and S.~Ray, ``{Cosmological models with
  a hybrid scale factor},''
  \href{http://dx.doi.org/10.1142/S0218271821400058}{{\em Int. J. Mod. Phys. D}
  {\bf 30} (2021) no.~16, 2140005}, \href{http://arxiv.org/abs/2106.04368}{{\tt
  arXiv:2106.04368 [gr-qc]}}.

\bibitem{Alhallak:2022szt}
M.~Alhallak, N.~Chamoun, and M.~S. Eldaher, ``{Salvaging power-law inflation
  through warming},''
  \href{http://dx.doi.org/10.1140/epjc/s10052-023-11667-9}{{\em Eur. Phys. J.
  C} {\bf 83} (2023) no.~6, 533}, \href{http://arxiv.org/abs/2212.04935}{{\tt
  arXiv:2212.04935 [astro-ph.CO]}}.

\bibitem{Varshney:2021mvx}
G.~Varshney, A.~Pradhan, and U.~K. Sharma, ``{Bianchi type-III THDE
  quintessence model with hybrid scale factor},''
  \href{http://dx.doi.org/10.1139/cjp-2023-0201}{{\em Can. J. Phys.} {\bf 102}
  (2024) no.~3, 199--209}, \href{http://arxiv.org/abs/2108.07157}{{\tt
  arXiv:2108.07157 [physics.gen-ph]}}.

\bibitem{LIGOScientific:2017ync}
B.~P. Abbott {\em et al.}, ``{Multi-messenger Observations of a Binary Neutron
  Star Merger},'' \href{http://dx.doi.org/10.3847/2041-8213/aa91c9}{{\em
  Astrophys. J. Lett.} {\bf 848} (2017) no.~2, L12},
  \href{http://arxiv.org/abs/1710.05833}{{\tt arXiv:1710.05833 [astro-ph.HE]}}.

\bibitem{Ezquiaga:2017ekz}
J.~M. Ezquiaga and M.~Zumalac{\'a}rregui, ``{Dark Energy After GW170817: Dead
  Ends and the Road Ahead},''
  \href{http://dx.doi.org/10.1103/PhysRevLett.119.251304}{{\em Phys. Rev.
  Lett.} {\bf 119} (2017) no.~25, 251304},
  \href{http://arxiv.org/abs/1710.05901}{{\tt arXiv:1710.05901 [astro-ph.CO]}}.

\bibitem{Odintsov:2019clh}
S.~D. Odintsov and V.~K. Oikonomou, ``{Inflationary Phenomenology of Einstein
  Gauss-Bonnet Gravity Compatible with GW170817},''
  \href{http://dx.doi.org/10.1016/j.physletb.2019.134874}{{\em Phys. Lett. B}
  {\bf 797} (2019)  134874}, \href{http://arxiv.org/abs/1908.07555}{{\tt
  arXiv:1908.07555 [gr-qc]}}.

\bibitem{Odintsov:2020zkl}
S.~D. Odintsov and V.~K. Oikonomou, ``{Swampland implications of
  GW170817-compatible Einstein-Gauss-Bonnet gravity},''
  \href{http://dx.doi.org/10.1016/j.physletb.2020.135437}{{\em Phys. Lett. B}
  {\bf 805} (2020)  135437}, \href{http://arxiv.org/abs/2004.00479}{{\tt
  arXiv:2004.00479 [gr-qc]}}.

\bibitem{Shoom:2021mdj}
A.~A. Shoom, P.~K. Gupta, B.~Krishnan, A.~B. Nielsen, and C.~D. Capano,
  ``{Testing the post-Newtonian expansion with GW170817},''
  \href{http://dx.doi.org/10.1007/s10714-023-03100-z}{{\em Gen. Rel. Grav.}
  {\bf 55} (2023) no.~4, 55}, \href{http://arxiv.org/abs/2105.02191}{{\tt
  arXiv:2105.02191 [gr-qc]}}.

\end{thebibliography}\endgroup

\end{document}